%% file: paper.tex
\documentclass[times,svgnames,envcountsame,orivec,babel]{llncs}
\usepackage[english]{babel}
\input{defs}
\input{theorem-eng}

\usepackage{marginnote}

\usepackage{todonotes}
\renewcommand{\todo}[2][]{\tikzexternaldisable\@todo[#1]{#2}\tikzexternalenable}
\newcommand{\tinytodo}[2][]{\todo[caption={#2}, size=\scriptsize,
  #1]{\renewcommand{\baselinestretch}{0.5}\selectfont#2\par}}
\renewcommand{\todo}[1]{}
\renewcommand{\tinytodo}[2][]{}

\usepackage{changes}
\renewcommand{\deleted}[1]{}

\usepackage{soul}
\usepackage{boxedminipage}
{\ifhmode\par\fi\noindent\begin{boxedminipage}{\linewidth}}
{\end{boxedminipage}}
\newtheorem{fact}[definition]{Fact}
\usepackage{mathtools,amssymb,amsmath,stmaryrd,MnSymbol}

\usepackage{subcaption}
\newcommand{\short}[1]{}
\newcommand{\full}[1]{#1}

\title{Rewriting Abstract Structures: \\[.5ex]
  Materialization Explained Categorically\thanks{Partially supported
    by AFOSR.}} \author{Andrea Corradini\inst{1} \and Tobias
  Heindel\inst{2} \and Barbara K\"onig\inst{3} \and \\ Dennis
  Nolte\inst{3} \and Arend~Rensink\inst{4}} \institute{%
  Universit\`a di Pisa, Italy\\
  \email{andrea@di.unipi.it}
  \and University of Hawaii, USA\\
  \email{heindel@hawaii.edu}
  \and Universit\"at Duisburg-Essen, Germany\\
  \email{\{barbara\_koenig,dennis.nolte\}@uni-due.de}
  \and University of Twente, Netherlands\\
  \email{arend.rensink@utwente.nl} }

\usepackage{microtype}
\usepackage[utf8]{inputenc}
\usepackage[T1]{fontenc}
\usepackage{xcolor}
\usepackage{etoolbox}
\usepackage{tikz-cd}
\usepackage{times}

\usetikzlibrary{external}
\usetikzlibrary{backgrounds}


\begin{document}

\maketitle

\begin{abstract}
  The paper develops an abstract (over-approximating) semantics for
  double-pushout rewriting of graphs and graph-like objects. The focus
  is on the so-called materialization of left-hand sides from abstract
  graphs, a central concept in previous work.  The first contribution
  is an accessible, general explanation of how materializations arise
  from universal properties and categorical constructions, in
  particular partial map classifiers, in a topos. Second, we introduce
  an extension by enriching objects with annotations and give a
  precise characterization of strongest post-conditions, which are
  effectively computable under certain assumptions.
\end{abstract}

% ============================================================================ %
\section{Introduction}
\label{sec:introduction}

Abstract interpretation \cite{c:abstract-interpretation}
is a fundamental static analysis technique %
that applies not only to conventional programs %
but also to general infinite-state systems. %
Shape analysis~\cite{srw:shape-analysis-3vl}, a specific instance of
abstract interpretation, %
pioneered an approach for analyzing pointer structures %
that keeps track of information about %
the ``heap topology'', % 
e.g., out-degrees or existence of certain paths. % 
One central idea of shape analysis %
is \emph{materialization}, %
which arises as companion operation to %
summarizing distinct objects that share relevant properties. %
Materialization, a.k.a.\ partial concretization, is %
also fundamental in verification approaches based on separation
logic~%
\cite{%
  cr:relational-inductive-shape-analysis,%
  cdhy:compositional-shape-analysis,%
  lrc:shape-analysis-unstructured-sharing%
}, where it is also known as
rearrangement~\cite{h:primer-separation-logic}, %
a special case of frame inference. %
Shape analysis%
---construed in a wide sense---%
has been adapted to graph transformation \cite{r:gra-handbook}, %
a general purpose modelling language for %
systems with dynamically evolving topology, %
such as network protocols and cyber-physical systems. %
Motivated by earlier work of shape analysis for graph transformation %
\cite{%
  sww:abstract-gts,%
  bw:partner-abstraction,%
  b:cluster-abstraction-thesis,%
  br:cluster-abstraction,%
  rz:neighbourhood-abstraction-groove,%
  r:canonical-graph-shapes%
},
we want to put the materialization operation on a new footing, %
widening the scope of shape analysis. %

A natural abstraction mechanism for transition systems % 
with graphs as states %
``summarizes'' all graphs over a specific \emph{shape graph}. %
Thus a single graph is used as abstraction for %
all graphs that can be mapped homomorphically into it. %
Further annotations on shape graphs, %
such as cardinalities of preimages of its nodes %
and general first-order formulas, 
enable fine-tuning of the granularity of abstractions. % 
While these natural abstraction principles %
have been successfully applied in previous work %
\cite{%
  sww:abstract-gts,%
  bw:partner-abstraction,%
  b:cluster-abstraction-thesis,%
  br:cluster-abstraction,%
  rz:neighbourhood-abstraction-groove,%
  r:canonical-graph-shapes%
}, % 
their companion materialization constructions are %
notoriously difficult to develop,
hard to understand, %
and are redrawn from scratch for every single setting. %
Thus, %
we set out to explain materializations based on mathematical principles, %
namely universal properties %
(in the sense of category theory). %
In particular, % 
partial map classifiers in the topos of graphs %
(and its slice categories) %
cover the purely structural aspects of materializations; %
this is related to final pullback complements~\cite{DT87}, %
a fundamental construction of graph rewriting~\cite{l:graph-rewritinginspancat,chhk:sesqui}. % 
Annotations of shape graphs are treated orthogonally via op-fibrations. %

The first milestones of %
a general framework for shape analysis of graph transformation %
and more generally rewriting of objects in a topos  are
the following:
\reversemarginpar
%%%%%%%%

\noindent{\large${\rhd}$} A rewriting formalism for graph
abstractions %
that lifts the rule-based rewriting from single graphs to
\emph{abstract graphs}; %
it is developed for (abstract) objects in a topos.
%%%%%%%%

\noindent{\large${\rhd}$} We characterize the materialization
operation for abstract objects in a topos in terms of partial map
classifiers, %
giving a sound and complete description of all occurrences of
right-hand sides of rules obtained by rewriting an abstract object.
\hfill{\(\to\text{\Sct.~}\ref{sec:materialization}\)}
%%%%%%%%

\noindent{\large${\rhd}$} We decorate abstract objects with
annotations from an ordered monoid and extend abstract rewriting to
abstract objects with annotations. %
For the specific case of graphs, %
we consider global annotations %
(counting the nodes and edges in a graph), %
local annotations %
(constraining the degree of a node), and %
path annotations %
(constraining the existence of paths between certain nodes).
%%%%%%%%
\hfill{\(\to\text{\Sct.~}\ref{sec:annotated-objects}\)}

\noindent{\large${\rhd}$} We show that abstract rewriting with
annotations is sound and, with additional assumptions, complete. %
Finally, we even derive strongest post-conditions for the case of %
graph rewriting with annotations.
\hfill{\(\to\text{\Sct.~}\ref{sec:abstract-rewriting-annotated}\)}

\smallskip

\noindent\emph{Related work:} The idea of shape graphs together
with shape constraints was pioneered in \cite{srw:shape-analysis-3vl}
where the constraints are specified in a three-valued logic. A similar
approach was proposed in \cite{sww:abstract-gts}, %
using first-order formulas as constraints.  In partner abstraction
\cite{b:partner-abstraction-thesis,bw:partner-abstraction}, %
cluster abstraction
\cite{b:cluster-abstraction-thesis,br:cluster-abstraction}, %
and neighbourhood abstraction
\cite{rz:neighbourhood-abstraction-groove} nodes are clustered
according to local criteria, such as their neighbourhood and the
resulting graph structures are enriched with counting constraints,
similar to our constraints. The idea of counting multiplicities of
nodes and edges is also found in canonical graph shapes
\cite{r:canonical-graph-shapes}. %
The uniform treatment of monoid annotations was introduced %
in previous
work~\cite{k:ver-mobile,ckn:graph-languages-type-graphs,ckn:graph-languages-type-graphs-journal},
in the context of type systems and with the aim of studying
decidability and closure properties, but not for abstract rewriting. %

% \vspace*{-.2cm}
% ============================================================================ %
\section{Preliminaries}
\label{sec:preliminaries}
% ============================================================================ %

% \vspace*{-.1cm} 
This paper presupposes familiarity with category theory and the topos
structure of graphs. Some concepts (in particular elementary topoi,
subobject and partial map classifiers, and slice categories) are
defined in \short{\bkchange{the full version of this paper
    \cite{chknr:materialization-arxiv}}}\full{App.~\ref{apx:topos}}. Furthermore
all proofs can be found in \short{\bkchange{the full
    version}}\full{App.~\ref{apx:proofs}}.\todo{App. reference. DONE}

The rewriting formalism for graphs and graph-like structures %
that we use throughout the paper is %
the double-pushout (DPO)
approach~\cite{cmrehl:algebraic-approaches}. %
Although it was originally introduced for
graphs~\cite{DBLP:conf/focs/EhrigPS73}, %
it is well-defined in any category~$\C$. %
However, %
certain standard results for graph rewriting require %
that the category~$\C$ has ``good'' properties. %
The category of graphs is an elementary topos---%
an extremely rich categorical structure---%
but weaker conditions on~\(\C\), for instance adhesivity, have been
studied %
\cite{ls:adhesive-journal,ehpp:adhesive-hlr,ehrig2013categorical}. %

% \vspace*{-.1cm}
\begin{definition}[Double-pushout rewriting]
  \label{def:dpo-rewriting}
  A \emph{production} in $\C$ is a span of monos
  $L\lat~I\rat R$ in $\C$; the objects \(L\) and \(R\) are called
  left- and right-hand side, respectively. \\
  \begin{minipage}[t]{0.74\linewidth}%
    A \emph{match} of a production $p\colon L\lat I \rat R$ to an
    object~\(X\) of~\(\C\) is a mono $m_L \colon L \rat X$ in $\C$. %
    The production~\(p\) rewrites \(X\) to~$Y$ at~$m_L$ (resp.\ the
    match \(m_L\) to the \emph{co-match} \(m_R \colon R\to Y\)) %
    if the production and the match %
    (and the co-match) extend to a diagram in $\C$, shown to the
    right, %
    such that both squares are pushouts.
  \end{minipage}%
  \begin{minipage}[t]{0.26\linewidth} % 
    \vspace*{-.2cm}
    \begin{center}
      {{ \tikzsetnextfilename{dpotikzcd}%
          \begin{tikzpicture}
            \node{\begin{tikzcd}[row sep=1em,column sep=1.2em,ampersand
                replacement=\&,baseline=4ex] L \ar[d%,tail
                ,"m_L"'] \& I
                \ar[l%,tail
                ] \ar[r%,tail
                ] \ar[d] \& R \ar[d,"m_R"]
                \\
                X
                \&
                C
                \ar[l%,tail
                ]
                \ar[r%,tail
                ]
                \ar[ur,phantom,"\color{gray}\scriptstyle\mathrm{(PO)}"]
                \ar[ul,phantom,"\color{gray}\scriptstyle\mathrm{(PO)}"]
                \&
                Y
              \end{tikzcd}};
          \end{tikzpicture} 
        }}
    \end{center}
  \end{minipage}
  
  \noindent In this case, %
  we write $X \oset{p,m_L}{\Longrightarrow} Y$ %
  (resp.\
  \((L\oset{m_L}{\rat} X) \oset{p}{\Rightarrow} (R\oset{m_R}{\to}
  Y)\)). %
  We also write $X \oset{p,m_L}{\Longrightarrow}$ \bkchange{if there
    exists an object $Y$ such that
    $X \oset{p,m_L}{\Longrightarrow} Y$} and
  $X \oset{p}{\Rightarrow} Y$ if the specific match $m_L$ is not
  relevant. %
\end{definition}

Given a
production $p$ and a match $m_L$, %
if there exist arrows \(X \gets C\) and \(C \gets I\) that make the
left-hand square of the diagram in
Def.~\ref{def:dpo-rewriting} %
a pushout square, %
the \emph{gluing condition} is satisfied. %

If~\(\C\) is an adhesive category 
(and thus also if it is a topos \cite{ls:toposes-adhesive}) and
the production consists of monos, %
then all remaining arrows of double-pushout diagrams of rewriting %
are monos~\cite{ls:adhesive-journal} %
and the result of rewriting---%
be it the object~\(Y\) or the co-match~\(m_R\)---%
is unique (up to a canonical isomorphism). %

% \vspace*{-.2cm} 
\subsection{Subobject Classifiers and Partial Map Classifiers of
  Graphs}
\label{sec:classifiers}

A standard category for graph rewriting that is also a topos is %
the category of edge-labelled, directed graphs that we shall use in examples, %
as recalled in the next definition. %
Note that due to the generality of the categorical framework, our results
also hold for various other forms of graphs, such as node-labelled
graphs, hypergraphs, graphs with scopes or graphs with second-order
edges.

\begin{definition}[Category of graphs]
  \label{def:category-graph}
  Let $\Lambda$ be a fixed set of \emph{edge labels}.  A
  \emph{($\Lambda$-labelled) graph} is a tuple %
  $G = (V_G,E_G,\sSrc_G,\sTgt_G,\sLab_G)$ where %
  $V_G$~is a finite set of \emph{nodes}, %
  $E_G$~is a finite set of \emph{edges}, %
  $\sSrc_G,\sTgt_G\colon E_G\to V_G$ are the \emph{source} and
  \emph{target mappings} and %
  $\sLab_G\colon E_G\to\Lambda$ is the \emph{labelling function}.
  
  Let $G,H$ be two $\Lambda$-labelled graphs. A \emph{graph morphism}
  $\phi\colon G\to H$ consists of two functions
  $\phi_V\colon V_G\to V_{H}$, $\phi_E\colon E_G\to E_{H}$, such
  that for each edge $e\in E_G$ we have %
  $\fSrc[H]{\phi_E(e)} = \phi_V(\fSrc[G]{e})$, %
  $\fTgt[H]{\phi_E(e)} = \phi_V(\fTgt[G]{e})$ and %
  $\fLab[H]{\phi_E(e)} = \fLab[G]{e}$. %
  If $\phi_V,\phi_E$ are both bijective, $\phi$ is an
  isomorphism. The category having ($\Lambda$-labelled) graphs as objects
  and graph morphisms as arrows is denoted by $\GR$.
\end{definition}
 
% \vspace*{-.1cm}
We shall often write~$\phi$ instead of~$\phi_V$ or~$\phi_E$ %
to avoid clutter. The graph morphisms in our diagrams will be
indicated by \bkchange{black and white nodes}\todo{Reviewer1: The
  paper refers to colours on page 3, but, like the appendices, these
  are absent from the version I read. \\ DN: Replaced the word
  \emph{colours} by \emph{filling}. \\ BK: I think ``black and white
  nodes'' is better. DONE} and thick edges. In the
category %

\noindent \begin{minipage}[t]{.56\linewidth}
   $\GR$, where the objects are labelled graphs over
  the label alphabet $\Lambda$, the subobject classifier
  $\mathtt{true}$ is displayed to the right %
  where every $\Lambda$-labelled edge
  represents several edges, one for each $\lambda \in \Lambda$. 
\end{minipage}%
\begin{minipage}[t]{.44\linewidth}
  \vspace*{-.05cm}
    $\quad \mathtt{true}:$  %
    \tikzsetnextfilename{LambdaTrueDomain}%
    \scalebox{0.9}{
    \begin{tikzpicture}[baseline={(g1.south)}]%
      \node[gnodeb] (g1) {} ;
      \draw[gedge,thick,overlay] %
      (g1) %
      .. controls +(45:\onemedunit) and +(135:\onemedunit) .. % 
      (g1) %
      node[pos=.8,left,outer sep=0pt] (labA) {\small\(\Lambda\)} % 
      coordinate [pos=.2] (labAopp);
      \node at (labAopp) {~};
      \node[glab] (ghost) at (-.45,.4) {~};
      \node[glab] (ghost2) at (.15,0) {~};
      \ghostgraphbox[l]{(g1) (ghost) (ghost2)};
    \end{tikzpicture}
    \ {\Large $\rightarrowtail$} 
\tikzsetnextfilename{LambdaTrueCodomain}
\begin{tikzpicture}[baseline={(g1.south)}]
  \node[gnodeb] (g1) at (1.3,0) {} ;
  \draw[gedge,thick,overlay] %
    (g1) %
    .. controls +(45:\onemedunit) and +(135:\onemedunit) .. % 
    (g1) %
    node[pos=.8,left,outer sep=0pt] (labA) {\small\(\Lambda\)} % 
    coordinate [pos=.2] (labAopp)
    ;
    %\node at (labA) {\small\(\Lambda\)};
    \node at (labAopp) {~};
    \draw[gedge,thick,overlay,densely dotted] %
    (g1) %
    .. controls +(135+180:\onemedunit) and +(45+180:\onemedunit) .. % 
    (g1) %
    node[pos=.8,left,outer sep=0pt] (labA') {\small\(\Lambda\)} % 
    coordinate [pos=.2] (labA'opp)
    ;
    %\node at (labA') {\small\(\Lambda\)};
    \node at (labA'opp) {~};
    \begin{scope}[shift={(1,0)}]
  \node[gnode] (h1) at (1.3,0) {} ;
    \draw[gedge,thick,overlay,densely dotted] %
    (h1) %
    .. controls +(45+180:\onemedunit) and +(135+180:\onemedunit) .. % 
    (h1) %
    node[pos=.8,right,outer sep=0pt] (labA') {\small\(\Lambda\)} % 
    coordinate [pos=.2] (labA'opp);
    \end{scope}
    \draw[gedge,bend right=15,<->,densely dotted,thick] (g1) to 
     node[pos=.5,auto] {\small\(\Lambda\)}
     (h1) % 
     ;
     \ghostgraphbox[l]{(g1) (h1) (labA') (labA)};
\end{tikzpicture}
  }
\end{minipage}

\vspace{0.1cm}

\noindent The subobject classifier
$\mathtt{true} \colon \mathbf{1} \rat \Omega$ from the terminal object
$\mathbf{1}$ to $\Omega$ allows us to single out a subgraph $X$ of a
graph $Y$, by mapping $Y$ to $\Omega$ in such a way that all elements
of $X$ are mapped to the image of $\mathtt{true}$\full{ (see also
Def.~\ref{def:subobclass} in
App.~\ref{apx:topos})}.\todo{App. reference. DONE}

\smallskip

Given arrows $\alpha,m$ as in the diagram in Definition~\ref{def:fpbc}, we can 
construct the
most general pullback,  called final pullback complement 
\cite{DT87,chhk:sesqui}. %

% \vspace*{-.1cm}
\begin{definition}[Final pullback complement] 
  \label{def:fpbc}
      A pair of arrows $I 
    \oset{\gamma}{\to} F 
    \oset{\beta}{\to} G$ is a \emph{final pullback complement (FPBC)} of % 
    another pair $I 
    \oset{\alpha}{\to} L \oset{m}{\to} G$ if \medskip
    
    \hspace*{-.8cm}
    \begin{minipage}[t]{0.67\linewidth}
    \begin{itemize}
    \item they induce a pullback square
    \item for each pullback square
      $G \oset{m}{\leftarrow} L \oset{\alpha'}{\leftarrow} I'
      \oset{\gamma'}{\to} F' \oset{\beta'}{\to} G$ and arrow
      $f \colon I' \to I$ such that $\alpha \circ f = \alpha'$, there
      exists a unique arrow $f' \colon F' \to F$ such that
      $\beta \circ f' = \beta'$ and
      $\gamma \circ f = f' \circ \gamma'$ both hold (see the
      diagram to the right). 
    \end{itemize}
  \end{minipage} \quad 
  \begin{minipage}[t]{0.33\linewidth}
    \vspace*{-.9cm}
    \begin{center}
    \tikzsetnextfilename{fpbcdef}
    \begin{tikzpicture}[baseline={([yshift=-3ex]current bounding box.north)}]
      \node{\begin{tikzcd}
      L
      \ar[d,"m"'] 
      &
      I 
      \ar[l,"\alpha"]
      \ar[d,"\gamma"]
      \ar[dl,phantom,"\color{gray}\scriptstyle\mathrm{(FPBC)}"]
      &
      I'
      \ar[ll,"\alpha'"',bend right=20,semithick,shift right=1]
      \ar[l,"f",thick]
      \ar[d,"\gamma'",semithick]
      \\
      G
      &
      F
      \ar[l,"\beta"']
      &F' 
      \ar[ll,"\beta'",bend left=20,semithick,shift left=1]
      \ar[l,"f'"',dashed]
    \end{tikzcd}};
  \end{tikzpicture}
  \!\!
  \end{center}
\end{minipage}
\end{definition}

% \vspace*{-.1cm} 
Final pullback complements and subobject classifiers are closely
related to partial map classifiers (see\full{
  Def.~\ref{def:partialmap} in App.~\ref{apx:topos}
  and}\todo{App. reference. DONE} \cite[Corollary~4.6]{DT87}): %
a category has FPBCs (over monos) and a subobject classifier if and
only if it has a partial map classifier. These exist in all elementary
topoi. \todo{Reviewer3: Since partial map classifiers play such a
  central role it would be god to include the definition of partial
  map classifiers in the main text.\\BK: I guess yes, but we do not
  have the space.}

\newcommand{\fpbcAndPartialMap}{ 
  Let $\C$ be a category with finite
  limits. Then the following are equivalent:
  \begin{description}
  \item[(1)] $\C$ has a subobject classifier
    $\mathtt{true} \colon \mathbf{1} \rightarrowtail \Omega$ and final
    pullback complements for each pair of arrows
    $I \oset{\alpha}{\to} L \oset{m}{\rat} G$ with
    $m$ mono;
  \item[(2)] $\C$ has a partial map classifier
    $(F: \C \to \C, \eta: Id \dotarrow F)$.
  \end{description}
}

\begin{proposition}[Final pullback complements, subobject and partial
  map classifiers]
  \label{prop:fpbc-soc}
  \fpbcAndPartialMap
\end{proposition}%

\subsection{Languages}

The main theme of the paper is %
``simultaneous'' rewriting of entire sets of objects of a category by
means of %
rewriting a single \emph{abstract} object that %
represents a collection of structures---%
the \emph{language} of the abstract object. %
The simplest example of an abstract structure is a plain object of a
category %
to which we associate the language of objects that can be mapped to
it; %
the formal definition is as follows (see also
\cite{ckn:graph-languages-type-graphs-journal}). %

\setcounter{footnote}{0}
\begin{definition}[Language of an object]
  \label{def:object-language}
  Let $A$ be an object of a category~$\C$. %
  Given another object~$X$, %
  we write $X\dashrightarrow A$ %
  whenever there exists an arrow from~$X$ to~$A$. We define the
  \emph{language}\footnote{Here we assume that $\C$ is essentially
    small, so that a language can be seen as a set instead of a proper class of
    objects.}  of~$A$, denoted by $\mathcal{L}(A)$, as
  $\mathcal{L}(A) = \{X \in \C \mid X\dashrightarrow A \}$.
\end{definition}

Whenever $X \in \mathcal{L}(A)$ holds, we will say that $X$ is
\emph{abstracted by}~$A$, and $A$ is called the \emph{abstract
  object}. %
In the following we will also need to characterize a class of
(co-)matches which are represented by a given (co-)match (which is a mono).

\short{\pagebreak}

\begin{definition}[Language of a mono]
  \label{def:mono-language}
  Let $\phi\colon L\rat A$ be a mono in $\C$.\todo{Reviewer1: In Def6, why do 
  the monos 
  "factor through" phi? I would say that they factor phi, but do not factor 
  through. \\ DONE}

  \noindent
  \begin{minipage}[t]{0.7\linewidth}
    The \emph{language} of $\phi$ is the set of monos $m$ with source
    $L$ \dnchange{that factor $\phi$} such that the square on the
    right is a pullback:
    \begin{eqnarray*}
      \mathcal{L}(\phi) & = & \{m\colon L\rat X \mid \exists
      (\psi\colon X\to A) \\
      && \qquad \mbox{such that square~(\ref{eq:pullback}) is
        a pullback} \}. 
    \end{eqnarray*}
  \end{minipage}
  \begin{minipage}[t]{0.3\linewidth}
    \vspace{-0.5cm}
    \begin{equation}
      \tikzsetnextfilename{pbsquareLLXA}
      \begin{tikzpicture}[baseline={([yshift=-.5ex]current bounding box.west)}]
        \node{%
          \begin{tikzcd}
            L \ar[tail,"\mathit{id}_L"',d] \ar[tail,"m",r] & X
            \ar[d,"\psi"]
            \\
            L \ar[tail,"\phi"',r] & A
            \ar[ul,phantom,"\color{gray}\scriptstyle\mathrm{(PB)}"]
          \end{tikzcd}%
        };
      \end{tikzpicture}
      \label{eq:pullback}
    \end{equation}
  \end{minipage}
\end{definition}

\noindent 
Intuitively, for any arrow $(L\oset{m}{\to} X) \in \mathcal{L}(\phi)$
we have $X\in\mathcal{L}(A)$ and $X$ has a distinguished subobject $L$
which corresponds precisely to the subobject $L \rat A$. In fact
$\psi$ restricts and co-restricts to an isomorphism between the images
of $L$ in $X$ and $A$. For graphs, no nodes or edges in
$X$ outside of $L$ are mapped by $\psi$ into the image of $L$ in $A$.
\todo{Reviewer3: So, the language of \(id_a\) is just a singleton
  containing \(id_a\)? \\ BK: I guess yes, but we do not have the
  space to explain this.}

% ============================================================================ %
\section{Materialization}
\label{sec:materialization}
% ============================================================================ %

Given a production $p: L\lat I\rat R $, an abstract object~$A$, and
\bkchange{a (possibly non-monic) arrow}\todo{Reviewer1: In the first
  sentence of Sec 3, do you mean a match of p to A instead of a match
  of L in A?  \\ BK: I rewrote this paragraph}
\bkchange{$\phi\colon L\to A$, we want to transform the abstract object 
  $A$}\todo{Reviewer3: Here is something fishy: In Definition 1 you
  allow only monic matches while phi does not need to be monic. And I
  could not find in the paper any rewriting of A, as I would have
  expected it. \\ BK: rewritten. Please check.} in order to
characterize all successors of objects in $\mathcal{L}(A)$, i.e., those obtained
by rewriting via~$p$ at a match compatible
with~$\phi$. \bkchange{(Note that requiring $\phi$ to be monic is not
  sound, since the left-hand side of p could be ``folded'' or
  ``fused'' in the abstraction.)} Roughly, we want to lift DPO
rewriting to the level of abstract objects.

For this, it is necessary to use the materialization construction,
defined categorically in \Sct.~\ref{sec:matCat}, that enables us to
concretize an instance of a left-hand side in a given abstract
object. This construction is refined in \Sct.~\ref{sec:matRew} where we
restrict to materializations that satisfy the gluing condition and can
thus be rewritten via~$p$. Finally in \Sct.~\ref{sec:RewMat} we present
the main result about materializations showing that we can fully
characterize the co-matches obtained by rewriting.

% \vspace*{-.2cm}
\subsection{Materialization Category and Existence of
  Materialization}
\label{sec:matCat}

From now on we assume~\(\C\) to be an elementary topos. \bkchange{We
  will now define the materialization, which, given an arrow
  $\phi\colon L\to A$, characterizes all objects $X$, abstracted over
  $A$, which contain a (mono) occurrence of the left-hand side
  compatible with $\phi$.}

\begin{definition}[Materialization]
  \label{def:materialization-category}
  Let $\phi\colon L\to A$ be an arrow in~$\C$. \\%
  The \emph{materialization category} for~$\phi$, %
  denoted~$\matcat{\phi}$, % 
  has as %
  
  \noindent
  \begin{minipage}[t]{0.67\linewidth}
    \begin{description}
    \item[objects] %
      all factorizations $L\rat X\to A$ of $\phi$ %
      whose first factor $L\rat X$ is a mono, %
      and as
    \item[arrows] %
      from a factorization $L\rat X\to A$ to another one
      $L\rat Y\to A$, %
      all arrows $f\colon X\to Y$ in~$\C$ such that %
      the diagram to the right comprises %
      a commutative triangle and a pullback square. \smallskip%
    \end{description}
  \end{minipage}
  \begin{minipage}[t]{0.3\linewidth}
    \vspace*{-.8cm}
    \[
      \tikzsetnextfilename{arrowsOfMatCat}%
      \begin{tikzpicture}
        \node{\begin{tikzcd}[row sep=1ex] L
            \ar[tail,"\mathit{id}_L"',dd] %
            \ar[tail,r] %
            \ar[dotted,drr,"\phi",bend left=30,near end,shift left=1]
            & X \ar[dd,"f" description] \ar[dr] %
            \\
            && A
            \\
            L \ar[tail,r] \ar[rru,dotted,"\phi"',bend right=30,near
            end,shift right=1] & Y \ar[ur]
            \ar[uul,phantom,"\color{gray}\scriptstyle\mathrm{(PB)}"]
          \end{tikzcd}};
      \end{tikzpicture}
    \]
  \end{minipage}

  \noindent
  If \matcat{\phi} has a terminal object it is denoted by
  $L\rat {\mat{\phi}} \to A$ and is called the \emph{materialization}
  of $\phi$. 
\end{definition}
Sometimes we will also call the object $\mat{\phi}$ the materialization of
$\phi$, omitting the arrows. 

\short{\pagebreak}

Since we are working in a topos by assumption, the slice category over
$A$ provides us with a convenient setting to construct
materializations. Note in particular that in the diagram in
Def.~\ref{def:materialization-category} above, the span
$X\leftarrowtail L\rat L$ is a partial map from $X$ to $L$ in the
slice category over $A$. Hence the materialization $\mat{\phi}$
corresponds to the partial map classifier for $L$ in this slice category.

\newcommand{\existenceOfMatInTop}{ Let~\(\phi\colon L \to~A\) be an
  arrow in~\(\C\), and %
  let \(\eta_\phi \colon \phi \to F(\phi)\), with
  $F(\phi) \colon \bar{A} \to A$,  be %
  the partial map classifier of~\(\phi\)  in %
  the slice category \(\C\!\downarrow\! A\) %
  (which also is a topos).\footnote{%
   This is by the Fundamental Theorem of topos theory  
   \cite[Theorem~2.31]{freyd1972aspects}. %
  }
  Then %
  $L \oset{\eta_\phi}{\to} \bar{A} \oset{F(\phi)}{\to} A$ is the
  materialization of~\(\phi\), hence $\mat{\phi} = \bar{A}$.  }

\todo{Reviewer3: Proposition 8 -> A diagram would be useful! \\ BK: we
  do not have space.}
\begin{proposition}[Existence of materialization]
  \label{prop:term-pmc}% 
  \existenceOfMatInTop
\end{proposition} % 

As a direct consequence of Prop.~\ref{prop:fpbc-soc} and
Prop.~\ref{prop:term-pmc} (and the fact that final pullback
complements in the slice category correspond to those in the base
category \cite{l:graph-rewritinginspancat}),
the terminal object of the
materialization category can be constructed for each arrow of a topos
by taking final pullback complements.

\newcommand{\constructOfMat}{ Let $\varphi \colon L \to A$ be an arrow
  of $\C$ and let
  $\mathtt{true}_A \colon A \rightarrowtail A \times \Omega$ be the
  subobject classifier in the slice category $\C \downarrow A$ from
  $\textit{id}_A \colon A \to A$ to the projection
  $\pi_1 \colon A \times \Omega \to A$\full{ (see
    Fact~\ref{prop:subobclass-slice} in App.~\ref{apx:topos})}.\todo{App. 
    reference. DONE}

  \noindent
  \begin{minipage}[t]{0.6\linewidth}
    Then the terminal object
    $L \oset{\eta_\phi}{\rat} {\mat{\phi}} \oset{\psi}{\to} A$ in the
    materialization category consists of the arrows
    $\eta_\phi$ and $\psi = \pi_1\circ \chi_{\eta_\phi}$,
    where
    $L \oset{\eta_\phi}{\rightarrowtail} {\mat{\phi}}
    \oset{\chi_{\eta_\phi}}{\to} A \times \Omega$ is the final
    pullback complement of
    $L \oset{\varphi}{\to} A \oset{\mathtt{true}_A}{\rightarrowtail} A
    \times \Omega$.
  \end{minipage}
  \begin{minipage}[t]{0.4\linewidth}
    \vspace{-0.7cm}
    \[
      \xymatrix{ L \ar@{->}[d]_{\phi} \ar@{ >.>}[r]^(.6){\eta_\phi
          \hspace*{.3cm}} & {\mat{\phi}}
        \ar@{.>}[d]^{\chi_{\eta_\phi}} \ar@{.>}[dr]^{\psi}&  \\
        A \ar@{ >.>}[r]_{\mathtt{true}_A \hspace*{.3cm}} & A \times
        \Omega \ar@{->}[r]_(.6){\pi_1}
        \itull{\mbox{\rm\footnotesize\color{gray}(FPBC)}} & A }
    \]
  \end{minipage}
}

\begin{corollary}[Construction of the materialization]
  \label{cor:term-constr} 
  \constructOfMat
\end{corollary}

\begin{example}
  \label{ex:construction-materialization}
  We construct the  materialization $L \oset{\eta_\phi}{\rat} {\mat{\phi}}
    \oset{\psi}{\to} A$ for the following morphism \\
    \noindent \begin{minipage}[t]{0.48\linewidth}
      $\varphi \colon L \to A$ of graphs with a single (omitted)
      label:
      \begin{center}
        \vspace*{-.3cm}
        \scalebox{.8}{
          \begin{tikzpicture}[x=3.2cm,y=-1.8cm,baseline=(g1.south)]
            \node[glab] (g0) at (-.2,0) {\large{$\varphi$:}} ;
            \node[gnode] (g1) at (0,0) {} ;
            \node[gnode] (g2) at (0.4,0) {} ;
            \draw[gedge,->,thick] (g1) to node[midway,above] {$\A$} (g2);
            \ghostgraphbox[g]{(g1) (g2)};
            
            \node[gnode] (h1) at (1.4,0) {} ;
            \draw[gedge,thick] (h1) .. controls +(45:1cm) and +(125:1cm) .. 
            node[above] (labA) {$\A$} (h1) ;
            \ghostgraphbox[h]{(h1)};
            
            \draw[-{>[scale=1.3]}] (.6,0) -- (1.2,0) ;
            
          \end{tikzpicture}
        }
      \end{center}

      \noindent In particular, the materialization is obtained as a final
      pullback complement as depicted to the right (compare with the
      corresponding diagram in Corollary~\ref{cor:term-constr}). Note
    \end{minipage}
    \begin{minipage}[t]{0.5\linewidth}
      \begin{center}
        \vspace*{-.4cm}
        \scalebox{.57}{
          \begin{tikzpicture}[x=3.2cm,y=-1.8cm,baseline=(f2.south)]
            \node[gnode] (a1) at (-.2,-1.5) {} ;
            \node[gnode] (a2) at (.2,-1.5) {} ;
            \draw[gedge,thick] (a1) to node[midway,above] (alab) {$\A$} (a2);
            \ghostgraphbox[a]{(a1) (a2) (alab)};

            \node[gnode] (i1) at (0,.5) {} ;
            \draw[gedge,thick] (i1) .. controls +(45:1cm) and +(125:1cm) .. 
            node[above] (labA) {$\A$} (i1) ;
            \ghostgraphbox[i]{(i1) (labA)};
            
            \node[gnode] (t1) at (1.3,-2) {} ;
            \node[gnode] (t2) at (1.7,-2) {} ;
            \node[gnode] (t3) at (1.5,-1.3) {} ;
            \draw[gedge,thick] (t1) to node[midway,above] {$\A$} (t2); 
            \draw[gedge,<->,dashed] (t1) to [bend angle=50,bend right] 
            node[glab,midway,below=.1cm] {$\A$} (t2);
            \draw[gedge,<->,dashed] (t1) to [bend angle=30,bend right]  
            node[glab,midway,left] {$\A$} (t3);
            \draw[gedge,<->,dashed] (t2) to [bend angle=30,bend left]   
            node[glab,midway,right] {$\A$} (t3);
            \draw[gedge,dashed] (t1) .. controls +(140:1cm) and +(220:1cm) .. 
            node[glab,left] {$\A$} (t1) ;
            \draw[gedge,dashed] (t2) .. controls +(40:1cm) and +(320:1cm) .. 
            node[glab,right] {$\A$} (t2) ;
            \draw[gedge,dashed] (t3) .. controls +(225:1cm) and +(305:1cm) .. 
            node[glab,below] {$\A$} (t3) ;
            \ghostgraphbox[t]{(t1) (t2) (t3)};

            \node[gnode] (f1) at (1.3,.5) {} ;
            \node[gnode] (f2) at (1.7,.5) {} ;
            \draw[gedge,thick] (f1) .. controls +(45:1cm) and +(125:1cm) .. 
            node[above] (labA) {$\A$} (f1) ;
            \draw[gedge,<->,dashed] (f1) to node[glab,midway,above] {$\A$} (f2);
            \draw[gedge,dashed] (f1) .. controls +(225:1cm) and +(305:1cm) .. 
            node[glab,below] {$\A$} (f1) ;
            \draw[gedge,dashed] (f2) .. controls +(225:1cm) and +(305:1cm) .. 
            node[glab,below] {$\A$} (f2) ;
            \ghostgraphbox[f]{(f1) (f2)};

            \draw[{>[scale=1.3]}-{>[scale=1.3]}] (.4,-1.5) -- 
            (1,-1.5); \node[glab] (m) at (.7,-1.65) {{\Large $\eta_{\varphi}$}};
            \draw[-{>[scale=1.3]}] (0,-1.1) -- 
            (0,-.2) ; \node[glab] (m) at (-.1,-.65) {{\Large $\varphi$}};
            \draw[-{>[scale=1.3]}] (1.5,-.7) -- 
            (1.5,-.2) ; \node[glab] (m) at (1.65,-.5) {{\Large 
            $\chi_{\eta_{\varphi}}$}};
            \draw[{>[scale=1.3]}-{>[scale=1.3]}] (.2,.5) -- 
            (1.1,.5) ; \node[glab] (m) at (.65,.7) {{\Large 
            $\mathtt{true}_A$}};
            \draw[-{>[scale=1.3]}] (1.8,-1.5) -- 
            (2.8,0.1); \node[glab] (m) at (2.35,-.9) {{\Large 
                                    $\psi$}};
            \draw[-{>[scale=1.3]}] (1.9,0.5) -- 
            (2.8,0.5); \node[glab] (m) at (2.35,.7) {{\Large 
                        $\pi_1$}};

            \node[gnode] (h1) at (3,0.5) {} ;
            \draw[gedge,thick] (h1) .. controls +(45:1cm) and +(125:1cm) .. 
            node[above] (labA) {$\A$} (h1) ;
            \ghostgraphbox[h]{(h1)};

            \node at (.7,-.5) {\LARGE\color{gray}(FPBC)};
          \end{tikzpicture}
        }
      \end{center}
    \end{minipage}

    \vspace{0.07cm}
    
    \noindent that edges which are not in the image of $\eta_\phi$
    resp.\ $\mathtt{true}_A$ are dashed.
\end{example}

This construction corresponds to the usual intuition behind
materialization: the left-hand side and the edges that are attached to
it are ``pulled out'' of the given abstract graph. \full{The concrete
  construction in the category $\GR$ is spelled out in
  App.~\ref{apx:materialization-graphs}.}\todo{App. reference. DONE}

We can summarize the result of our constructions in the
following\todo{Reviewer3: I would have liked to see a short discussion
  of this motivation already in Subsection 2.2 or at the beginning of
  Section 3, before all the "materialization business" starts. \\ BK:
  I moved the explanation to the start of Sct.~\ref{sec:matCat}.}
proposition:

% , which states that the materialization characterizes all objects $X$,
% abstracted over $A$, which contain a (mono) occurrence of the
% left-hand side compatible with $\phi$.

\newcommand{\languageMaterialization}{ Let $\phi\colon L\to A$ be an
  arrow in $\C$ and let $L\oset{\eta_\phi}{\rat} {\mat{\phi}} \to A$ be
  the corresponding materialization. Then we have
  \[ \mathcal{L}(L\oset{\eta_\phi}{\rat} {\mat{\phi}}) =
    \{L\oset{m_L}{\rat} X \mid \exists \psi\colon (X\to A).\ (\phi
    = \psi\circ m_L) \}. \]
}

\begin{proposition}[Language of the materialization]
  \label{prop:language-materialization}
  \languageMaterialization
\end{proposition}

\subsection{Characterizing the Language of Rewritable Objects}
\label{sec:matRew}

\deleted{As observed earlier,} \dnchange{A} match obtained through the 
materialization of
the left-hand side of a production from a given object may not allow a
DPO rewriting step because of the gluing condition.\todo{Reviewer3: ''As 
observed earlier`` - Please give some references! \\ DN: Removed "As observed 
earlier`` since the problem was only stated implicit up until now. DONE}
We illustrate this problem with an example.

\begin{example}
  \label{ex:POCexample}
  Consider the materialization $L \rat {\mat{\phi}} \to A$ from 
  Example~\ref{ex:construction-materialization} and the~pro-\\
  \noindent \begin{minipage}[t]{0.46\linewidth}
    duction~$L~\lat~I~\rat~R$ shown in the diagram to the right.  It
    is easy to see that the pushout complement of morphisms
    $I \rat L \rat {\mat{\phi}}$ does not exist. \\
    
  \noindent Nevertheless there exist factorizations $L \rat X \to A$ 
  abstracted by
  ${\mat{\phi}}$ that could be rewritten using the production.
  \end{minipage}
  \begin{minipage}[t]{0.52\linewidth}
    \vspace*{-.15cm}
    \begin{center}
      \scalebox{.64}{
        \begin{tikzpicture}[x=3.2cm,y=-1.8cm,baseline=(b2.south)]
          \node[gnode] (a1) at (-.2,-1.5) {} ;
          \node[gnodeb] (a2) at (.2,-1.5) {} ;
          \draw[gedge,thick] (a1) to node[midway,above] (alab) {$\A$} (a2);
          \ghostgraphbox[a]{(a1) (a2) (alab)};
  
          \node[gnodeb] (i1) at (1,-1.5) {};
          \ghostgraphbox[i]{(i1)};
  
          \node[gnode] (b1) at (1.8,-1.5) {} ;
          \node[gnodeb] (b2) at (2.2,-1.5) {} ;
          \draw[gedge] (b2) to node[midway,above] (blab) {$ $} (b1);
          \ghostgraphbox[b]{(b2) (b1) (blab)};
          
          \node[gnode] (t1) at (-.2,-.3) {} ;
          \node[gnodeb] (t2) at (.2,-.3) {} ;
          \node[gnode] (t3) at (0,.3) {} ;
          \draw[gedge,thick] (t1) to node[midway,above] {$\A$} (t2); 
          \draw[gedge,<->] (t1) to [bend angle=50,bend right] 
          node[glab,midway,below=.1cm] {$\A$} (t2);
          \draw[gedge,<->] (t1) to [bend angle=30,bend right]  
          node[glab,midway,left] {$\A$} (t3);
          \draw[gedge,<->] (t2) to [bend angle=30,bend left]   
          node[glab,midway,right] {$\A$} (t3);
          \draw[gedge] (t1) .. controls +(140:1cm) and +(220:1cm) .. 
          node[glab,left] {$\A$} (t1) ;
          \draw[gedge] (t2) .. controls +(40:1cm) and +(320:1cm) .. 
          node[glab,right] {$\A$} (t2) ;
          \draw[gedge] (t3) .. controls +(180:1cm) and +(260:1cm) .. 
          node[glab,left] {$\A$} (t3) ;
          \ghostgraphbox[t]{(t1) (t2) (t3)};
  
          \node[glab] (c1) at (1,0) {\large $?$};
          \ghostgraphbox[c]{(c1)};
          
          \draw[{>[scale=1.3]}-{>[scale=1.3]}] (.9,-1.5) -- 
          (.4,-1.5) ;
          \draw[{>[scale=1.3]}-{>[scale=1.3]}] (1.1,-1.5) -- 
          (1.6,-1.5) ;
  
          \draw[{>[scale=1.3]}-{>[scale=1.3]}] (0,-1.3) -- 
          (0,-.65) ;
          \draw[-{>[scale=1.3]},dashed] (1,-1.3) -- 
          (1,-.2) ;
  
          \draw[-{>[scale=1.3]},dashed] (.9,0) -- 
          (.5,0) ;
  
          \draw[decorate,decoration={brace,amplitude=10pt,mirror,raise=4pt},yshift=0pt]
          (-.4,-1.7) -- (-.4,.7) node
          [black,rotate=-90,midway,yshift=-.8cm] {\footnotesize
            $L \rat {\mat{\phi}}$};
          
          \draw[decorate,decoration={brace,amplitude=10pt,raise=4pt},yshift=0pt]
          (-.4,-1.7) -- (2.4,-1.7) node 
          [black,midway,yshift=.8cm] 
          {\footnotesize $L\lat I\rat R$};
        \end{tikzpicture}
      }
    \end{center}
  \end{minipage}
\end{example}

In order to
take the existence of pushout complements into account, we consider a
subcategory of the materialization category.

\begin{definition}[Materialization subcategory of rewritable objects]
  \label{def:materialization-category-rewritable}
  Let $\phi\colon L\to A$ be an arrow of $\C$ and let
  $\phi_L \colon I\rat L$ be a mono (corresponding to the left leg of
  a production). 
  The \emph{materialization subcategory of rewritable objects} for
  $\phi$ and $\phi_L$, denoted $\matcat[\phi_L]{\phi}$, is the full
  subcategory of $\matcat{\phi}$ containing as objects all factorizations
  $L\oset{m}{\rat} X\to A$ of $\phi$, where $m$ is a mono and
  $I\oset{\phi_L}{\rat} L\oset{m}{\rat} X$ has a pushout
  complement.
  
  Its terminal element, if it exists, is denoted by
  $L\oset{n_L}{\rat}\rmat{\phi}{\phi_L} \to A$ and is called the
  \emph{rewritable materialization}.
\end{definition}

We will show that this subcategory of the materialization category has
a terminal object.

\newcommand{\rewritableMat}{ Let $\phi\colon L\to A$ be an arrow and
  let $\phi_L\colon I\rat L$ be a mono of $\C$.  Then the
  \emph{rewritable materialization of $\phi$ w.r.t.~$\phi_L$} exists
  and can be constructed as the following factorization
  $L\oset{n_L}{\rat} \rmat{\phi}{\phi_L} \oset{\psi\circ
    \alpha}{\longrightarrow} A$ of $\phi$. In the left diagram, $F$ is
  obtained as the final pullback complement of
  $I \oset{\phi_L}{\rat} L \rat {\mat{\phi}}$, where
  $L \rat {\mat{\phi}} \oset{\psi}{\to} A$ is the materialization of
  $\phi$ (Def.~\ref{def:materialization-category}).  Next in the right
  diagram
  $L \oset{n_L}{\rat} \rmat{\phi}{\phi_L} \oset{\beta}{\lat} F$ is the
  pushout of the span $L \oset{\phi_L}{\lat} I \rat F$ and $\alpha$ is
  the resulting mediating arrow.

  \noindent
  \begin{minipage}{0.42\textwidth}
    \begin{equation}
      \xymatrix{
        &  L \ar[dl]_{\phi} \ar@{ >->}[d] & I \ar@{ >->}[d]
        \ar@{ >->}[l]_{\phi_L}\\
        A & {\mat{\phi}} \ar[l]_{\psi} & F
        \itul{\rm\footnotesize\color{gray}
          (FPBC)}
        \ar@{>->}[l] }
      \label{eq:fpbc-al-i}
    \end{equation} 
  \end{minipage}
  \quad
  \begin{minipage}{0.54\textwidth} 
    \begin{equation}
      \xymatrix{
        & L \ar[dl]_{\phi}   \ar@{ >->}[d] & L \ar@{ >->}[d]_{n_L}
        \ar@{ >->}[l]_{\mathit{id}_L} & I \ar@{ >->}[d]
        \ar@{ >->}[l]_{\phi_L} \\
        A & {\mat{\phi}} \ar[l]_{\psi} & {\rmat{\phi}{\phi_L}}
        \ar@{>.>}[l]_(.5){\alpha} 
        & F \ar@{ >->}[l]_(.4){\beta} \ar@/^1pc/@{ >->}[ll]
        \itull{\rm\footnotesize\color{gray} \hspace*{0.1cm}(PO)}
      }
      \label{eq:po-l-fpbc}
    \end{equation} 
  \end{minipage}
%  \vspace*{.1cm}
}  

\begin{proposition}[Construction of the rewritable materialization]
  \label{prop:mat-rewritable}
  \rewritableMat
\end{proposition}

\begin{example}
  We come back to the running example (see Ex.~\ref{ex:POCexample})
  and, as in Prop.~\ref{prop:mat-rewritable}, determine the final
  pullback complement $I \rat F \rat {\mat{\phi}}$ of
  $I \oset{\phi_L}{\rat} L \rat \mat{\phi}$ (see diagram below left)
  and obtain $\rmat{\phi}{\phi_L}$ by taking the pushout over
  $L \lat I \rat F$ (see diagram below right).

  \begin{minipage}{0.42\textwidth}
    \begin{center}
      \scalebox{.68}{
        \begin{tikzpicture}[x=3.2cm,y=-1.8cm,baseline=(a2.south)]
          \node[gnode] (a1) at (-.2,-1.5) {} ; 
          \node[gnodeb] (a2) at (.2,-1.5) {} ; 
          \draw[gedge,thick] (a1) to node[midway,above]
          (alab) {$\A$} (a2); \ghostgraphbox[a]{(a1) (a2) (alab)};
        
          \node[gnodeb] (i1) at (1,-1.5) {};
          \ghostgraphbox[i]{(i1)};
        
          \node[gnode] (t1) at (-.2,-.3) {} ; 
          \node[gnodeb] (t2) at (.2,-.3) {} ; 
          \node[gnode] (t3) at (0,.3) {} ;
          \draw[gedge,thick] (t1) to node[midway,above] {$\A$} (t2);
          \draw[gedge,<->] (t1) to [bend angle=50,bend right]
          node[glab,midway,below=.1cm] {$\A$} (t2); \draw[gedge,<->]
          (t1) to [bend angle=30,bend right] node[glab,midway,left]
          {$\A$} (t3); \draw[gedge,<->] (t2) to [bend angle=30,bend
          left] node[glab,midway,right] {$\A$} (t3); \draw[gedge] (t1)
          .. controls +(140:1cm) and +(220:1cm) ..  node[glab,left]
          {$\A$} (t1) ; \draw[gedge] (t2) .. controls +(40:1cm) and
          +(320:1cm) ..  node[glab,right] {$\A$} (t2) ; \draw[gedge]
          (t3) .. controls +(225:1cm) and +(305:1cm) ..  node[below]
          {$\A$} (t3) ; \ghostgraphbox[t]{(t1) (t2) (t3)};
        
          \node[gnodeb] (f2) at (1.1,-.3) {} ; 
          \node[gnode] (f3) at (.9,.3) {} ;
          \node[glab] (ghost1) at (1.3,-.4) {};
          \node[glab] (ghost2) at (.8,.6) {};
          \draw[gedge,<->] (f2) to [bend angle=30,bend left]
          node[glab,midway,right] {$\A$} (f3); 
          \draw[gedge] (f2) .. controls +(40:1cm) and
          +(320:1cm) ..  node[glab,right] {$\A$} (f2) ; 
          \draw[gedge] (f3) .. controls +(225:1cm) and +(305:1cm) ..  
          node[below] {$\A$} (f3) ; 
          \graphboxwhite[f]{(f2) (f3) (ghost1) (ghost2)};
          \draw[{>[scale=1.3]}-{>[scale=1.3]}] (.9,-1.5) -- (.4,-1.5)
          ;
        
          \draw[{>[scale=1.3]}-{>[scale=1.3]}] (0,-1.3) -- (0,-.65) ;
          \draw[{>[scale=1.3]}-{>[scale=1.3]}] (1,-1.3) -- (1,-.65) ;
        
          \draw[{>[scale=1.3]}-{>[scale=1.3]}] (.7,0) -- (.4,0) ;
        
          \draw[decorate,decoration={brace,amplitude=10pt,mirror,raise=4pt},yshift=0pt]
          (-.4,-1.7) -- (-.4,.7) node
          [black,rotate=-90,midway,yshift=-.8cm] {\footnotesize
            $L \rat {\mat{\phi}}$};
        
          \draw[decorate,decoration={brace,amplitude=10pt,raise=4pt},yshift=0pt]
          (-.4,-1.7) -- (1.1,-1.7) node [black,midway,yshift=.8cm]
          {\footnotesize $L\lat I$};
        
          \node at (.5,-.8) {\large \color{gray}(FPBC)}; \node at (1,.9) {\large
            $F$};
        \end{tikzpicture}
      }
    \end{center}
  \end{minipage}
  \qquad
  \begin{minipage}{0.47\textwidth}
    \begin{center}
      \vspace*{1mm}
      \scalebox{.68}{
        \begin{tikzpicture}[x=3.2cm,y=-1.8cm,baseline=(a2.south)]
          \node[gnode] (a1) at (-.2,-1.5) {} ; \node[gnodeb] (a2) at
          (.2,-1.5) {} ; \draw[gedge,thick] (a1) to node[midway,above]
          (alab) {$\A$} (a2); \ghostgraphbox[a]{(a1) (a2) (alab)};
        
          \node[gnodeb] (i1) at (1,-1.5) {};
          \ghostgraphbox[i]{(i1)};
        
          \node[gnode] (t1) at (-.2,-.3) {} ; \node[gnodeb] (t2) at
          (.2,-.3) {} ; \node[gnode] (t3) at (0,.3) {} ;
          \node[glab] (ghost1) at (.4,-.4) {};
          \node[glab] (ghost2) at (-.1,.6) {};
          \draw[gedge,thick] (t1) to node[midway,above] (labB) {$\A$}
          (t2); \draw[gedge] (t3) .. controls +(225:1cm) and
          +(305:1cm) ..  node[below] (labA) {$\A$} (t3) ;
          \draw[gedge] (t2) .. controls +(40:1cm) and
          +(320:1cm) ..  node[glab,right] {$\A$} (t2) ; 
          \draw[gedge,<->] (t2) to [bend angle=30,bend
                    left] node[glab,midway,right] {$\A$} (t3);
          \graphboxwhite[t]{(t1) (t2) (t3) (ghost1) (ghost2)};
        
          \node[gnodeb] (f2) at
          (1.1,-.3) {} ; \node[gnode] (f3) at (.9,.3) {} ;
          \draw[gedge] (f3) .. controls +(225:1cm) and
          +(305:1cm) ..  node[below] (labA) {$\A$} (f3) ;
          \draw[gedge,<->] (f2) to [bend angle=30,bend
                    left] node[glab,midway,right] {$\A$} (f3);
          \draw[gedge] (f2) .. controls +(40:1cm) and
          +(320:1cm) ..  node[glab,right] {$\A$} (f2) ; 
        
          \draw[{>[scale=1.3]}-{>[scale=1.3]}] (.9,-1.5) -- (.4,-1.5)
          ;
        
          \draw[{>[scale=1.3]}-{>[scale=1.3]}] (0,-1.3) -- (0,-.65) ;
          \draw[{>[scale=1.3]}-{>[scale=1.3]}] (1,-1.3) -- (1,-.65) ;
        
          \draw[{>[scale=1.3]}-{>[scale=1.3]}] (.8,0) -- (.5,0) ;
        
          \draw[decorate,decoration={brace,amplitude=10pt,raise=16pt},yshift=0pt]
          (1.1,-1.7) -- (1.1,.7) node
          [black,rotate=-90,midway,yshift=1.2cm] {\footnotesize
            $I \rat F$};
        
          \draw[decorate,decoration={brace,amplitude=10pt,raise=4pt},yshift=0pt]
          (-.4,-1.7) -- (1.1,-1.7) node [black,midway,yshift=.8cm]
          {\footnotesize $L\lat I$};
        
          \node at (.5,-.8) {\large \color{gray}(PO)}; \node at (.05,1)
          {\large $\rmat{\phi}{\phi_L}$};
        \end{tikzpicture}
      }
    \end{center}
  \end{minipage}
\end{example}

It remains to be shown that $L\rat \rmat{\phi}{\phi_L}\to A$
represents every factorization which can be rewritten.  As before we
obtain a characterization of the rewritable objects, including the
match, as the language of an arrow.

\newcommand{\languageRewritableMaterialization}{ Assume there is a
  production $p\colon L\oset{\phi_L}{\lat} I\oset{\phi_R}{\rat} R$ and
  let $L\oset{n_L}{\rat}\rmat{\phi}{\phi_L}$ be the match for the
  rewritable materialization for $\phi$ and $\phi_L$. Then we have
  \[ \mathcal{L}(L\oset{n_L}{\rat} \rmat{\phi}{\phi_L}) =
    \{L\oset{m_L}{\rat} X\mid \exists \psi\colon (X\to A).\ (\phi =
    \psi \circ m_L \land X\oset{p,m_L}{\Longrightarrow}) \}. \]
  \todo{Reviewer3: Should not there be a symbol right of the double
    arrow? If you use this notation, however, to denote a kind of
    ''implicit existential quantification``, you should mention it
    here or directly after Definition 1. \\ BK: I defined this in
    Def.~\ref{def:dpo-rewriting}.} }

\begin{proposition}[Language of the rewritable materialization]
  \label{prop:language-rewriting-materialization}
  \languageRewritableMaterialization
\end{proposition}

\subsection{Rewriting Materializations}
\label{sec:RewMat}

In the next step we will now rewrite the rewritable materialization
$\rmat{\phi}{\phi_L}$ with the match
$L\oset{n_L}{\rat}\rmat{\phi}{\phi_L}$, resulting in a co-match
$R\rat B$. In particular, we will show that this co-match represents
all co-matches that can be obtained by rewriting an object $X$ of
$\mathcal{L}(A)$ at a match compatible with $\phi$. We first start
with an example.

\begin{example}
  We can rewrite the materialization 
  $L \rat \rmat{\phi}{\phi_L} \to A$ as follows:
  \begin{center}
    \scalebox{.68}{
      \begin{tikzpicture}[x=3.2cm,y=-1.8cm,baseline=(b2.south)]
        \node[gnode] (a1) at (-.2,-1.5) {} ;
        \node[gnodeb] (a2) at (.2,-1.5) {} ;
        \draw[gedge,thick] (a1) to node[midway,above] (alab) {$\A$} (a2);
        \ghostgraphbox[a]{(a1) (a2) (alab)};

        \node[gnodeb] (i1) at (1,-1.5) {};
        \ghostgraphbox[i]{(i1)};

        \node[gnode] (b1) at (1.8,-1.5) {} ;
        \node[gnodeb] (b2) at (2.2,-1.5) {} ;
        \draw[gedge,thick] (b2) to (b1);
        \ghostgraphbox[b]{(b1) (b2) };
        
        \node[gnode] (t1) at (-.2,-.3) {} ;
        \node[gnodeb] (t2) at (.2,-.3) {} ;
        \node[gnode] (t3) at (0,.3) {} ;
        \draw[gedge,thick] (t1) to node[midway,above] {$\A$} (t2); 
        \draw[gedge] (t3) .. controls +(225:1cm) and +(305:1cm) .. 
        node[below] {$\A$} (t3) ;
        \draw[gedge] (t2) .. controls +(40:1cm) and
                  +(320:1cm) ..  node[glab,right] {$\A$} (t2) ; 
        \draw[gedge,<->] (t2) to [bend angle=30,bend
                            left] node[glab,midway,right] {$\A$} (t3);
        \ghostgraphbox[t]{(t1) (t2) (t3)};

        \node[gnode] (r1) at (1.8,-.3) {} ;
        \node[gnodeb] (r2) at (2.2,-.3) {} ;
        \node[gnode] (r3) at (2,.3) {} ;
          \node[glab] (ghost3) at (2.4,-.4) {};
          \node[glab] (ghost4) at (1.75,.6) {};
        \draw[gedge,thick] (r2) to node[midway,above] {$\A$} (r1); 
        \draw[gedge] (r3) .. controls +(225:1cm) and +(305:1cm) .. 
        node[below] {$\A$} (r3) ;
        \draw[gedge] (r2) .. controls +(40:1cm) and
                  +(320:1cm) ..  node[glab,right] {$\A$} (r2) ; 
        \draw[gedge,<->] (r2) to [bend angle=30,bend
                            left] node[glab,midway,right] {$\A$} (r3);
        \graphboxwhite[r]{(r1) (r2) (r3) (ghost3) (ghost4) };

          \node[gnodeb] (f2) at (1.05,-.3) {} ; 
          \node[gnode] (f3) at (.85,.3) {} ;
          \node[glab] (ghost1) at (1.25,-.4) {};
          \node[glab] (ghost2) at (.75,.6) {};
          \draw[gedge,<->] (f2) to [bend angle=30,bend left]
          node[glab,midway,right] {$\A$} (f3); 
          \draw[gedge] (f2) .. controls +(40:1cm) and
          +(320:1cm) ..  node[glab,right] {$\A$} (f2) ; 
          \draw[gedge] (f3) .. controls +(225:1cm) and +(305:1cm) ..  
          node[below] {$\A$} (f3) ; 
          \graphboxwhite[f]{(f2) (f3) (ghost1) (ghost2)};

        \draw[{>[scale=1.3]}-{>[scale=1.3]}] (.9,-1.5) -- 
        (.4,-1.5) ;
        \draw[{>[scale=1.3]}-{>[scale=1.3]}] (1.1,-1.5) -- 
        (1.6,-1.5) ;

        \draw[{>[scale=1.3]}-{>[scale=1.3]}] (0,-1.3) -- 
        (0,-.65) ;
        \draw[{>[scale=1.3]}-{>[scale=1.3]}] (1,-1.3) -- 
        (1,-.65) ;

        \draw[{>[scale=1.3]}-{>[scale=1.3]}] (.6,0) -- 
        (.4,0) ;

        \draw[{>[scale=1.3]}-{>[scale=1.3]}] (1.4,0) -- 
        (1.6,0) ;
        
        \draw[{>[scale=1.3]}-{>[scale=1.3]}] (2,-1.3) -- 
        (2,-.65) ;
        
        \draw[decorate,decoration={brace,amplitude=10pt,mirror,raise=4pt},yshift=0pt]
        (-.4,-1.7) -- (-.4,.6) node 
        [black,rotate=-90,midway,yshift=-.8cm] 
        {\footnotesize $L \rat \rmat{\phi}{\phi_L}$};
        
        \node at (1,.9) {\large $C$};
        \node at (2,.9) {\large $B$};
        \node at (.5,-.8) {\large \color{gray}(PO)};
        \node at (1.5,-.8) {\large \color{gray}(PO)};
      \end{tikzpicture}
    }
  \end{center}
\end{example}

\newcommand{\materializationRewriting}{ Let a match $n_L\colon L\rat \tilde{A}$ 
and a production $p\colon L\lat I\rat R$
  be given. Assume that $\tilde{A}$ is rewritten along the match 
  $n_L$, i.e.,
  $(L \oset{n_L}{\rat} \tilde{A}) \oset{p}{\Rightarrow}
  (R\oset{n_R}{\rat} B)$.  Then
  \[
    \mathcal{L}(R\oset{n_R}{\rat} B) = \{ R\oset{m_R}{\rat}
    Y\mid \exists (L\oset{m_L}{\rat} X)\in
    \mathcal{L}(L\oset{n_L}{\rat} \tilde{A}).\
    \big((L\oset{m_L}{\rat} X) \oset{p}{\Rightarrow}
    (R\oset{m_R}{\rat} Y)\big)\}
  \]
}

%\vspace*{-.3cm}
\begin{proposition}[Rewriting abstract matches]
  \label{prop:rewriting-materializations}
  \materializationRewriting
\end{proposition}

If we combine
Prop.~\ref{prop:language-rewriting-materialization} and
Prop.~\ref{prop:rewriting-materializations}, we 
obtain the following corollary that characterizes the co-matches
obtained from rewriting a match compatible with $\phi\colon L\to A$.

\newcommand{\corollaryCoMatch}{ Let $\phi\colon L\to A$ and a
  production $p\colon L\oset{\phi_L}{\lat} I\oset{\phi_R}{\rat} R$ be
  given. Assume that $\rmat{\phi}{\phi_L}$ is obtained as the
  rewritable materialization of $\phi$ and $\phi_L$ with match
  $L\oset{n_L}{\rat} \rmat{\phi}{\phi_L}$ (see
  Prop.~\ref{prop:mat-rewritable}). Furthermore let
  $(L \oset{n_L}{\rat} \rmat{\phi}{\phi_L}) \oset{p}{\Rightarrow}
  (R\oset{n_R}{\rat} B)$.  Then
  \begin{eqnarray*}
    \mathcal{L}(R\oset{n_R}{\rat} B) & = & \{ R\oset{m_R}{\rat}
    Y\mid \exists (L\oset{m_L}{\rat} X), (X\oset{\psi}{\to}
    A).\ \big(\phi = \psi\circ m_L \land \\
    && \qquad (L\oset{m_L}{\rat} X)
    \oset{p}{\Rightarrow} (R\oset{m_R}{\rat} Y)\big)\}
  \end{eqnarray*}
}

\begin{corollary}[Co-match language of the rewritable materialization]
  \label{cor:rewriting-materializations}
  \corollaryCoMatch
\end{corollary}

This result does not yet enable us to construct post-conditions. While
the set of co-matches is fully characterized, this can only be
achieved by fixing the right-hand side $R$ and thus ensuring that
\bkchange{exactly one occurrence of $R$ is
  represented.}\todo{Reviewer1: In the last paragraph on page 8, why
  is it important that R is present exactly once? \\ BK: I am not
  entirely sure what to do. I rewrote the sentence a bit.} However, as
soon as we forget about the co-match, this effect is gone and can only
be retrieved by adding annotations, which will be introduced next.

% ============================================================================ %
\section{Annotated Objects}
\label{sec:annotated-objects}
% ============================================================================ %

We now endow objects with annotations, thus making object languages
more expressive. In particular we will use ordered monoids in order to
annotate objects. Similar annotations have already been studied in
\cite{k:ver-mobile} in the context of type systems and in
\cite{ckn:graph-languages-type-graphs-journal} with the aim of
studying decidability and closure properties, but not for abstract
rewriting.

\begin{definition}[Ordered monoid]
  An \emph{ordered monoid} $(\mathcal{M},+,\leq)$ consists of a set
  $\mathcal{M}$, a partial order $\le$ and a binary operation $+$ such
  that $(\mathcal{M},+)$ is a monoid with unit $0$ (which is the
    bottom element wrt.\ $\le$) and the partial order is compatible
  with the monoid operation. In particular $a\le b$ implies
  $a+c\le b+c$ and $c+a\le c+b$ for all $a,b,c\in \mathcal{M}$.
  An ordered monoid is commutative if $+$ is commutative.

  A tuple $(\mathcal{M},+,-,\le)$, where $(\mathcal{M},+,\leq)$ is an
  ordered monoid and $-$ is a binary operation on $\mathcal{M}$, is
  called an \emph{ordered monoid with subtraction}.

  We say that subtraction is \emph{well-behaved} whenever for all
  $a,b\in \mathcal{M}$ it holds that $a-a = 0$ and  $(a-b)+b = a$
  whenever $b\le a$.
\end{definition}

For now subtraction is just any operation, without specific
requirements. Later we will concentrate on specific subtraction
operations and demand that they are well-behaved.

In the following we will consider only commutative monoids.

\begin{definition}[Monotone maps and homomorphisms]
  Let $\mathcal{M}_1$, $\mathcal{M}_2$ be two ordered monoids. A map
  $h\colon \mathcal{M}_1\to \mathcal{M}_2$ is called \emph{monotone}
  if $a\le b$ implies $h(a)\le h(b)$ for all $a,b\in
  \mathcal{M}_1$. \bkchange{The category of ordered monoids with
    subtraction and monotone maps is called $\Mon$.}

  \bkchange{A monotone map} $h$ is called a \emph{homomorphism} if
  $h(0)=0$ and $h(a+b)= h(a)+h(b)$. If $\mathcal{M}_1,\mathcal{M}_2$
  are ordered monoids with subtraction, we say that $h$ preserves
  subtraction if $h(a-b) = h(a)-h(b)$.
\end{definition}

\begin{example}
  \label{ex:counting-monoid}
  Let $n \in \mathbb{N}\backslash\{0\}$ and take
  $\mathcal{M}_n = \{0,1,\dots,n,*\}$ (zero, one, $\dots$, $n$, many)
  with $0 \leq 1 \leq \dots\leq n\leq *$ and addition as (commutative)
  monoid operation with the proviso that $a+b=*$ if the sum is larger
  than $n$. In addition $a+* = *$ for all $a\in
  \mathcal{M}_n$. Subtraction is truncated subtraction where $a-b = 0$
  if $a \le b$. Furthermore $*-a = *$ for all $a\in \mathbb{N}$.
  It is easy to see that subtraction is well-behaved.
\end{example}
 
Given a set $S$ and an ordered monoid (with subtraction)
$\mathcal{M}$, it is easy to check that also $\mathcal{M}^S$ is an
ordered monoid (with subtraction), where the elements are functions
from $S$ to $\mathcal{M}$ and the partial order, the monoidal
operation and the subtraction are taken pointwise.

The following path monoid is useful if we want to annotate a graph
with information over which paths are present. Note that due to the
\bkchange{fusion of nodes and edges} caused by the abstraction, a path
in the abstract graph does not necessarily imply the existence of a
corresponding path in a concrete graph.  Hence annotations based on
such a monoid\bkchange{, which provide information about the existence
  of paths,} can yield useful additional information.\todo{Reviewer1:
  What does folding mean?  And why ''Hence'' in the last sentence? Why
  does the statement follow? \\ BK: I rewrote this part.}

\begin{example}
  \label{ex:path-monoid}
  Given a graph $G$, we denote by $E_G^+\subseteq V_G\times V_G$ the
  transitive closure of the edge relation $E^{\to}_G = 
  \{(\sSrc_G(e),\sTgt_G(e)) \mid e \in E_G\}$. The \emph{path monoid}
  $\mathcal{P}_G$ of $G$ has the carrier set $\mathcal{P}(E_G^+)$. The
  partial order is simply inclusion and the monoid operation is
  defined as follows: given $P_0,P_1\in \mathcal{P}_G$, we have
  \begin{eqnarray*}
    P_0+P_1 & = & \{(v_0,v_n)\mid \exists v_1,\dots,v_{n-1} \colon
    (v_i,v_{i+1})\in P_{j_i}, \\ & & \qquad \qquad \qquad \dnchange{j_i \in 
    \{0,1\}}, 
    j_{i+1} = 
    1-j_i, 
    i\in\{0,\dots,n-1\} 
    \dnchange{\text{ and } n \in \mathbb{N}}\}.
  \end{eqnarray*}
  That is, new paths can be formed by concatenating alternating path
  fragments from $P_0,P_1$. It is obvious to see that $+$ is
  commutative and one can also show associativity. $P=\emptyset$ is
  the unit. \todo{Reviewer3: I think there are two conditions missing here: (i) 
  \(j_0=0\) and \\ (ii) \(n \in \mathbb{N}\) \\ BK:
  no, $j_0=0$ is wrong, this would not give us commutativity \\ DN: You're 
  right. Fixed it with \(j_0 \in \{0,1\} \).}
  Subtraction simply returns the first parameter: $P_0-P_1 = P_0$.
\end{example}
  
We will now formally define annotations for objects via a functor from
a given category to $\Mon$.

\begin{definition}[Annotations for objects]
  Given a category $\C$ and a functor $\mathcal{A} \colon \C \to \Mon$,
  an \emph{annotation based on $\mathcal{A}$} for an object $X \in \C$
  is an element $a \in \mathcal{A}(X)$.  We write $\mathcal{A}_\phi$,
  instead of $\mathcal{A}(\phi)$, for the action of functor
  $\mathcal{A}$ on a $\C$-arrow $\phi$.  We assume that for each
  object $X$ there is a \emph{standard annotation} based on
  $\mathcal{A}$ that we denote by $s_X$, thus
  $s_X \in \mathcal{A}(X)$.
\end{definition}

It can be shown quite straightforwardly that the forgetful functor
mapping an annotated object $X[a]$, with $a \in \mathcal{A}(X)$,
  to $X$ is an op-fibration (or co-fibration
\cite{j:categorical-logic-type-theory}), arising via the Grothendieck
construction.

Our first example is an annotation of graphs with global
multiplicities, counting nodes and edges, where the action of the
functor is to sum up those multiplicities.

\begin{example}
  \label{ex:global-annot}
  Given $n\in\mathbb{N}\backslash\{0\}$, we define the functor
  $\mathcal{B}^n:\GR \to \Mon$: For every graph $G$,
  $\mathcal{B}^n(G) = \mathcal{M}_n^{V_G\cup E_G}$. For every graph
  morphism $\varphi \colon G \to H$ and $a \in \mathcal{B}^n(G)$, we
  have %\\
  $\mathcal{B}^n_{\varphi}(a) \in \mathcal{M}_n^{V_{H} \cup E_{H}}$
  with:
  \[ \mathcal{B}^n_{\varphi}(a)(y) = \sum\limits_{\varphi(x)=y}^{}
    a(x), \quad 
    \textit{where } x \in (V_G \cup E_G) \textit{ and } y \in (V_{H} \cup 
    E_{H}). \]
  Therefore an annotation based on a functor $\mathcal{B}^n$
  associates every item of a graph with a number (or the top value
  $*$). We will call such \deleted{kind of} annotations 
  \emph{multiplicities}.\todo{Reviewer1: ''such kind of annocations`` should be 
  ''such annotations`` \\ DONE}
  Furthermore the action of the functor on a morphism transforms a
  multiplicity by summing up (in $\mathcal{M}_n$) the values of all
  items of the source graph that are mapped to the same item of the
  target graph.

  For a graph $G$, its \emph{standard multiplicity}
  $s_G \in \mathcal{B}^n(G)$ is defined as the function which maps
  every node and edge of $G$ to $1$.
\end{example}

As another example we consider local annotations which record the
out-degree of a node and where the action of the functor is to take
the supremum instead of the sum.

\begin{example}
  \label{ex:local-annot}
  Given $n\in\mathbb{N}\backslash\{0\}$, we define the functor
  $\mathcal{S}^n:\GR \to \Mon$ as follows: For every graph $G$,
  $\mathcal{S}^n(G) = \mathcal{M}_n^{V_G}$.  For every graph morphism
  $\varphi \colon G \to H$ and $a \in \mathcal{S}^n(G)$, we have %\\
  $\mathcal{S}^n_{\varphi}(a) \in \mathcal{M}_n^{V_{H}}$ with:
    \[ \mathcal{S}^n_{\varphi}(a)(w) = \bigvee\limits_{\varphi(v)=w}^{}
      a(v), \quad \textit{where } v \in V_G \textit{ and }
      w \in V_{H}.  \]
  For a graph $G$, its \emph{standard annotation}
  $s_G \in \mathcal{S}^n(G)$ is defined as the function which maps
  every node of $G$ to its out-degree (or $*$ if the out-degree is
  larger than $n$).
\end{example}

Finally, we consider annotations based on the path monoid (see
Ex.~\ref{ex:path-monoid}).

\begin{example}
  \label{ex:path-annot}
  We define the functor $\mathcal{T}\colon\GR \to \Mon$ as follows:
  For every graph $G$, $\mathcal{T}(G) = \mathcal{P}_G$. For every
  graph morphism $\varphi \colon G \to H$ and $P \in \mathcal{T}(G)$,
  we have $\mathcal{T}_{\varphi}(P) \in \mathcal{P}_{H}$ with:
  \[ \mathcal{T}_{\varphi}(P) = \{(\phi(v),\phi(w))\mid (v,w)\in
    P\}. \]
  For a graph $G$, its \emph{standard annotation}
  $s_G \in \mathcal{T}(G)$ is the transitive closure of the edge
  relation, i.e., $s_G = E_G^+$.
\end{example}

In the following we will consider only annotations satisfying certain
properties in order to achieve soundness and completeness.\todo{Reviewer1: On 
page 11 some properties of annotations are given. How strong are they? How 
widely supported? What goes wrong without them? \\ BK: good question
\dots}

\begin{definition}[Properties of annotations]
  \label{def:prop-annotations}
  Let $\mathcal{A}:\C \to \Mon$ be an annotation functor, together
  with standard annotations. In this setting we say that 
  \begin{itemize}
  \item the \emph{homomorphism property} holds if whenever $\phi$ is a
    mono, then $\mathcal{A}_\phi$ is a monoid homomorphism, preserving
    also subtraction.
  \item the \emph{adjunction property} holds if whenever
    $\phi\colon A\rat B$ is a mono, then
    \begin{itemize}
    \item $\mathcal{A}_\phi\colon \mathcal{A}(A)\to \mathcal{A}(B)$
      has a right adjoint
      $\mathit{red}_\phi\colon \mathcal{A}(B)\to \mathcal{A}(A)$,
      i.e., $\mathit{red}_\phi$ is monotone and satisfies
      $a \le \mathit{red}_\phi(\mathcal{A}_\phi(a))$ for
      $a\in \mathcal{A}(A)$ and
      $\mathcal{A}_\phi(\mathit{red}_\phi(b))\le b$ for
      $b\in \mathcal{A}(B)$.\footnotemark 
    \item $\mathit{red}_\phi$ is a monoid homomorphism that preserves
      subtraction.
    \item it holds that $\mathit{red}_\phi(s_B) = s_A$, where
        $s_A,s_B$ are standard annotations.
    \end{itemize}
  \end{itemize}
  \footnotetext{This amounts to saying that the forgetful functor is a 
  bifibration when we restrict to monos, see       
  \cite[Lem.~9.1.2]{j:categorical-logic-type-theory}.}
  Furthermore, assuming that $\mathcal{A}_\phi$ has a right adjoint
  $\mathit{red}_\phi$, we say that 
  \begin{itemize}
  \item 
    \begin{minipage}[t]{0.8\linewidth}
      the \emph{pushout property} holds, whenever for each pushout as
      shown in the diagram to the right, with all arrows monos where
      $\eta = \psi_1 \circ \phi_1 = \psi_2 \circ \phi_2$, it holds
      that for every $d\in \mathcal{A}(D)$:\full{\footnotemark} 
      \[ d = \mathcal{A}_{\psi_1}(\mathit{red}_{\psi_1}(d)) +
        (\mathcal{A}_{\psi_2}(\mathit{red}_{\psi_2}(d)) -
        \mathcal{A}_{\eta}(\mathit{red}_{\eta}(d))). \] 
    \end{minipage}
    \begin{minipage}[t]{0.2\linewidth}
      \vspace{-0.7cm}
      \[
        \xymatrix{ A \ar@{>->}[r]^{\phi_2} \ar@{>.>}[dr]_{\eta}
          \ar@{>->}[d]_{\phi_1} & C
          \ar@{>->}[d]^{\psi_2} \\
          B \ar@{>->}[r]_{\psi_1} & D }
      \] \ 
    \end{minipage}
    \noindent We say that the \emph{pushout property for standard
      annotations} holds if we replace $d$ by $s_D$,
    $\mathit{red}_\eta(d)$ by $s_A$, $\mathit{red}_{\psi_1}(d)$ by
    $s_B$ and $\mathit{red}_{\psi_2}(d)$ by
    $s_C$.\full{\footnotetext{Note that the brackets below are
        essential, for instance in $\mathcal{M}_3$ we have
        $2+(2-1) = 3 \neq * = (2+2)-1$.}}
  \item 
    \begin{minipage}[t]{0.8\linewidth}
    the \emph{Beck-Chevalley property} holds if whenever the square shown to the 
    right is a pullback with $\phi_1$,
    $\psi_2$ mono, then it holds for every $b\in\mathcal{A}(B)$~that
    \[ \mathcal{A}_{\phi_2}(\mathit{red}_{\phi_1}(b)) =
      \mathit{red}_{\psi_2}(\mathcal{A}_{\psi_1}(b)). \]
    \end{minipage}
    \begin{minipage}[t]{0.2\linewidth}
      \vspace{-0.7cm}
      \[
        \xymatrix{
          A \ar[r]^{\phi_2} \ar@{>->}[d]_{\phi_1} & C \ar@{>->}[d]^{\psi_2} \\
          B \ar[r]_{\psi_1} & D \itul{\rm\footnotesize\color{gray}(PB)}
        }
      \]
    \end{minipage}
  \end{itemize}
\end{definition}

Note that the annotation functor from Ex.~\ref{ex:global-annot}
satisfies all properties above, whereas the functors from
Ex.~\ref{ex:local-annot} and~\ref{ex:path-annot} satisfy both the
homomorphism property and the pushout property for standard
annotations, but do not satisfy all the remaining requirements
\short{(for more details see the full version
  \cite{chknr:materialization-arxiv})}\full{(see
  Lem.~\ref{lem:global-annot-properties},
  \ref{lem:local-annot-properties} and~\ref{lem:path-annot-properties}
  in App.~\ref{apx:proofs})}.\todo{App. reference. DONE}

We will now introduce objects with two annotations, giving lower and
upper bounds.\todo{Reviewer1: Say what the 
upper and lower bounds are for. \\ BK: does anybody have a good idea
what to write her?}

\begin{definition}[Doubly annotated object]
  \label{def:doubly-annotated-object}
  Given a topos $\C$ and a functor
  $\mathcal{A}\colon \C \to \mathbf{Mon}$, a \emph{doubly annotated
    object} $A[a_1,a_2]$ is an object $A$ of $\C$ with two annotations
  $a_1,a_2\in \mathcal{A}(A)$.
  An arrow $\phi\colon A[a_1,a_2]\to B[b_1,b_2]$, also called a
  \emph{legal arrow}, is a $\C$-arrow $\phi\colon A\to B$ such that
  $\mathcal{A}_\phi(a_1)\ge b_1$ and $\mathcal{A}_\phi(a_2) \le b_2$.

  The \emph{language of a doubly annotated object} $A[a_1,a_2]$ (also
  called the language of objects which are abstracted by $A[a_1,a_2]$)
  is defined as follows:
  \[ \mathcal{L}(A[a_1,a_2]) = \{ X\in \C\mid \mbox{there exists a
      legal arrow $\phi\colon X[s_X,s_X]\to A[a_1,a_2]$}\} \]
\end{definition}

Note that legal arrows are closed under composition
\cite{ckn:graph-languages-type-graphs}. Examples of doubly annotated
objects are given in Ex.~\ref{ex:annot-rewriting-step} for global
annotations from Ex.~\ref{ex:global-annot} (providing upper and lower
bounds for the number of nodes resp.\ edges in the preimage of a given
element). Graph elements without annotation are annotated by $[0,*]$
by default.

\begin{definition}[Isomorphism property]
  \label{def:isomorphism-property}
  An annotation functor $\mathcal{A}\colon \C\to \mathbf{Mon}$,
  together with standard annotations, satisfies the \emph{isomorphism
    property} if the following holds: whenever
  $\phi\colon X[s_X,s_X]\to Y[s_Y,s_Y]$ is legal, then $\phi$ is an
  isomorphism, i.e.,  $\mathcal{L}(Y[s_Y,s_Y])$ contains only $Y$
  itself (and objects isomorphic to $Y$).
\end{definition}

% ============================================================================ %
\section{Abstract Rewriting of Annotated Objects}
\label{sec:abstract-rewriting-annotated}
% ============================================================================ %

We will now show how to actually rewrite annotated objects. The
challenge is both to find suitable annotations for the materialization
and to ``rewrite'' the annotations.

\subsection{Abstract Rewriting and Soundness}
\label{sec:abstract-rewriting-soundness}

We first describe how the annotated rewritable materialization is
constructed and then we investigate its properties.

\begin{definition}[Construction of annotated rewritable
  materialization]
  \label{def:annotated-rewritable-materialization}
  Let $p\colon L\oset{\phi_L}{\lat} I\oset{\phi_R}{\rat} R$ be a
  production and let $A[a_1,a_2]$ be a doubly annotated object.
  Furthermore let $\phi\colon L\to A$ be an arrow.

  We first construct the factorization
  $L \oset{n_L}{\rat} \rmat{\phi}{\phi_L} \oset{\psi}{\to} A$,
  obtaining the rewritable materialization $\rmat{\phi}{\phi_L}$ from
  Def.~\ref{def:materialization-category-rewritable}. Next, let~$M$
  contain all maximal\footnote{``Maximal'' means maximality with
    respect to the interval order
    $(a_1,a_2)\sqsubseteq (a'_1,a'_2) \iff a'_1\le~a_1, a_2\le a'_2$.}
  elements of the set
  \[ \{(a'_1,a'_2)\in \mathcal{A}(\rmat{\phi}{\phi_L})^2 \mid
    \mathcal{A}_{n_L}(s_L)\le a'_2, a_1 \le \mathcal{A}_\psi(a'_1),
    \mathcal{A}_\psi(a'_2) \le a_2 \}. \] Then the doubly annotated
  objects $\rmat{\phi}{\phi_L}[a'_1,a'_2]$ with $(a'_1,a'_2)\in M$ are
  the annotated rewritable materializations for $A[a_1,a_2]$, $\phi$
  and $\phi_L$.
\end{definition}
Note that in general \dnchange{there can be several such
  materializations, differing by}\todo{Reviewer1: ''there are several
  materializations, differing for`` should be ''there can be several
  materializations, differing by``.  \\DN: Replaced it. DONE} the
annotations only, or possibly none.  The definition of $M$ ensures
that the upper bound $a'_2$ of the materialization covers the
annotations arising from the left-hand side. We cannot use a
corresponding condition for the lower bound, since the materialization
might contain additional structures, hence the arrow $n_L$ is only
``semi-legal''. A more symmetric condition will be studied in
Sec.~\ref{sec:completeness}.
 
\newcommand{\annotMatMorphism}{ Given a production 
  $p\colon L\oset{\phi_L}{\lat} I\oset{\phi_R}{\rat} R$, let $L\oset{m_L}{\rat} 
  X$ be the match of $L$ in an
  object $X$ such that $X\oset{p,m_L}{\Longrightarrow}$, i.e., $X$ can
  be rewritten. Assume that $X$ is abstracted by $A[a_1,a_2]$,
  witnessed by $\psi$. Let $\phi = \psi\circ m_L$ and let
  $L\oset{n_L}{\rat} \rmat{\phi}{\phi_L} \oset{\psi'}{\to} A$ the the
  corresponding rewritable materialization. Then there exists an arrow
  $\zeta_A$ and a pair of annotations $(a'_1,a'_2)\in M$ for
  $\rmat{\phi}{\phi_L}$ (as described in
  Def.~\ref{def:annotated-rewritable-materialization}) such that the
  diagram below commutes and the square is a pullback in the
  underlying category. Furthermore the triangle consists of legal
  arrows. This means in particular that $\zeta_A$ is legal.
  \[
    \xymatrix{
      L[s_L,s_L] \ar@{>->}_{\mathit{id}_L}[d] \ar@{>->}[r]^{m_L} 
      & X[s_X,s_X] \ar[d]^{\zeta_A} \ar[r]^{\psi} & A[a_1,a_2] \\
      L[s_L,s_L] \ar@{>->}[r]_(.45){n_L} & {\rmat{\phi}{\phi_L}}[a'_1,a'_2] 
      \itull{\rm\footnotesize\color{gray} \hspace*{-.5cm} 
      (PB)\phantom{~~~~~~~~~}}
      \ar[ru]_{\psi'} &
    }
  \]
}

\begin{proposition}[Annotated rewritable materialization is terminal]
  \label{prop:morphism-with-mult-into-mat}
  \annotMatMorphism
\end{proposition}

\dnchange{Having performed}\todo{Reviewer1: ''Once we have performed`` should 
be ''Having performed`` \\ DN: Replaced it. DONE} the materialization, we will 
now show how to
rewrite annotated objects. Note that we cannot simply take pushouts
in the category of annotated objects and legal arrows, since this
would result in taking the supremum of annotations, when instead we
need the sum (subtracting the annotation of the interface $I$,
analogous to the inclusion-exclusion principle).

\begin{definition}[Abstract rewriting step $\leadsto$]
  \label{def:abstract-rewriting}
  Let $p\colon L\oset{\phi_L}{\lat} I\oset{\phi_R}{\rat} R$ be a
  production and let $A[a_1,a_2]$ be an annotated abstract
  object. Furthermore let $\phi\colon L\to A$ be a match of a
  left-hand side, let $n_L\colon L\rat \rmat{\phi}{\phi_L}$ be the
  match obtained via materialization and let $(a'_1,a'_2)\in M$ (as in
  Def.~\ref{def:annotated-rewritable-materialization}).

  Then $A[a_1,a_2]$ can be transformed to $B[b_1,b_2]$ via
  $p$ if there are arrows such that the
  two squares below are pushouts in the base category and 
   $b_1,b_2$ are defined as:
   \[ b_i = \mathcal{A}_{\phi_B}(c_i) + (\mathcal{A}_{n_R}(s_R) -
    \mathcal{A}_{n_R\circ \phi_R}(s_I)) \qquad \mbox{for
      $i\in\{1,2\}$} \]
  where    
  $c_1,c_2$ are maximal annotations such that:
  \[ a'_1 \le \mathcal{A}_{\phi_A}(c_1) + (\mathcal{A}_{n_L}(s_L) -
    \mathcal{A}_{n_L\circ \phi_L}(s_I)) \quad
    \mathcal{A}_{\phi_A}(c_2) + (\mathcal{A}_{n_L}(s_L) -
    \mathcal{A}_{n_L\circ \phi_L}(s_I)) \le a'_2 \]
  \[
    \xymatrix{ L[s_L,s_L] \ar@{>->}[d]_{n_L} & I[s_I,s_I]
      \ar@{>->}[l]_{\phi_L} \ar@{>->}[r]^{\phi_R} \ar@{>->}[d]^{n_I} &
      R[s_R,s_R]
      \ar@{>->}[d]^{n_R} \\
      {\rmat{\phi}{\phi_L}}[a'_1,a'_2] & C[c_1,c_2]
      \ar@{>->}[l]_(.45){\phi_A} \ar@{>->}[r]^(.55){\phi_B} &
      B[b_1,b_2] }
  \]  
  In this case we write
  $A[a_1,a_2] \oset{p,\phi}{\leadsto} B[b_1,b_2]$ and say that
  $A[a_1,a_2]$ makes an \emph{abstract rewriting step} to
  $B[b_1,b_2]$.
\end{definition}

We will now show soundness of abstract rewriting,\todo{Reviewer1: On page 13 
you say you will ''show soundness of abstract rewriting``. But, sadly, nothing 
is actually ''shown``. \\ DN: Probably refers to the missing proofs in the 
final version?} i.e., whenever an
object $X$ is abstracted by $A[a_1,a_2]$ and $X$ is rewritten to $Y$,
then there exists an abstract rewriting step from $A[a_1,a_2]$ to
$B[b_1,b_2]$ such that $Y$ is abstracted by $B[b_1,b_2]$.

\smallskip

\noindent\textbf{\emph{Assumption}}: In the following we will require that
the homomorphism property as well as the pushout property for standard
annotations hold (cf. Def.~\ref{def:prop-annotations}).

\newcommand{\soundnessProp}{ Relation $\leadsto$ is sound in the
  following sense: Let $X\in \mathcal{L}(A[a_1,a_2])$ (witnessed via a
  legal arrow $\psi\colon X[s_X,s_X]\to A[a_1,a_2]$) where
  $X\oset{p,m_L}{\Longrightarrow} Y$. Then there exists an abstract
  rewriting step
  $A[a_1,a_2]\oset{p,\psi\circ m_L}{\leadsto} B[b_1,b_2]$ such that
  $Y\in \mathcal{L}(B[b_1,b_2])$.  }

\begin{proposition}[Soundness for $\leadsto$]
  \label{prop:soundness}
  \soundnessProp
\end{proposition}

%\vspace*{-.5cm}
\subsection{Completeness}
\label{sec:completeness}

The conditions \bkchange{on the annotations} that we imposed so far
are too weak to guarantee completeness, that is the fact that every
object represented by $B[b_1,b_2]$ can be obtained by rewriting an
object represented by $A[a_1,a_2]$. \todo{Reviewer1: The first
  sentence of Sec 5.2 should be: ''The condition that we have imposed
  thus far -- namely, that every object ...
  represented by A[\(a_1,a_2\)] -- is too weak to guarantee completeness.`` \\
  DN: That's not the condition that we imposed so far but rather the
  description of the completeness. So I guess the sentence is right as
  it is? \\ BK: added ``on the annotations''} This can be clearly seen
by the fact that the requirements hold also for the singleton monoid
and, as discussed before, the graph structure of $B$ is insufficient
to characterize the successor objects or graphs.

Hence we will now strengthen our requirements in order to obtain
completeness.

\smallskip

\noindent \textbf{\emph{Assumption:}} In addition to the assumptions
of \Sct.~\ref{sec:abstract-rewriting-soundness}, we will need that
subtraction is well-behaved and that the adjunction property, the
pushout property, the Beck-Chevalley property
(Def.~\ref{def:prop-annotations}) and the isomorphism property
\bkchange{(Def.~\ref{def:isomorphism-property})} hold.

\smallskip

The global annotations from Ex.~\ref{ex:global-annot} satisfy all
these properties. In particular, given an injective graph morphism
$\phi\colon G\rat H$ the right adjoint
$\mathit{red}_\phi : \mathcal{M}_n^{V_H\cup E_H} \to
\mathcal{M}_n^{V_G\cup E_G}$ to $\mathcal{B}^n_\phi$ is defined as
follows: given an annotation $b\in \mathcal{M}_n^{V_H\cup E_H}$,
$\mathit{red}_\phi(b)(x) = b(\phi(x))$, i.e., $\mathit{red}_\phi$
simply provides a form of reindexing\full{ (see also
  Lem.~\ref{lem:global-annot-properties} in
  App.~\ref{apx:proofs})}.\todo{App.  reference. DONE}

\noindent We will now modify the abstract rewriting relation and
allow only those abstract annotations for the materialization that
reduce to the standard annotation of the left-hand~side.

\begin{definition}[Abstract rewriting step $\hookrightarrow$]
  \label{def:abstract-rewriting-variant}
  Given $\phi\colon L\to A$, assume that $B[b_1,b_2]$ is
  constructed from $A[a_1,a_2]$ via the construction described in
  Def.~\ref{def:annotated-rewritable-materialization}
  and~\ref{def:abstract-rewriting}, with the modification that the set
  of annotations from which the set of maximal annotations $M$ of the
  materialization $\rmat{\phi}{\phi_L}$ are taken, is replaced by:
  \begin{eqnarray*}
    && \{(a'_1,a'_2)\in \mathcal{A}(\rmat{\phi}{\phi_L})^2 \mid 
    \mathit{red}_{n_L}(a'_i) = s_L,
    i\in\{1,2\}, %\\
    %&& \qquad\qquad
    a_1\le \mathcal{A}_\psi(a'_1),
    \mathcal{A}_\psi(a'_2) \le a_2\}. 
  \end{eqnarray*}
  In this case we write
  $A[a_1,a_2] \oset{p,\phi}{\hookrightarrow} B[b_1,b_2]$.
\end{definition}

Due to the adjunction property we have
  $\mathcal{A}_{n_L}(s_L) =
  \mathcal{A}_{n_L}(\mathit{red}_{n_L}(a'_2)) \le a'_2$ and hence the
  set $M$ of annotations of Def.~\ref{def:abstract-rewriting-variant}
  is a subset of the corresponding set of
  Def.~\ref{def:abstract-rewriting}.

\begin{example}
  \label{ex:annot-rewriting-step}
  We give a small example of an abstract rewriting step (a more
  extensive, worked example can be found in \short{the full version
    \cite{chknr:materialization-arxiv}}\full{App.~\ref{apx:worked-example}}).\todo{App.
     reference. DONE}
  Elements without annotation are annotated by $[0,*]$ by default and
  those with annotation $[0,0]$ are omitted. Furthermore elements in
  the image of the match and co-match are annotated by the standard
  annotation $[1,1]$ to specify the concrete occurrence of the
  left-hand and right-hand side.
  
\vspace*{-.3cm}
  \begin{center}
    \scalebox{.69}{
      \begin{tikzpicture}[x=3.2cm,y=-1.8cm,baseline=(b2.south)]
        \node[gnodeb] (a1) at (0,-1.5) {} ;
        \node (a1lab) at (0,-1.25) {$[1,1]$} ;
        \draw[gedge,thick] (a1) .. controls +(330:1cm) and +(30:1cm) .. 
        node[midway,right] {$C\ [1,1]$} (a1) ;
        \ghostgraphbox[a]{(a1)};

        \node[gnodeb] (i1) at (1.75,-1.5) {} ;
        \node (i1lab) at (1.75,-1.25) {$[1,1]$} ;
        \ghostgraphbox[i]{(i1)};

        \node[gnodeb] (b1) at (3.05,-1.5) {} ;
        \node[gnode] (b2) at (3.45,-1.5) {} ;
        \node (b2lab) at (3.6,-1.5) {$[1,1]$} ;
        \node (b1lab) at (3.05,-1.25) {$[1,1]$} ;
        \draw[gedge] (b1) to [bend angle=30,bend left] node[midway,above] 
        (blaa) {$\qquad \quad A\ [1,1]$} (b2);
        \draw[gedge] (b2) to [bend angle=30,bend left] node[midway,below] 
        (blab) {$\qquad \quad B\ [1,1]$} (b1);
        \ghostgraphbox[b]{(b1) (b2) (blab)};
        
        \node[gnode] (t1) at (-.2,0) {} ;
        \node[gnodeb] (t2) at (.2,0) {} ;
        \node (t2lab) at (.2,.25) {$[1,1]$} ;
        \draw[gedge] (t1) to [bend angle=30,bend left] node[midway,above] 
        {$D$} (t2); 
        \draw[gedge] (t2) to [bend angle=40,bend left] node[midway,above] 
        {$D$} (t1);
        \draw[gedge] (t1) .. controls +(150:1cm) and +(210:1cm) .. 
        node[above=.15cm] {$D$} (t1) ;
        \draw[gedge] (t2) .. controls +(60:1cm) and +(120:1cm) .. 
        node[midway,above] {$D$} (t2) ;
        \draw[gedge,thick] (t2) .. controls +(330:1cm) and +(30:1cm) .. 
        node[midway,right] {$C\ [1,1]$} (t2) ;
        \ghostgraphbox[t]{(t1) (t2)};

        \node[gnode] (c1) at (1.55,0) {} ;
        \node[gnodeb] (c2) at (1.95,0) {} ;
        \node (c2lab) at (1.95,.25) {$[1,1]$} ;
        \draw[gedge] (c1) to [bend angle=30,bend left] node[midway,above] 
        {$D$} (c2); 
        \draw[gedge] (c2) to [bend angle=40,bend left] node[midway,above] 
        {$D$} (c1);
        \draw[gedge] (c1) .. controls +(150:1cm) and +(210:1cm) .. 
        node[above=.15cm] {$D$} (c1) ;
        \draw[gedge] (c2) .. controls +(60:1cm) and +(120:1cm) .. 
        node[midway,above] {$D$} (c2) ;
        \ghostgraphbox[c]{(c1) (c2)};

        \node[gnode] (h1) at (2.85,0) {} ;
        \node[gnodeb] (h2) at (3.25,0) {} ;
        \node[gnode] (h3) at (3.65,0) {} ;
        \node (h2lab) at (3.25,.25) {$[1,1]$} ;
        \node (h3lab) at (3.8,0) {$[1,1]$} ;
        \draw[gedge] (h1) to [bend angle=30,bend left] node[midway,above] 
        {$D$} (h2); 
        \draw[gedge] (h2) to [bend angle=40,bend left] node[midway,above] 
        {$D$} (h1);
        \draw[gedge] (h1) .. controls +(150:1cm) and +(210:1cm) .. 
        node[above=.15cm] {$D$} (h1) ;
        \draw[gedge] (h2) .. controls +(60:1cm) and +(120:1cm) .. 
        node[midway,above] {$D$} (h2) ;
        \draw[gedge] (h2) to [bend angle=30,bend left] node[midway,above] 
        (hlaa) {$\qquad \quad A\ [1,1]$} (h3);
        \draw[gedge] (h3) to [bend angle=30,bend left] node[midway,below] 
        (hlab) {$\qquad \quad B\ [1,1]$} (h2);
        \ghostgraphbox[h]{(h1) (h2) (h3)};

        \node[gnodeb] (f1) at (-1.2,-1.5) {} ;
        \node (f1lab) at (-1.2,-1.25) {$[1,*]$} ;
        \draw[gedge,thick] (f1) .. controls +(330:1cm) and +(30:1cm) .. 
        node[midway,right] {$C\ [1,1]$} (f1) ;
        \draw[gedge] (f1) .. controls +(150:1cm) and +(210:1cm) .. 
        node[midway,left] {$D$} (f1) ;
        \ghostgraphbox[f]{(f1) (f1lab)};

        \draw[{>[scale=1.3]}-{>[scale=1.3]}] (1.6,-1.5) -- 
        (.7,-1.5) ; \node[glab] (m) at (1.1,-1.65) {{\large 
                                $\varphi_L$}};
        \draw[{>[scale=1.3]}-{>[scale=1.3]}] (1.9,-1.5) -- 
        (2.9,-1.5) ; \node[glab] (m) at (2.4,-1.65) {{\large 
                                        $\varphi_R$}};

        \draw[{>[scale=1.3]}-{>[scale=1.3]}] (1.2,0) -- 
        (.9,0) ; \node[glab] (m) at (1.05,-.15) {{\large 
                                        $\varphi_A$}};
        \draw[{>[scale=1.3]}-{>[scale=1.3]}] (2.1,0) -- 
        (2.5,0) ; \node[glab] (m) at (2.3,-.15) {{\large 
                                                $\varphi_B$}};

        \draw[{>[scale=1.3]}-{>[scale=1.3]}] (0,-1) -- 
        (0,-.55) ; \node[glab] (m) at (-.125,-.8) {{\large 
                                                $n_L$}};
        \draw[{>[scale=1.3]}-{>[scale=1.3]}] (1.75,-1) -- 
        (1.75,-.55) ; \node[glab] (m) at (1.64,-.8) {{\large 
                                                        $n_I$}};
        \draw[{>[scale=1.3]}-{>[scale=1.3]}] (3.25,-1) -- 
        (3.25,-.65) ; \node[glab] (m) at (3.375,-.825) {{\large 
                                                        $n_R$}};

        \draw[-{>[scale=1.3]}] (-.5,-.55) -- (-1,-1) ;
        \draw [-{>[scale=1.3]}] (-.15,-1.5) -- (-.5,-1.5);
           \node[glab] (m) at (-.3,-1.65) {{\large 
                                          $\varphi$}};
        \draw[decorate,decoration={brace,amplitude=10pt,raise=4pt},yshift=0pt]
        (-1.55,-1.9) -- (3.75,-1.9) node 
        [black,midway,yshift=.8cm] 
        {$A \leftarrow L\lat I\rat R$};
      \end{tikzpicture}
    }
  \end{center}
\end{example}

The variant of abstract rewriting introduced in
Def.~\ref{def:abstract-rewriting-variant} can still be proven to be
sound, assuming the extra requirements stated above.

\newcommand{\soundnessPropVariant}{
  Relation $\hookrightarrow$ is sound in the sense of
  Prop.~\ref{prop:soundness}.
}

\begin{proposition}[Soundness for $\hookrightarrow$]
  \label{prop:soundness-variant}
  \soundnessPropVariant
\end{proposition}

Using the assumptions we can now show completeness.

\newcommand{\completenessProp}{If
  $A[a_1,a_2]\oset{p,\phi}{\hookrightarrow} B[b_1,b_2]$ and 
  $Y\in \mathcal{L}(B[b_1,b_2])$, then there exists
  $X\in \mathcal{L}(A[a_1,a_2])$ (witnessed via a legal arrow
  $\psi\colon X[s_X,s_X]\to A[a_1,a_2]$) such that
  $X\oset{p,m_L}{\Longrightarrow} Y$ and $\phi = \psi\circ m_L$.
}

\begin{proposition}[Completeness for $\hookrightarrow$]
  \label{prop:completeness}
  \completenessProp
\end{proposition}

Finally, we can show that annotated graphs of this kind are expressive
enough to construct a strongest post-condition. If we would allow
several annotations for objects, as in
\cite{ckn:graph-languages-type-graphs}, we could represent the
language with a single (multiply) annotated object.

\newcommand{\corollaryStrongestPost}{ Let $A[a_1,a_2]$ be an annotated
  object and let $\phi\colon L\to A$. We obtain (several) abstract
  rewriting steps
  $A[a_1,a_2] \oset{p,\phi}{\hookrightarrow} B[b_1,b_2]$, where we
  always obtain the same object $B$. ($B$ is dependent on $\phi$, but
  not on the annotation.) Now let
  $N = \{(b_1,b_2) \mid A[a_1,a_2]
  \oset{p,\phi}{\hookrightarrow} B[b_1,b_2]\}$.  Then
  \begin{eqnarray*}
    \bigcup_{(b_1,b_2)\in N} \mathcal{L}(B[b_1,b_2]) & = & \{Y \mid \exists (X\in
    \mathcal{L}(A[a_1,a_2]), \mbox{witnessed by $\psi$}),
    (L\oset{m_L}{\rat} X).\ \\[-0.3cm]
    && \qquad (\phi = \psi\circ m_L \land
    X\oset{p,m_L}{\Longrightarrow} Y)\}
  \end{eqnarray*}
}

\begin{corollary}[Strongest post-condition]
  \label{cor:strongest-postcondition}
  \corollaryStrongestPost
\end{corollary}

% \tinytodo{B: Introduce a notation for $X\in \mathcal{L }(A[a_1,a_2])$,
%   \mbox{witnessed by $\psi$ }.}

% ============================================================================ %
\section{Conclusion}
\label{sec:conclusion}
% ============================================================================ %

We have described a rewriting framework for %
abstract graphs that also applies to objects in any topos, %
based on existing work for
graphs~\cite{sww:abstract-gts,bw:partner-abstraction,%
b:cluster-abstraction-thesis,br:cluster-abstraction,%
rz:neighbourhood-abstraction-groove,r:canonical-graph-shapes}.
In particular, %
we have given a blueprint for materialization %
in terms of the universal property of partial map classifiers. %
This is a first theoretical milestone towards %
shape analysis as a general static analysis method %
for rule-based systems with graph-like objects as states. %
Soundness and completeness results for the %
rewriting of abstract objects with annotations in an ordered monoid %
provide an effective verification method %
for the special case of graphs\full{ (see also
  App.~\ref{apx:worked-example})}.\todo{App. reference. DONE} %
We plan to implement the materialization construction and the
computation of rewriting steps of abstract graphs in \dnchange{a} %\deleted{the}
prototype
tool. %\deleted{{\scalebox{0.87}{\textsf{DrAGoM}}}}.
\todo{Reviewer1: Please 
include a URL for DrAGoM. \\ DN: The current state of DrAGoM is not ready for 
publication. Therefore I removed the name to emphasize that the tool is future 
work.}

The extension of annotations with logical formulas is 
the natural next step, %
which will lead to a more flexible and versatile specification language, % 
as described in previous work \cite{srw:shape-analysis-3vl,sww:abstract-gts}. %
The logic can possibly be developed in full generality %
using the framework of nested application conditions %
\cite{hp:nested-constraints,lo:tableau-graph-properties}
that applies to objects in adhesive categories. %
This logical approach might even reduce %
the proof obligations for annotation functors. 
Another topic for future work is %
the integration of %
widening or similar approximation techniques, %
which collapse abstract objects and ideally lead to finite
abstract transition systems %
that (over-)approximate the typically infinite transitions systems %
of graph transformation systems. %

\bibliography{references}
\bibliographystyle{plain}

\short{\end{document}}

\appendix

\newpage

\section{Definitions and Results about Topoi}
\label{apx:topos}

In this section, we present some known definitions and results related to
elementary topoi,  for the convenience of the reader.  \\

\noindent \begin{minipage}{0.78\textwidth}
  \begin{definition}[Subobject classifier]
    \label{def:subobclass}
    Let $\C$ be a category where $\mathbf{1}$ is the terminal object
    and for each object $X \in \C$ let
    $\mathbf{!}_X \colon X \to \mathbf{1}$ be the unique arrow from
    $X$ into the terminal object. A mono
    $\mathtt{true} \colon \mathbf{1} \rat \Omega$ is a subobject
    classifier if for every mono $i \colon X \rat Y$ in $\C$ there
    exists a unique arrow $\chi_i \colon Y \to \Omega$ such that the
    diagram to the right is a pullback. In this case object $\Omega$
    is called the \emph{truth value object}. \\
  \end{definition}
\end{minipage} 
\begin{minipage}{0.23\textwidth}
  \vspace*{-.2cm}
  \begin{center}
    \scalebox{1}{
      $$
      \xymatrix{ X \ar@{ >->}[r]^i \ar@{->}[d]_{\textbf{!}_X} &
        Y \ar@{.>}[d]^{\chi_i} \\
        \mathbf{1} \ar@{ >->}[r]_{\mathtt{true}} & \Omega
        \itul{\rm\footnotesize\color{gray}(PB)} }
      $$ 
    }
  \end{center}
\end{minipage}

In $\mathbf{Set}$ the subobject classifier
  $\mathtt{true}$ is simply the embedding of $\{1\}$ into the
  two-element set $\{0,1\}$. A subset $X\subseteq Y$ can be
  characterized via its characteristic function
  $\chi_X\colon Y\to\{0,1\}$.

\smallskip

The notion of elementary topos \cite{j:elephant-1} is used in
logic and it abstracts from the structure of the category of sets.

\begin{definition}[Elementary topos]
  An elementary topos is a category which has finite limits, is
  cartesian closed and has a subobject classifier.
\end{definition}

We will often omit the qualifier ``elementary'' and
simply talk about topoi. Every elementary topos 
has so-called \emph{partial map classifiers} \cite{CL2}.

\bigskip

  \begin{definition}[Partial map classifier]
    \label{def:partialmap}
   Let $\C$ be a category with pullbacks. A partial map
       $(m,f) \colon X \rightharpoonup Y$ in $\C$ is a span
       $X \oset{m}{\leftarrowtail} Z \oset{f}{\to} Y$ where $m \colon Z 
       \rightarrowtail X$ is a mono. A \emph{partial} \\
\noindent \begin{minipage}{0.73\textwidth}
    \vspace*{.05cm}
    \emph{map classifier} $(F,\eta)$ is a functor $F \colon \C \to \C$ together 
    with a
    natural transformation $\eta \colon \textit{Id}_{\C} \dotarrow F$ such
    that for each object $Y$ of $\C$ with the component
    $\eta_Y \colon Y \rightarrowtail F(Y)$ the following holds: for
    each partial map $(m,f) \colon X \rightharpoonup Y$ there exists a
    unique arrow $\varphi(m,f) \colon X \to
    F(Y)$ such that the diagram to the right is a pullback. 
\end{minipage}
\begin{minipage}{0.22\textwidth}
  \vspace*{-.4cm}
  \begin{center}
    \scalebox{1}{
      $$
      \xymatrix{
        Z \ar@{ >->}[r]^{m} \ar@{->}[d]_{f} & X \ar@{.>}[d]^{\varphi(m,f)} \\ 
        Y \ar@{ >->}[r]^{\eta_Y} & F(Y)
        \itull{\rm\footnotesize\color{gray}\ (PB)}}$$ 
    }
  \end{center}
\end{minipage}
\end{definition}

In $\mathbf{Set}$ the functor $F$ enriches each set $Y$ with an
  additional element $\star$, i.e., $F(Y) = Y + \{\star\}$. Then a
  partial map $p\colon X\rightharpoonup Y$ corresponds to a total map
  $p'\colon X\to F(Y)$ such $p'(x) = p(x)$ if $p(x)$ is defined and
  $p'(x) = \star$ otherwise.

  \begin{example}
    We now consider a more involved example in the category
      $\GR$.  Let the partial map $(m,f) \colon G \rightharpoonup H$
    (depicted below left) and a corresponding span
    $G \oset{m}{\leftarrowtail} P \oset{f}{\to} H$ (depicted below on
    the right) be given. We use a single edge label, which is
      omitted.

\smallskip \bigskip
    
    \noindent\begin{minipage}{0.40\textwidth}
    \vspace*{0.3cm}
    \begin{tabular}{rcl}
      \scalebox{.8}{
        \begin{tikzpicture}[x=3.2cm,y=-1.8cm,baseline=(g1.center)]
          \node[glab] (g0) at (-.3,0) {\large{$(m,f)$:}} ;
          \node[gnode] (g1) at (0,0) {} ;
          \node[gnode] (g2) at (0.3,0) {} ;
          \draw[gedge,->,thick] (g1) to node[midway,above] {$\A$} 
          (g2);
          \draw[gedge] (g1) .. controls +(45:1cm) and +(125:1cm) .. 
                        node[above] (labA) {$\A$} (g1) ;
          \ghostgraphbox[g]{(g1) (g2)};
        \end{tikzpicture} }
        & \hspace*{-.4cm}
        \scalebox{1.2}{
          $$
          \xymatrix{
           \ar@^{-{>}}[r] & } $$ 
        }
        & \hspace*{-.3cm}
      \scalebox{.8}{
        \begin{tikzpicture}[x=3.2cm,y=-1.8cm,baseline=(h1.center)]
          \node[gnode] (h1) at (1.4,0) {} ;
          \node[gnode] (h2) at (1.7,0) {} ;
          \draw[gedge,->,thick] (h1) to node[midway,above] {$\A$} (h2);
          \ghostgraphbox[h]{(h1)};
        \end{tikzpicture} } 
    \end{tabular}
\end{minipage} \quad
\begin{minipage}{0.60\textwidth}
    \begin{center}
      \vspace*{-.3cm}
      \scalebox{.8}{
        \begin{tikzpicture}[x=3.2cm,y=-1.8cm,baseline=(g1.south)]
          \node[gnode] (g1) at (0,0) {} ;
          \node[gnode] (g2) at (0.3,0) {} ;
          \draw[gedge,->,thick] (g1) to node[midway,above] {$\A$} (g2);
          \draw[gedge] (g1) .. controls +(45:1cm) and +(125:1cm) .. 
          node[above] (labA) {$\A$} (g1) ;
          \ghostgraphbox[g]{(g1) (g2)};
          
          \node[gnode] (h1) at (.9,0) {} ;
          \node[gnode] (h2) at (1.2,0) {} ;
          \draw[gedge,->,thick] (h1) to node[midway,above] {$\A$} (h2);
          \ghostgraphbox[h]{(h1)};
          
          \node[gnode] (i1) at (1.8,0) {} ;
          \node[gnode] (i2) at (2.1,0) {} ;
          \draw[gedge,->,thick] (i1) to node[midway,above] {$\A$} (i2);
          \ghostgraphbox[h]{(i1)};

          \draw[{>[scale=1.3]}-{>[scale=1.3]}] (.7,0) -- (0.5,0) ;
          \draw[-{>[scale=1.3]}] (1.4,0) -- (1.6,0) ;
          
          \node[glab] (m) at (.6,-.15) {{\large $m$}};
          \node[glab] (f) at (1.5,-.15) {{\large $f$}};
          
          \draw[decorate,decoration={brace,amplitude=10pt,raise=4pt},yshift=0pt]
          (-.1,-.3) -- (2.2,-.3) node [black,midway,yshift=.8cm]
          {\footnotesize $G\lat P \to H$};
        \end{tikzpicture}
      }
    \end{center}
\end{minipage} \\ \newpage

\noindent 
The partial map classifier object $F(H)$ alongside the component of the natural 
transformation $\eta_{H}\colon H \rat F(H)$ is depicted below:

  \medskip
      \begin{center}
        \vspace*{-.4cm}
        \scalebox{.7}{
          \begin{tikzpicture}[x=3.2cm,y=-1.8cm,baseline=(f2.south)]
            \node[gnode] (a1) at (-.2,-1.5) {} ;
            \node[gnode] (a2) at (.2,-1.5) {} ;
            \draw[gedge,thick] (a1) to node[midway,above] (alab) {$\A$} (a2);
            \ghostgraphbox[a]{(a1) (a2) (alab)};

            \node[gnode] (i1) at (-.2,.2) {} ;
            \node[gnode] (i2) at (.2,.2) {} ;
            \draw[gedge,thick] (i1) to node[midway,above] (ilab) {$\A$} (i2);
            \ghostgraphbox[i]{(i1) (i2) (ilab)};
            
            \node[gnode] (t1) at (1.0,-.3) {} ;
            \node[gnode] (t2) at (1.4,-.3) {} ;
            \node[gnode] (t3) at (1.2,.4) {} ;
            \draw[gedge,thick] (t1) to node[midway,above] {$\A$} (t2); 
            \draw[gedge,<->,dashed] (t1) to [bend angle=50,bend right] 
            node[glab,midway,below=.1cm] {$\A$} (t2);
            \draw[gedge,<->,dashed] (t1) to [bend angle=30,bend right]  
            node[glab,midway,left] {$\A$} (t3);
            \draw[gedge,<->,dashed] (t2) to [bend angle=30,bend left]   
            node[glab,midway,right] {$\A$} (t3);
            \draw[gedge,dashed] (t1) .. controls +(140:1cm) and +(220:1cm) .. 
            node[glab,left] {$\A$} (t1) ;
            \draw[gedge,dashed] (t2) .. controls +(40:1cm) and +(320:1cm) .. 
            node[glab,right] {$\A$} (t2) ;
            \draw[gedge,dashed] (t3) .. controls +(225:1cm) and +(305:1cm) .. 
            node[glab,below] {$\A$} (t3) ;
            \ghostgraphbox[t]{(t1) (t2) (t3)};

            \node[gnode] (f1) at (1.0,-1.5) {} ;
            \node[gnode] (f2) at (1.4,-1.5) {} ;
            \draw[gedge] (f1) .. controls +(45:1cm) and +(125:1cm) .. 
            node[above] (labA) {$\A$} (f1) ;
            \draw[gedge,->,thick] (f1) to node[glab,midway,above] {$\A$} (f2);
            \ghostgraphbox[f]{(f1) (f2)};

            \draw[{>[scale=1.3]}-{>[scale=1.3]}] (.4,-1.5) -- 
            (.8,-1.5); \node[glab] (m) at (.6,-1.65) {{\large $m$}};
            \draw[-{>[scale=1.3]}] (0,-1.1) -- 
            (0,-.2) ;  \node[glab] (f) at (-.1,-.6) {{\large $f$}};
            \draw[-{>[scale=1.3]}] (1.2,-1.1) -- 
            (1.2,-.6) ; \node[glab] (m') at (1.5,-.85) {{\large 
            $\varphi(m,f)$}};
            \draw[{>[scale=1.3]}-{>[scale=1.3]}] (.4,.2) -- 
            (.9,.2) ; \node[glab] (f') at (.65,.35) {{\large $\eta_{H}$}};
            
            \node at (.65,-.65) {\LARGE\color{gray}(PB)};
          \end{tikzpicture}
        }
      \end{center}
\end{example}

We will now consider slice categories in connection with subobject
classifiers.

\bigskip

\noindent \begin{minipage}{0.68\textwidth}
  \begin{definition}[Slice category]
    The slice category $\C \downarrow A$ of a category $\C$ over an
    object $A \in \C$ has the arrows $f \in \C$ such that
    $\mathtt{cod}(f)= A$ as objects. An arrow $g\colon f \to f'$ in
    $\C \downarrow A$, with $f \colon X \to A$ and
    $f' \colon Y \to A$, is an arrow $g \colon X \to Y \in \C$ such
    that the diagram
    to the right commutes.\\
  \end{definition}
\end{minipage} \quad
\begin{minipage}{0.29\textwidth}
  \vspace*{-.4cm}
  \begin{center}
    \scalebox{1.0}{
      $$
      \xymatrix{
        X \ar@{->}[rr]^g \ar@{->}[dr]_f & & Y \ar@{->}[dl]^{f'} 
        \\ 
        & A & }
      $$ 
    }
  \end{center}
\end{minipage}

The existence of a subobject classifier in a slice category over a
topos directly follows from the following theorem \cite{mm:sheaves}.

\begin{theorem}[Slice category over a topos is a topos \cite{mm:sheaves}]
  \label{thm:slice-topos}
  For any object $A$ in a topos $\C$, the slice category $\C 
  \downarrow A$ of objects over $A$ is also a topos.
\end{theorem}

In particular, the subobject classifier in the slice category can be
constructed as follows.

\bigskip

\noindent \begin{minipage}{0.66\textwidth}
  \begin{fact}[Subobject classifier in slice category
    \cite{mm:sheaves}]
    \label{prop:subobclass-slice}
    Let $\C$ be a topos with subobject classifier 
    $\mathtt{true} \colon \mathbf{1} \rat \Omega$ and truth value object 
    $\Omega$. For any object $A \in \C$ let $A \times \Omega$ be the 
    product with projections $\pi_1 \colon A \times \Omega \to A$ 
    and $\pi_2 \colon A \times \Omega \to \Omega$. Then a subobject classifier 
    $\mathtt{true}_A$ of the slice category $\C \downarrow A$ is the 
    unique mono $\mathtt{true}_A \colon A \rightarrowtail A \times 
    \Omega$ such that the diagram to the right commutes. 
  \end{fact}
\end{minipage} \quad
\begin{minipage}{0.3\textwidth}
  \vspace*{-.5cm}
  \begin{center}
    \scalebox{.9}{ 
      \xymatrix{
        & A \ar@{ >->}@/_1pc/[ddl]_{\textit{id}_A} 
        \ar@{->}@/^1pc/[ddr]^{\mathtt{true}\,\circ\, \textbf{!}} 
        \ar@{ >.>}[d]^{\mathtt{true}_A} & \\
        & A \times \Omega \ar@{->}[dl]^{\pi_1} 
        \ar@{->}[dr]_{\pi_2} & \\
        A & & \Omega
      }
    }
  \end{center}
\end{minipage}

\begin{example}
  In order to provide an example for a subobject classifier in
    a slice category, we consider again the category $\GR$.
  Let $A = $ \raisebox{-7pt}{$\scalebox{0.7}{\input{ex-edge.tex}}$} be the base graph
  for the slice category $\GR\downarrow A$ of graph morphisms into
  $A$. The subobject classifier
  $\mathtt{true}_A \colon A \rightarrowtail A \times \Omega$ for this
  slice category is the following graph morphism:
  
  \smallskip 
        \begin{center}
          \scalebox{.8}{
            \begin{tikzpicture}[x=3.2cm,y=-1.8cm,baseline=(g1.south)]
              \node[glab] (g0) at (-.3,0) {\large{$\mathtt{true}_A$:}} ;
              \node[gnodeb] (g1) at (0,0) {} ;
              \node[gnodeb] (g2) at (0.4,0) {} ;
              \draw[gedge,->,thick] (g1) to node[midway,above] {$\A$} (g2);
              \ghostgraphbox[g]{(g1) (g2)};
              
              \node[gnodeb] (h1) at (1.4,-.25) {} ;
              \node[gnodeb] (h2) at (1.8,-.25) {} ;
              \node[gnode] (h3) at (1.4,.25) {} ;
              \node[gnode] (h4) at (1.8,.25) {} ;
              \draw[gedge,->,thick] (h1) to node[midway,above] {$\A$} (h2);
              \draw[gedge,->,dashed] (h1) to node[midway,above] {$\A$} (h4);
              \draw[gedge,->,dashed] (h3) to node[midway,above] {$\A$} (h2);
              \draw[gedge,->,dashed] (h3) to node[midway,above] {$\A$} (h4);
              \draw[gedge,->,dashed] (h1) to[bend left=50] node[midway,above] 
              {$\A$} (h2) ;
              \ghostgraphbox[h]{(h1) (h2) (h3) (h4)};
              
              \draw[{>[scale=1.3]}-{>[scale=1.3]}] (.6,0) -- (1.2,0) ;
              
            \end{tikzpicture}
          }
        \end{center}
\end{example}

\newpage
\section{Worked Example}
\label{apx:worked-example}

In the following, we give an example for the computation of a postcondition. We 
specify an online-shop scenario using an annotated abstract graph with the 
following edge label semantics: \todo{DN: In the submitted version there still 
were $m$ denoting the upper bounds of the worked example. I replaced them by 
$*$ now.}

\begin{itemize}
\item[\textbf{C}$\colon$] The \textbf{c}onnection of a customer node
  to to the online-shop.
\item[\textbf{M}$\colon$] The \textbf{m}arket relation describing which items 
  are purchasable in the shop.
\item[\textbf{P}$\colon$] The \textbf{p}ossession relation describing which 
  items are purchased by a customer.
\item[\textbf{\$}$\colon$] One \textbf{\$}-coin of the currency used by 
  customers to buy items in the shop. 
\end{itemize}
Now, we would like to model the following situation: Exactly one of
many customers has established a connection to an online-shop. At
least one of the customers has a \$-coin to purchase items and the
online-shops have an arbitrary number of items available. A customer
can be in possession of an arbitrary number of items. Graphs modelling
this specification can for instance be part of the language described
by the following annotated abstract graph $A[a_1,a_2]$:

\begin{center}
  \begin{tabular}{cc}
    $A[a_1,a_2]$ &= 
    \begin{tikzpicture}[x=1.5cm,y=-1.2cm,baseline=(1.south)]
      \node[glab] (top) at (0,-.1) {} ;
      \node[gnode] (1) at (0,0) {} ; 
      \node[glab,left] (lab1) at (1.west) {$[1,*]$} ;
      \node[gnode] (2) at (1,0) {} ; 
      \node[glab,below] (lab2) at (2.south) {$[1,1]$} ;
      \node[gnode] (3) at (2,0) {} ; 
      \node[glab,right] (lab3) at (3.east) {$[0,*]$} ;

      \draw[gedge] (1) to node[arlab,above] {$\mathit{C}\ [1,1]$} (2) ;
      \draw[gedge] (2) to node[arlab,above] {$\mathit{M}\ [0,*]$} (3) ;
      \draw[gedge] (1) to[bend right=50] node[glab,below] {$\mathit{P}\ [0,*]$} 
      (3) ;
      \draw[gedge] (1) .. controls +(45:1cm) and +(125:1cm) .. 
            node[arlab,above] (lab0) {$\mathit{\$}\ [1,*]$} (1) ;
    \end{tikzpicture}
  \end{tabular}
\end{center}
The following graph transformation production $\rho \colon L \lat I \rat R$
specifies, that a customer, who is in possession of at least one
\$-coin and who is connected to the online shop, can purchase one of
the items in stock in exchange for the currency. The production morphisms
are indicated by the node positions:

\begin{center}
  \scalebox{.95}{
    \begin{tabular}{rl}
      $\rho = {}$ &
      \begin{tikzpicture}[x=1.2cm,y=-1.2cm,baseline=(1.south)]
        \begin{scope}[shift={(-3.5,0)}]
          \node[gnode] (1) at (0,0) {} ; 
          \node[glab,below] (lab1) at (1.south) {} ;
          \node[gnode] (2) at (1,0) {} ; 
          \node[glab,below] (lab2) at (2.south) {} ;
          \node[gnode] (3) at (2,0) {} ; 
          \node[glab,below] (lab3) at (3.south) {} ;
          \node[glab] (ghost) at (0,.74) {} ; 
          \node[glab] (ghost2) at (-.1,0) {} ; 
          
          \draw[gedge] (1) to node[arlab,above] {$\mathit{C}$} (2) ;
          \draw[gedge] (2) to node[arlab,above] {$\mathit{M}$} (3) ;
          \draw[gedge] (1) .. controls +(45:1cm) and +(125:1cm) .. 
          node[arlab,above] (lab0) {$\mathit{\$}$} 
          (1) ;
          \graphbox[l]{(lab0) (3) (1) (ghost) (ghost2) (lab1) (lab3)}
        \end{scope}
        \begin{scope}
          
          \node[gnode] (1) at (0,0) {} ; 
          \node[glab,below] (lab1) at (1.south) {} ;
          \node[gnode] (2) at (1,0) {} ; 
          \node[glab,below] (lab2) at (2.south) {} ;
          \node[gnode] (3) at (2,0) {} ; 
          \node[glab,below] (lab3) at (3.south) {} ;
          \node[glab] (ghost) at (0,.74) {} ;
          \node[glab] (ghost2) at (0,-.74) {} ;
          
          \draw[gedge] (1) to node[arlab,above] {$\mathit{C}$} (2) ;
          
          \graphbox[i]{ (3) (1) (ghost) (ghost2) (lab1) (lab3)}
        \end{scope}
        \begin{scope}[shift={(3.5,0)}]
          
          \node[gnode] (1) at (0,0) {} ; 
          \node[glab,above] (lab1) at (1.north) {} ;
          \node[gnode] (2) at (1,0) {} ; 
          \node[glab,below] (lab2) at (2.south) {} ;
          \node[gnode] (3) at (2,0) {} ; 
          \node[glab,above] (lab3) at (3.north) {} ;
          \node[glab] (ghost2) at (0,-.74) {} ;
          
          \draw[gedge] (1) to node[arlab,above] {$\ \ \mathit{C}$} (2) ;
          \draw[gedge] (1) to[bend right=50] node[glab,below] (lab4) 
          {$\mathit{P}$} 
          (3) ;
          \graphbox[r]{(3) (lab1) (lab3) (1)(lab4) (ghost2)}
        \end{scope}
        \draw[->] \substy{(i.west)}{0} -- \substy{(l.east)}{0} ;
        \draw[->] \substy{(i.east)}{0} -- \substy{(r.west)}{0} ;
      \end{tikzpicture}
    \end{tabular}
  }
\end{center}
Please note, that there exists only one possibility to map the
left-hand side graph $L$ of the production $\rho$ into the abstract
graph $A$. We now depict the rewritable abstract graph
$\rmat{\phi}{\phi_L}[a_1',a_2']$ consisting of the \blue{abstract graph
  $A$ (upper part)}, the \red{left-hand side graph $L$ (lower part)} and the additional edges
introduced in the construction of Prop.~\ref{prop:mat-rewritable}
alongside a maximal pair of annotations $(a_1',a_2') \in M$ conforming
to Definition~\ref{def:abstract-rewriting-variant}:

\begin{center}
  \begin{tabular}{cc}
    $\rmat{\phi}{\phi_L}[a_1',a_2']$ &= \
    \begin{tikzpicture}[x=1.5cm,y=-1.2cm,baseline=(labX)]
      \node[glab] (top) at (0,-.1) {} ;
      \node[gnodeblue] (1) at (0,0) {} ; 
      \node[gnodeblue] (2) at (2.5,0) {} ; 
      \node[gnodeblue] (3) at (5,0) {} ; 
      
      \blue{
        \node[glab,left] at (1.west) {$[0,*]$} ;
        \node[glab,above] at (2.north) {$[0,0]$} ;
        \node[glab,right] at (3.east) {$[0,*]$} ;
        \draw[gedge] (1) to node[arlab,above] {$\mathit{C}\,[0,0]$} (2) ;
        \draw[gedge] (2) to node[arlab,above] {$\mathit{M}\,[0,*]$} (3) ;
        \draw[gedge] (1) to[bend right=15] node[glab,below] 
        {$\mathit{P}\,[0,*]$} 
        (3) ;
        \draw[gedge] (1) .. controls +(45:1cm) and +(125:1cm) .. 
        node[arlab,above] (lab0) {$\mathit{\$}\,[0,*]$} (1) ;
      }
      
      \node[gnoder] (b1) at (0,3) {} ; 
      \node[gnoder] (b2) at (2.5,3) {} ; 
      \node[gnoder] (b3) at (5,3) {} ; 
      
      \red{
        \node[glab,left=.5mm] at (b1.west) {$[1,1]$} ;
        \node[glab,below=.5mm] at (b2.south) {$[1,1]$} ;
        \node[glab,right] at (b3.east) {$[1,1]$} ;
        \draw[gedge] (b1) to node[arlab,above] {$\mathit{C}\,[1,1]$} (b2) ;
        \draw[gedge] (b2) to node[arlab,above] {$\mathit{M}\,[1,1]$} (b3) ;
        \draw[gedge] (b1) .. controls +(55:1cm) and +(115:1cm) .. 
        node[arlab,above] (lab0) {$\mathit{\$}\,[1,1]$} (b1) ;
      }
      
      \draw[gedge] (b1) to[bend right=25] node[glab,below] {$\mathit{C}\,[0,0]$} 
      (b2) ;
      \draw[gedge] (b2) to[bend right=25] node[glab,below] 
      {$\mathit{M}\,[0,*]$} 
      (b3) ;
      \draw[gedge] (b1) to[bend right=35] node[glab,above] 
      {$\mathit{P}\,[0,*]$} 
      (b3) ;
      \draw[gedge] (b1) .. controls +(295:1cm) and +(245:1cm) .. 
      node[arlab,below] {$\mathit{\$}\,[0,*]$} (b1) ;
      \draw[gedge] (b1) to[bend left=45] node[glab,right] (labX) 
      {$\mathit{\$}\,[0,*]$} 
      (1) ;
      \draw[gedge] (1) to[bend left=45] node[glab,right] {$\mathit{\$}\,[0,*]$} 
      (b1) ;
      \draw[gedge] (b1) to node[arlab,pos=0.35, right] {$\mathit{C}\,[0,0]$} 
      (2) 
      ;
      \draw[gedge] (1) to node[arlab,pos=0.8, right] 
      {$\mathit{C}\,[0,0]$} (b2) 
      ;
      \draw[gedge] (b2) to node[arlab,pos=0.7, below right] 
      {$\mathit{M}\,[0,*]$} (3) ;
      \draw[gedge] (2) to node[arlab,pos=0.7, above right] 
      {$\mathit{M}\,[0,*]$} (b3) ;
      \draw[gedge] (b1) to node[arlab,pos=0.63, below=1.5mm] 
      {$\mathit{P}\,[0,*]$} (3) ;
      \draw[gedge] (1) to node[arlab,pos=0.45, above=1.5mm] 
      {$\mathit{P}\,[0,*]$} (b3) ;
    \end{tikzpicture}
  \end{tabular}
\end{center}
All elements in $\rmat{\phi}{\phi_L}[a_1',a_2']$ annotated with $[0,0]$
cannot be the target of a legal morphism and therefore can be removed
to simplify the graphical representation. If a node annotated with
$[0,0]$ is removed this way, all incident edges are removed as well
independently of their annotation. We apply the production $\rho$ to the
simplified rewritable abstract graph $\rmat{\phi}{\phi_L}[a_1',a_2']$
(shown below to the left) resulting in the abstract graph $B[b_1,b_2]$
(shown below, to the right):

\begin{center}
  \begin{tabular}{ccc}
    \begin{tikzpicture}[x=1.5cm,y=-1.2cm,baseline=(labX)]
      \node[glab] (top) at (0,-.1) {} ;
      \node[gnodeblue] (1) at (0,0) {} ; 
      \node[gnodeblue] (3) at (2.25,0) {} ; 

      \blue{
        \node[glab,left] (labl1) at (1.west) {$[0,*]$} ;
        \node[glab,right] (labr1) at (3.east) {$[0,*]$} ;
        \draw[gedge] (1) to[bend right=15] node[glab,below] 
        {$\mathit{P}\,[0,*]$} 
        (3) ;
        \draw[gedge] (1) .. controls +(45:1cm) and +(125:1cm) .. 
        node[arlab,above] (lab0) {$\mathit{\$}\,[0,*]$} (1) ;
      }
      
      \node[gnoder] (b1) at (0,2) {} ; 
      \node[gnoder] (b2) at (1,2) {} ; 
      \node[gnoder] (b3) at (2.25,2) {} ; 
      
      \red{
        \node[glab,left=.5mm] (labl2) at (b1.west) {$[1,1]$} ;
        \node[glab,below=.5mm] (labr2) at (b2.south) {$[1,1]$} ;
        \node[glab,right] at (b3.east) {$[1,1]$} ;
        \draw[gedge] (b1) to node[arlab,above] {$\mathit{C}\,[1,1]$} (b2) ;
        \draw[gedge] (b2) to node[arlab,above] {$\mathit{M}\,[1,1]$} (b3) ;
        \draw[gedge] (b1) .. controls +(55:1cm) and +(115:1cm) .. 
        node[arlab,above] {$\mathit{\$}\,[1,1]$} (b1) ;
      }

      \draw[gedge] (b2) to[bend right=15] node[glab,below] 
      {$\mathit{M}\,[0,*]$} 
      (b3) ;
      \draw[gedge] (b1) to[bend right=60] node[glab,above] 
      {$\mathit{P}\,[0,*]$} 
      (b3) ;
      \draw[gedge] (b1) .. controls +(295:1cm) and +(245:1cm) .. 
      node[arlab,below] (lab1) {$\mathit{\$}\,[0,*]$} (b1) ;
      \draw[gedge] (b1) to[bend left=45] node[glab,right] (labX) 
      {$\mathit{\$}\,[0,*]$} 
      (1) ;
      \draw[gedge] (1) to[bend left=45] node[glab,right] {$\mathit{\$}\,[0,*]$} 
      (b1) ;
      
      \draw[gedge] (b2) to node[arlab,pos=0.7, below right] 
      {$\mathit{M}\,[0,*]$} (3) ;
      \draw[gedge] (b1) to node[arlab,pos=0.24, below right=-.9mm] 
      {$\mathit{P}\,[0,*]$} (3) ;
      \draw[gedge] (1) to node[arlab,pos=0.85, above right=-.5mm] 
      {$\mathit{P}\,[0,*]$} (b3) ;
      \node[glab] (labP) at (0,3.08) {$ $};
      \interfacebox[l]{(top) (lab0) (lab1) (labl1) (labl2) (labr1) (labr2) (labP)}
      \node[glab] at (1,3.5) {{\large $\rmat{\phi}{\phi_L}[a_1',a_2']$}};
    \end{tikzpicture}
    &
    \huge{$\Rightarrow_\rho$}
    &
    \begin{tikzpicture}[x=1.5cm,y=-1.2cm,baseline=(labX)]
      \node[glab] (top) at (0,-.1) {} ;
      \node[gnodeblue] (1) at (0,0) {} ; 
      \node[gnodeblue] (3) at (2.25,0) {} ; 
      
      \blue{
        \node[glab,left] (labl1) at (1.west) {$[0,*]$} ;
        \node[glab,right] (labr1) at (3.east) {$[0,*]$} ;
        \draw[gedge] (1) to[bend right=15] node[glab,below] 
        {$\mathit{P}\,[0,*]$} 
        (3) ;
        \draw[gedge] (1) .. controls +(45:1cm) and +(125:1cm) .. 
        node[arlab,above] (lab0) {$\mathit{\$}\,[0,*]$} (1) ;
      }
      
      \node[gnoder] (b1) at (0,2) {} ; 
      \node[gnoder] (b2) at (1,2) {} ; 
      \node[gnoder] (b3) at (2.25,2) {} ; 

      \red{
        \node[glab,left=.5mm] (labl2) at (b1.west) {$[1,1]$} ;
        \node[glab,below=.5mm] (labr2) at (b2.south) {$[1,1]$} ;
        \node[glab,right] at (b3.east) {$[1,1]$} ;
        \draw[gedge] (b1) to node[arlab,above] {$\mathit{C}\,[1,1]$} (b2) ;
      }
      
      \dgreen{
        \draw[gedge] (b1) to[bend right=70] node[glab,below] (labP) 
        {$\mathit{P}\,[1,1]$} 
        (b3) ;
      }
      
      \draw[gedge] (b2) to[bend right=15] node[glab,below] 
      {$\mathit{M}\,[0,*]$} 
      (b3) ;
      \draw[gedge] (b1) to[bend right=60] node[glab,above] 
      {$\mathit{P}\,[0,*]$} 
      (b3) ;
      \draw[gedge] (b1) .. controls +(295:1cm) and +(245:1cm) .. 
      node[arlab,below] (lab1) {$\mathit{\$}\,[0,*]$} (b1) ;
      \draw[gedge] (b1) to[bend left=45] node[glab,right] (labX) 
      {$\mathit{\$}\,[0,*]$} 
      (1) ;
      \draw[gedge] (1) to[bend left=45] node[glab,right] {$\mathit{\$}\,[0,*]$} 
      (b1) ;
      
      \draw[gedge] (b2) to node[arlab,pos=0.7, below right] 
      {$\mathit{M}\,[0,*]$} (3) ;
      \draw[gedge] (b1) to node[arlab,pos=0.24, below right=-.9mm] 
      {$\mathit{P}\,[0,*]$} (3) ;
      \draw[gedge] (1) to node[arlab,pos=0.85, above right=-.5mm] 
      {$\mathit{P}\,[0,*]$} (b3) ;
      \interfacebox[l]{(top) (lab0) (lab1) (labl1) (labl2) (labr1) (labr2) (labP)}
      \node[glab] at (1,3.5) {{\large $B[b_1,b_2]$}};
    \end{tikzpicture}
  \end{tabular}
\end{center}

\noindent We can use the postcondition for an invariant check of the graph 
language $\mathcal{L}(A[a_1,a_2])$ with respect to the production 
$\rho$. In fact, the annotated abstract graph $B[b_1,b_2]$ specifies \\
\noindent \begin{minipage}{0.75\textwidth} \vspace*{.05cm}  (a part of) the
  strongest postcondition and therefore the graph
  $G \in \mathcal{L}(B[b_1,b_2])$, shown to the right, is a witness
  for the fact that the graph language $\mathcal{L}(A[a_1,a_2])$ is not
    closed under production application of $\rho$ since
    $G \notin \mathcal{L}(A[a_1,a_2])$ due to a missing $\$$-coin edge in $G$, 
    which is required by $A[a_1,a_2]$. 
\end{minipage} 
\begin{minipage}{0.28\textwidth}
  \vspace*{-.2cm}
  \begin{center}
    \begin{tikzpicture}[x=1.5cm,y=-1.2cm,baseline=(1.south)]
      \node[glab] (top) at (0,-.1) {} ;
      \node[gnode] (1) at (0,0) {} ; 
      \node[glab,left] (lab1) at (1.west) {} ;
      \node[gnode] (2) at (.75,0) {} ; 
      \node[glab,below] (lab2) at (2.south) {} ;
      \node[gnode] (3) at (1.5,0) {} ; 
      \node[glab,right] (lab3) at (3.east) {} ;
      
      \draw[gedge] (1) to node[arlab,above] {$\mathit{C}$} (2) ;
      \draw[gedge] (1) to[bend right=50] node[glab,below] {$\mathit{P}$} 
      (3) ;
    \end{tikzpicture}
  \end{center}
\end{minipage}

\section{Construction of the Materialization in the Category of Graphs}
\label{apx:materialization-graphs}

In this chapter, we specify the concrete construction of the
terminal object $L\rat \mat{\phi} \to A$ in the
materialization category for the base category $\GR$. Afterwards we
prove that our construction is correct.

\begin{definition}[Construction of the materialization]
  \label{def:terminal-object}
  Let $L = (V_L, E_L, \sSrc_L, \sTgt_L, \sLab_L)$ and
  $A =(V_A, E_A, \sSrc_A, \sTgt_A, \sLab_A)$ be two graphs over a given edge 
  label alphabet $\Lambda$ and let $\phi\colon L\to A$ be a fixed graph 
  morphism.
  
  First we define the function
  $\psi_V \colon (V_L \cupdot V_A) \to V_A$ which maps the nodes of
  $L$ and $A$ to the nodes of $A$ with respect to $\varphi$:
  \begin{align*}
    \psi_{V} (x) &=
    \begin{cases}
      \varphi_{V} (x) & \text{if}\ x \in V_L \\
      x & \text{otherwise}
    \end{cases}
  \end{align*}
  We construct $\mat{\phi} = (V,E,\sSrc,\sTgt,\sLab)$ in the
  following way:
  \begin{align*}
    V &= V_L \cupdot V_A \\
    E &= E_L \cupdot \{ (e,s,t,l) \in E_A \times V \times V \times
    \Lambda \mid
    \\
    & \qquad \qquad \qquad \qquad \sSrc_A(e) = \psi_V (s) \land \sTgt_A(e)
    = \psi_V
    (t) \land \sLab_A (e) = l \} 
  \end{align*}
    \vspace*{-1cm} 
  \noindent\begin{align*}
    \sSrc &\colon\ E \to V & &
    \ \sSrc(x) =
    \begin{cases}
      s & \text{if}\ x = (e,s,t,l) \\
      \sSrc_L (x) & \text{otherwise} \\
    \end{cases} && \\
    \sTgt &\colon\ E \to V & &
    \ \sTgt(x) =
    \begin{cases}
      t & \text{if}\ x = (e,s,t,l) \\
      \sTgt_L (x) & \text{otherwise} \\
    \end{cases} && \\
    \sLab &\colon\ E \to \Lambda & &
    \quad \sLab(x) =
    \begin{cases}
      l & \text{if}\ x = (e,s,t,l) \\
      \sLab_L (x) & \text{otherwise}
    \end{cases} &&
  \end{align*}
  
  \noindent \begin{minipage}{0.65\textwidth}
  This concludes the construction of the graph $\mat{\phi}$. We
  now define the embedding graph morphism
  $\alpha \colon L \to \mat{\phi}$ where $\alpha(x) = x$ to get
  the diagram shown to the right. \\
  \end{minipage} \quad
  \begin{minipage}{0.3\textwidth}
    \vspace*{-.4cm}
    \begin{center}
      \begin{tikzpicture}[shorten >=1pt, node distance=15mm and 15mm, on
        grid]
        \draw node (L) {\(L\)} node [right=of L] (X)
        {\(\mat{\phi}\)} node [right=of X] (A) {\(A\)};
        
        \begin{scope}[->]
          \draw[>->] (L) to node [midway,above] {\(\alpha\)} (X); \draw
          (L) to [bend right] node [midway, below] {\(\varphi\)} (A);
        \end{scope}
      \end{tikzpicture}
    \end{center}
  \end{minipage}
  
  \noindent To get a valid factorization $L\rat \mat{\phi} \to A$ of
  $\phi$, we define the morphism $\psi \colon \mat{\phi} \to A$
  with $\psi = (\psi_V, \psi_E)$ where $\psi_E \colon E \to E_A$ is
  given by:
  \begin{align*}
    \psi_{E} (x) &=
    \begin{cases}
      e & \text{if}\ x = (e,s,t,l) \\
      \phi_E (x) & \text{otherwise}\ (\text{i.e., } x \in E_L)
    \end{cases}
  \end{align*}

  \noindent \begin{minipage}{0.65\textwidth}
  Obviously $\psi \circ \alpha = \phi$ holds. The object
    $L\oset{\alpha}{\rat} \mat{\phi} \oset{\psi}{\to}
    A$ is a factorization of $L \oset{\phi}{\to} A$ and the 
    diagram shown to the right commutes. \\
  \end{minipage} \quad
  \begin{minipage}{0.3\textwidth}
    \vspace*{-.4cm}
      \begin{center}
        \begin{tikzpicture}[shorten >=1pt, node distance=15mm and 15mm, on
          grid]
          \draw node (L) {\(L\)} node [right=of L] (X)
          {\(\mat{\phi}\)} node [right=of X] (A) {\(A\)};
        
          \begin{scope}[->]
            \draw[>->] (L) to node [midway,above] {\(\alpha\)} (X); \draw
            (L) to [bend right] node [midway, below] {\(\varphi\)} (A);
            \draw (X) to node [midway,above] {\(\psi\)} (A);
          \end{scope}
        \end{tikzpicture}
      \end{center}
  \end{minipage}
\end{definition}

Next, we prove that the above constructed object
$L\oset{\alpha}{\rat} \mat{\phi} \oset{\psi}{\to} A$ is the
terminal object in the materialization category.

\begin{proof}
  Given the factorization
  $L\oset{\alpha}{\rat} \mat{\phi} \oset{\psi}{\to} A$ of
  $L \oset{\phi}{\to} A$ from Definition \ref{def:terminal-object}
  with $\varphi = \psi \circ \alpha$. The morphism
  $\alpha \colon L \to \mat{\phi}$ is the embedding morphism from $L$
  into $\mat{\phi}$ and by the construction of $\mat{\phi}$ there
  exists a second embedding morphism $\gamma \colon A \to \mat{\phi}$
  with $\text{img}(\alpha) \cap \text{img}(\gamma) = \emptyset$ and
  \begin{align*}
    \gamma(x) &= 
       \begin{cases}
         x & \text{if}\ x \in V_A \\
         (x, \sSrc_A(x), \sTgt_A(x), \sLab_A(x)) & \text{if}\ x \in E_A \\
       \end{cases}
  \end{align*}

  \noindent It is easy to see that $\gamma$ is well-defined.

  \noindent \begin{minipage}{0.6\textwidth}
  Let $L\oset{\beta}{\rat} G \oset{g}{\to} A$ be another 
  factorization of $L \oset{\phi}{\to} A$ with $\varphi = g \circ \beta$. If 
  the object $L\oset{\alpha}{\rat} \mat{\phi} \oset{\psi}{\to} 
  A$ 
  is the terminal object in the materialization category, there must exist a
  unique graph morphism $f \colon G \to \mat{\phi}$ such that the 
  diagram to the right commutes and the square is a pullback. \\
  \end{minipage} \quad
  \begin{minipage}{0.3\textwidth}
  \begin{tikzpicture}[shorten >=1pt, node distance=15mm and 18mm, on grid]
    \draw
    node (a) at (0.9,-0.7) { \color{gray}\footnotesize (PB) } 
    node (L) at (0,0) {\(L\)}
    node [right=of L] (G) {\(G\)}
    node [below=of L] (L2) {\(L\)}
    node [right=of L2] (X) {\(\mat{\phi}\)}
    node [right=of G] (A) {\(A\)};
    
    \begin{scope}[->]
      \draw[>->] (L) to node [midway,above] {\(\beta\)} (G);
      \draw[->] (G) to node [midway,above] {\(g\)} (A);
      \draw[>->] (L) to node [midway,left] {\(id\)} (L2);
      \draw[>->] (L2) to node [midway,above] {\(\alpha\)} (X);
      \draw (L) to [bend angle=45,bend left] node [midway, above] {\(\varphi\)} 
      (A);
      \draw (A.south) to [bend left] node [midway, below] {\(\gamma\)} 
      (X.east);
      \draw[->,dashed] (G) to node [midway,right] {\(f\)} (X);
      \draw (X.north east) to node [midway,above] {\(\psi\)} (A.south west);
    \end{scope}
  \end{tikzpicture} \\
\end{minipage}

\noindent Define $f = (f_V,f_E)$ in the following way:
\begin{align*}
  f_V &\colon\ V_G \to V_{\mat{\phi}} &&
  f_V(x) =
  \begin{cases}
    \alpha_V \circ \beta_{V}^{-1} (x) & \text{if}\ x \in
    \text{img}(\beta_V)
    \\
    \gamma_V \circ g_V (x) & \text{otherwise} \\
  \end{cases} \\
  f_E &\colon\ E_G \to E_{\mat{\phi}} &&
  f_E(x) =
  \begin{cases}
    \alpha_E \circ \beta_{E}^{-1} (x) & \text{if}\ x \in
    \text{img}(\beta_E)
    \\
    \bigl(g_E (x), f_V(\sSrc_G (x)), f_V(\sTgt_G (x)), \sLab_G(x)
    \bigr) & \text{otherwise}
    \\
  \end{cases}
\end{align*}

\noindent Note that since $\beta$ is an injection, the element
$\beta^{-1}(x)$ is unique whenever $x$ is in the image of $\beta$.

\medskip

\hrule

\medskip

\noindent We will next prove that $f$ preserves the structure of $G$,
i.e., that it is a well-defined graph morphism. We need to prove that
the following three properties hold for every edge $x\in E_G$:
\begin{align}
  f_V(\sSrc_G(x)) &= \sSrc_{\mat{\phi}}(f_E(x))\label{eq4} \\
  f_V(\sTgt_G(x)) &= \sTgt_{\mat{\phi}}(f_E(x))\label{eq5} \\
  \sLab_G(x) &= \sLab_{\mat{\phi}}(f_E(x))\label{eq6}
\end{align} There are the following two cases:

\medskip

\noindent Case 1: Suppose $x\in \text{img}(\beta)$. Then there exists
$y\in L$ such that $x = \beta_E(y)$. In this case we obtain
\begin{align*}
  f_V(\sSrc_G(x)) &= \alpha_V(\beta_V^{-1}(\sSrc_G(x))) =
  \alpha_V(\beta_V^{-1}(\sSrc_G(\beta_E(y)))) \\
  &= \alpha_V(\beta_V^{-1}(\beta_V(\sSrc_L(y)))) =
  \alpha_V(\sSrc_L(y)) =
  \sSrc_{\mat{\phi}}(\alpha_E(y)) \\
  &= \sSrc_{\mat{\phi}}(\alpha_E(\beta_E^{-1}(x))) =
  \sSrc_{\mat{\phi}}(f_E(x)) \\
  \sLab_G(x) &= \sLab_G(\beta_E(y)) = \sLab_L(y) =
  \sLab_{\mat{\phi}}(\alpha_E(y)) \\
  & = \sLab_{\mat{\phi}}(\alpha_E(\beta_E^{-1}(\beta_E(y)))) =
  \sLab_{\mat{\phi}}(\alpha_E(\beta_E^{-1}(x))) \\
  &= \sLab_{\mat{\phi}}(f_E(x))
\end{align*}
The case of the target function ($\sTgt$) is equivalent to the source
function ($\sSrc$).

\medskip

\noindent Case 2: Whenever $x \notin \text{img}(\beta)$, we get that
\[ f_E(x) = \bigl(g_E (x), f_V(\sSrc_G (x)), f_V(\sTgt_G(x)),
  \sLab_G(x) \bigr). \]
  
\noindent Since $x \notin \text{img}(\beta)$ we obtain the following
equations:
  
  \begin{align*}
    \sSrc_{\mat{\phi}}(f_E(x)) &= \sSrc_{\mat{\phi}}\bigl((g_E (x),
    f_V(\sSrc_G
    (x)), f_V(\sTgt_G(x)), \sLab_G(x)) \bigr) \\
    &= f_V(\sSrc_G(x)) \\
    \sLab_{\mat{\phi}}(f_E(x)) &= \sLab_{\mat{\phi}}\bigl((g_E (x),
    f_V(\sSrc_G (x)), f_V(\sTgt_G(x)), \sLab_G(x)) \bigr) \\
    &= \sLab_G(x)
  \end{align*}

  Again, the case of the target function is equivalent to the case of
  the source function.
    
  \noindent Therefore $f \colon G \to \mat{\phi}$ is a graph
  morphism.

  \medskip \hrule \medskip
  
  \noindent We now prove that the following three properties hold for
  $f$:
  \begin{align}
    \psi \circ f &= g\label{eq1} \\
    f \circ \beta &= \alpha \label{eq2} \\
    \forall x \in G,\ x \notin \text{img}(\beta) &\implies f(x) \notin
    \text{img}(\alpha)\label{eq3}
  \end{align}

  Properties~\eqref{eq2} and~\eqref{eq3} together ensure that every
  element of $\text{img}(\alpha)$ has a unique preimage under $f$,
  which -- together with the commutativity of the square -- guarantees
  that it is a pullback.

  \medskip
  
  \noindent Proof of \eqref{eq1}: Assume $x \in
  \text{img}(\beta)$. Since
  $\psi \circ \alpha = \varphi = g \circ \beta$ we get:
  \begin{align*}
    (\psi \circ f)(x) = \psi(f(x)) = \psi(\alpha(\beta^{-1}(x))) =
    \varphi(\beta^{-1}(x)) = g(\beta(\beta^{-1}(x))) = g(x)
  \end{align*}
  Assume $x \notin \text{img}(\beta)$. Then $x$ is either a node or an
  edge of
  $G$. \\
  First we assume that $x \in V_G$ and $x \notin \text{img}(\beta_V)$.
  Since $\psi_V \circ \gamma_V = id_V$ we get:
  \begin{align*}
    (\psi_V \circ f_V)(x) = \psi_V(f_V(x)) = \psi_V(\gamma_V(g_V(x)))
    = id_V(g_V(x)) = g_V(x)
  \end{align*}
  Now assume $x \in E_G$ and $x \notin \text{img}(\beta_E)$:
  \begin{align*}
    (\psi_E \circ f_E)(x) = \psi_E\bigl((g_E (x), f_V(\sSrc_G (x)),
    f_V(\sTgt_G(x)), \sLab_G(x)) \bigr) = g_E(x)
  \end{align*}
  Proof of \eqref{eq2}: Since $\beta$ is a mono, we get that for all
  $x \in L$, there exists a unique $y \in \text{img}(\beta)$ such that
  $\beta(x) = y$ and $\beta^{-1}(y) = x$. By the construction of $f$,
  the following equation holds:
  \begin{align*}
    (f \circ \beta)(x) = f(\beta(x)) = f(y) = (\alpha \circ
    \beta^{-1})(y) = \alpha(\beta^{-1}(y)) = \alpha(x)
  \end{align*}
  Proof of \eqref{eq3}: Let $x \in G$ be given and
  $x \notin \text{img}(\beta)$. Then $x$ is either a node or an edge
  of $G$. First we assume that $x \in V_G$. Then
  $f_V(x) = \gamma_V \circ g_V (x)$. By the construction of
  $\mat{\phi}$ it follows that
  $\text{img}(\alpha) \cap \text{img}(\gamma) = \emptyset$ and
  therefore we get that $f_V(x) \notin \text{img}(\alpha)$.
  
  \noindent Now assume $x \in E_G$ and
  $f_E(x) = \bigl(g_E (x), f_V(\sSrc_G (x)), f_V(\sTgt_G(x)),
  \sLab_G(x) \bigr)$. By the construction of $\mat{\phi}$ we
  have that all edges of $E_{\mat{\phi}}$ are either of the
  form $(e,s,t,l)$, with $(e,s,t,l) \notin \text{img}(\alpha)$ or an
  edge from $E_L$ and therefore in $\text{img}(\alpha)$. We get that
  $f_E(x) \notin \text{img}(\alpha)$.

  \medskip \hrule \medskip
  
  To prove that $f$ is unique, we show that any other morphism
  $f'\colon G\to \mat{\phi}$, satisfying the properties
  \eqref{eq1}, \eqref{eq2} or \eqref{eq3}, equals $f$. We show
  equality by checking that $f(x) = f'(x)$ for all $x\in G$.

  \medskip
    
  \noindent Case 1: Suppose $x \in \text{img}(\beta)$. Then there
  exists an element $y\in L$ such that $\beta(y) = x$ and we obtain:
  \begin{align*}
    f'(x) = f'(\beta(y)) \oset{\eqref{eq2}}{=} \alpha(y) =
    \alpha(\beta^{-1}(x)) = f(x)
  \end{align*}

  \medskip

  \noindent Case 2: Suppose $x \notin \text{img}(\beta)$ and $x$ is a
  node of $G$ (e.g. $x \notin \text{img}(\beta_V)$). If
  $f'_V(x) \in V_L = \text{img}(\alpha_V)$, we would get that
  $x \in \text{img}(\beta_V)$, due to property~\eqref{eq3}, which is a
  contradiction. We can hence conclude that $f'_V(x) \in V_A$, which
  implies $\gamma_V(\psi_V(f'_V(x))) = f'_V(x)$, and furthermore:
  \begin{align*}
    f'_V(x) &= \gamma_V(\psi_V(f'_V(x))) \oset{\eqref{eq1}}{=}
    \gamma_V(g_V(x)) = f_V(x)
  \end{align*}
    
  \medskip
    
  \noindent Case 3: Suppose $x \notin \text{img}(\beta)$ and $x$ is an
  edge of $G$ (e.g. $x \notin \text{img}(\beta_E)$). If
  $f'_E(x) \in E_L = \text{img}(\alpha_E)$, we would get that
  $x \in \text{img}(\beta_E)$, due to property~\eqref{eq3}, which is a
  contradiction. We can hence conclude that $f'_E(x) \in E_A$, which
  implies that $f_E(x)$ must be of the form
  $(e,s,t,l) \in E_{\mat{\phi}}$. We will now show that
  \[ (e,s,t,l) = (g_E (x), f_V(\sSrc_G (x)), f_V(\sTgt_G (x)),
    \sLab_G(x)), \] which implies $f'_E(x) = f_E(x)$.

  \begin{align*}
    g_E(x) & \oset{\eqref{eq1}}{=} \psi_E(f'_E(x)) =
    \psi_E(e,s,t,l) = e \\
    f_V(\sSrc_G(x)) &= \sSrc_{\mat{\phi}}(f_E(x)) =
    \sSrc_{\mat{\phi}}((e,s,t,l)) = s \\
    f_V(\sTgt_G(x)) &= \sTgt_{\mat{\phi}}(f_E(x)) =
    \sTgt_{\mat{\phi}}((e,s,t,l)) = t \\
    \sLab_G(x) &= \sLab_{\mat{\phi}}(f_E(x)) =
    \sLab_{\mat{\phi}}((e,s,t,l)) = l
  \end{align*}

  \medskip \hrule \medskip

  Hence the graph morphism $f \colon G \to \mat{\phi}$ exists
  and it is unique for all factorizations
  $L\oset{\beta}{\rat} G \oset{g}{\to} A$ of
  $L \oset{\phi}{\to} A$ with $\varphi = g \circ \beta$. Therefore
  the constructed object
  $L\oset{\alpha}{\rat} \mat{\phi} \oset{\psi}{\to}
  A$ is the terminal object in the materialization category. \qed
\end{proof}

\newpage

\section{Proofs}
\label{apx:proofs}

\subsection{Preliminaries}

The following result is known, we give the proof sketch for the
convenience of the reader, since the construction plays an important
role in this paper.\todo{DN: Fixed some more typos in the proof}\\

\begin{proposition_for}{prop:fpbc-soc}{\ (Final pullback complements,
    subobject and partial map classifier~\cite{DT87}).}
  \fpbcAndPartialMap
\end{proposition_for}

\begin{proof}[Sketch] We just report the corresponding constructions
  from~\cite{DT87}, omitting the proofs of the relevant properties.
  \begin{description}
  \item[\emph{(1) $\Rightarrow$ (2)}] The component
    $\eta_Y: Y \rightarrowtail F(Y)$ of the natural transformation 
    \deleted{$F$}\dnchange{$\eta$}
    at object $Y \in \C$ is obtained as the final pullback complement
    of
    $Y \oset{!_Y}{\to} \mathbf{1}
    \oset{\mathtt{true}}{\rightarrowtail} \Omega$, as shown in the
    left part of (\ref{eq:FPBC-classifiers}).
  \item[\emph{(2) $\Rightarrow$ (1)}] We first observe that, given a
    partial map classifier \deleted{$F$}\dnchange{$(F,\eta)$}, the subobject 
    classifier is obtained
    as $\mathbf{1} \oset{\eta_{\mathbf{1}}}{\rat} F(\mathbf{1})$.

    Next we show how to construct a final pullback complement: Given
    $I \oset{\alpha}{\to} L \oset{m}{\rightarrowtail} G$, consider the
    components of the natural transformation at $I$ and $L$, and arrow
    $F(\alpha) \colon F(I) \to F(L)$, as in the right part of
    (\ref{eq:FPBC-classifiers}). The mono
    $L \oset{m}{\rightarrowtail} G$ can be seen as a partial map
    $G \oset{m}{\leftarrowtail} L \oset{id_L}{\rat} L$ from $G$ to
    $L$, and this induces a unique arrow $\phi(m,id_L)$ making the
    square a pullback. Now let $G \oset{h}{\leftarrow} P \to F(I)$ be
    the pullback of
    $G \oset{\phi(m,id_L)}{\longrightarrow} F(L)
    \oset{F(\alpha)}{\longleftarrow} F(I)$. It is easy to see that
    there is an induced mono (mediating arrow)
    $n: I \rightarrowtail P$ and it can be shown that
    $I \oset{n}{\rightarrowtail} P \oset{h}{\to} G$ is the final
    pullback complement of
    $I \oset{\alpha}{\to} L \oset{m}{\rightarrowtail} G$.
  \end{description}
  \qed
\end{proof}
\begin{equation}
  \label{eq:FPBC-classifiers}
  \tikzsetnextfilename{fpbcclassifiersleftdiag}
  \begin{tikzpicture}[baseline={([yshift=-.5ex]current bounding box.west)}]
    \node{\begin{tikzcd}%
	Y \ar[d,"\mathbf{!}_Y"'] \ar[tail,r,"\eta_{Y}"] & F(Y)
	\ar[d,"\chi_{\eta_{Y}}"]
	\\
	\mathbf{1} \ar[tail,r,"\mathtt{true}"' ] &
	\ar[ul,phantom,"\color{gray}\scriptstyle\mathrm{(FPBC)}"] \Omega
      \end{tikzcd}};
  \end{tikzpicture}
  \quad
  \tikzsetnextfilename{fpbcclassifiersrightdiag}
  \begin{tikzpicture}[baseline={([yshift=-.5ex]current bounding box.west)}]
    \node{\begin{tikzcd}[row sep=1em,column sep=1.5em]
	&
	L
	\ar[tail,dd,"m" description]
	\ar[tail,dr,"\eta_L"' near end]
	&&
	I
	\ar[ll,"\alpha" description,near start] 
	\ar[tail,dr,"\eta_I"]
	\ar[tail,dashed,dd,"n"', near start]
	\\
	L
	\ar[tail,ur,"id_L"]
	\ar[tail,dr,"m"'] 
	&&
	F(L)
	&&
	F(I)
	\ar[ll,"F(\alpha)" description]
	\\
	&
	G
	\ar[ur,"{\phi(m,id_L)}"' near end]
	&&
	P
	\ar[ll,"h" description, near start] %
	\ar[ur]
      \end{tikzcd}};
  \end{tikzpicture}
\end{equation}

\subsection{Materialization}

\begin{proposition_for}{prop:term-pmc}{\ (Existence of materialization
    in a topos).}
  \existenceOfMatInTop
\end{proposition_for}

\begin{proof} 
  Let $L \oset{m}{\rightarrowtail} X \oset{\alpha}{\to} A$ be
  an object of $\matcat{\phi}$, i.e., a factorization such that
  $\varphi = \alpha \circ m$.  Note that this defines a partial map
  $(m,\mathit{id}_L) \colon \alpha \rightharpoonup \phi$ in
  $\C \downarrow A$ consisting of the span
  $\alpha \oset{m}{\leftarrowtail} \phi
  \oset{\mathit{id}_L}{\to} \phi$.  Since
  $\eta_\phi: \phi \to F(\phi)$ is the component of the partial map
  classifier, there exists a unique arrow
  $\varphi(m,\mathit{id}_L) \colon X \to {\mat{\phi}}$ from
  $\alpha \colon X \to A$ to $F(\phi) \colon {\mat{\phi}} \to A$ for which the
  left square in the following diagram is a pullback and the right
  triangle commutes. The latter holds since $\varphi(m,\mathit{id}_L)$
  is an arrow in the slice category.
  \begin{center}
    \scalebox{1}{
      $$
      \xymatrix{
        L \ar@/^1.3pc/[rr]|{\bx{\phi}}
        \ar@{ >->}[r]^{m} \ar@{>->}[d]_{\mathit{id}_L} & X 
        \ar@{.>}[d]^{\varphi(m,\mathit{id}_L)} \ar@{->}[r]^{\alpha} & A \\ 
        L \ar@{ >->}[r]_(.4){\eta_\phi} & {\mat{\phi}}
        \itul{\rm\footnotesize\color{gray}(PB)} 
        \ar@/_1.3pc/@{->}[ur]_{F(\phi)} & }$$ 
    }
  \end{center}
  \qed
\end{proof}

\begin{corollary_for}{cor:term-constr}{\ (Construction of the materialization).}
  \constructOfMat
\end{corollary_for}

\begin{proof}
  Straightforward from
Propositios~\ref{prop:fpbc-soc} and~\ref{prop:term-pmc} (and the fact that 
final pullback
complements in the slice category correspond to those in the base
category \cite{l:graph-rewritinginspancat}). \qed
\end{proof}

\begin{proposition_for}{prop:language-materialization}{\ (Language of
    the materialization).}
  \languageMaterialization
\end{proposition_for}

\begin{proof}
  We show that the two sets are included into each other:

  \begin{itemize}
  \item ($\subseteq$) Given the materialization
    $L\oset{\eta_\phi}{\rat} {\mat{\phi}} \oset{g}{\to} A$ of a
    $\mathbf{C}$-arrow $\phi\colon L\to A$, let $L \oset{m_L}{\rat} X$
    be a mono in the language
    $\mathcal{L}(L\oset{\eta_\phi}{\rat} {\mat{\phi}})$, i.e.,
    $(L \oset{m_L}{\rat} X) \in \mathcal{L}(L\oset{\eta_\phi}{\rat}
    {\mat{\phi}})$. Spelling out Definition~\ref{def:mono-language}
    we obtain the following commuting diagram where the square is a
    pullback:
    \begin{center}
      \scalebox{1}{
        $$
        \xymatrix{
          L \ar@{ >->}[r]^{m_L} \ar@{>->}[d]_{\mathit{id}_L} & X 
          \ar@{->}[d]^{f} \ar@{.>}[dr]^{\psi} & \\ 
          L \ar@/_1.3pc/[rr]|{\bx{\phi}} \ar@{ >->}[r]^(.4){\eta_\phi} & 
          {\mat{\phi}} \itul{\rm\footnotesize\color{gray}(PB)} 
          \ar@{->}[r]^g & A}$$ 
      }
    \end{center}
    Then we define $\psi = g\circ f \colon X \to A$ and observe that
    the following equation holds:
    \[ \varphi = g \circ \eta_\phi = g \circ \eta_\phi \circ
      \mathit{id}_L = g \circ f \circ m_L = \psi \circ m_L \]

  \item ($\supseteq$) Let the mono $L \oset{m_L}{\rat} X$ be a
    factorization of the $\mathbf{C}$-arrow $\phi\colon L\to A$,
    i.e., there exists an arrow $\psi \colon X \to A$ such that
    $\phi = \psi \circ m_L$. By terminality of the materialization
    $L\oset{\eta_\phi}{\rat} {\mat{\phi}} \to A$ there exists an arrow
    $X \to \mat{\phi}$ such that the following diagram commutes and the
    square is a pullback:
    \begin{center}
      \scalebox{1}{
        $$
        \xymatrix{
          L \ar@{ >->}[r]^{m_L} \ar@{>->}[d]_{\mathit{id}_L} & X 
          \ar@{.>}[d] \ar@{->}[dr]^{\psi} & \\ 
          L \ar@/_1.3pc/[rr]|{\bx{\phi}} \ar@{ >->}[r]^(.4){\eta_\phi} & 
          {\mat{\phi}} \itul{\rm\footnotesize\color{gray}(PB)} 
          \ar@{->}[r] & A}$$ 
      }
    \end{center} Therefore $(L \oset{m_L}{\rat} X) \in 
    \mathcal{L}(L\oset{\eta_\phi}{\rat} {\mat{\phi}})$ holds. \qed
  \end{itemize}
\end{proof}

\begin{proposition_for}{prop:mat-rewritable}{\ (Construction of the
    rewritable materialization).}
  \rewritableMat
\end{proposition_for}

\begin{proof}
  First note that in diagram (\ref{eq:fpbc-al-i}), $F$ is obtained as
  the final pullback complement of
  $I \oset{\phi_L}{\rat} L \rat {\mat{\phi}}$, where
  $L \rat {\mat{\phi}} \oset{\psi}{\to} A$ is the materialization of
  $\phi$ (Def.~\ref{def:materialization-category}). Arrow $I \rat F$
  is monic because it is reflected, while $F \rat {\mat{\phi}}$ is
  monic by properties of final pullback complements since
  $\phi_L \colon I \rat L$ is monic (see \cite{chhk:sesqui}).

  Next in diagram (\ref{eq:po-l-fpbc})
  $L \oset{n_L}{\rat} \rmat{\phi}{\phi_L} \oset{\beta}{\lat} F$ is the
  pushout of the span $L \oset{\phi_L}{\lat} I \rat F$. Since the
  right square is a pushout and the outer square commutes, there is a
  unique arrow $\alpha \colon \rmat{\phi}{\phi_L} \to {\mat{\phi}} $
  making the diagram commute. Note that arrow
  $L\oset{n_L}{\rat} \rmat{\phi}{\phi_L}$ is indeed monic, as pushouts
  preserve monos in a topos, and $\alpha$ is monic because topoi
  have effective unions. Therefore the rewritable materialization
  $L\oset{n_L}{\rat} \rmat{\phi}{\phi_L} \oset{\psi\circ
    \alpha}{\longrightarrow} A$ is an object of $\matcat{\phi}$, and
  clearly it is also an object of the subcategory
  $\matcat[\phi_L]{\phi}$, as by Diagram~(\ref{eq:po-l-fpbc})
  $I\oset{\phi_L}{\rat} L\oset{n_L}{\rat} \rmat{\phi}{\phi_L}$ has a pushout
  complement.
  
  We next prove that the left square of Diagram~(\ref{eq:po-l-fpbc})
  is a pullback, to show that $\alpha$ is the unique arrow from the
  rewritable materialization to the materialization in
  $\matcat{\phi}$.  Let the diagram below to the right be given.
	
  \bigskip
  
  \noindent \begin{minipage}{0.65\textwidth} We already know that the
    inner square commutes and therefore
    $\eta_{\phi} \circ \textit{id}_L = \alpha \circ n_L$. We will now
    show that the pullback property for the inner square holds,
    e.g. for any other object $X$ and two arrows $f \colon X \to L$
    and $g \colon X \to \rmat{\phi}{\phi_L}$ where the outer square commutes,
    there exists a unique arrow $h \colon X \to L$ such that
    $f = \textit{id}_L \circ h$ and $g = n_L \circ h$. It is clear
    that $h = f$ by this assumption. Since $\alpha$ is a mono, it is a
    left-cancellative arrow e.g. for any two arrows
    $f_1,f_2 \colon X \to \rmat{\phi}{\phi_L}$ we get that
    $\alpha \circ f_1 = \alpha \circ f_2$ implies $f_1 = f_2$.
  \end{minipage} 
  \begin{minipage}{0.38\textwidth}
    \scalebox{.9}{
      \xymatrix{
        X \ar@{.>}[dr]^{h} 
        \ar@/^1.1pc/@{->}[drr]^{f} 
        \ar@/_1.1pc/@{->}[ddr]_{g} & 
        & \\
        & L \ar@{ >->}[d]_{n_L} 
        \ar@{ >->}[r]^{\textit{id}_L} & L 
        \ar@{ >->}[d]^{\eta_{\phi}} \\
        & {\rmat{\phi}{\phi_L}} \ar@{>->}[r]_{\alpha} &{\mat{\phi}}
        \itul{\rm\footnotesize\color{gray}(PB) 
          \hspace*{0.3cm}}
      }
    }
  \end{minipage}
  % \vspace*{-.5cm}
  
  \smallskip
  
  We obtain the following equation:
  $\alpha \circ g = \eta_{\phi} \circ f = \eta_{\phi} \circ
  \textit{id}_L \circ h = \alpha \circ n_L \circ h$, which implies
  that $g = n_L \circ h$ since $\alpha$ is a mono.  Hence the inner
  square is a pullback.  Now let $L \oset{p}{\rat} X \oset{q}{\to} A$
  be an object of $\matcat[\phi_L]{\phi}$, i.e., a factorization of
  $\phi$ such that the pushout complement of
  $I \oset{\phi_L}{\rat} L \oset{p}{\rat} X$ exists, and let
  $I \rat C \rat X$ be such a pushout complement.  Then the following
  diagram (\ref{eq:pb-l-al-c}) commutes, where
  $g \colon X \to {\mat{\phi}}$ is the unique arrow making the left
  square a pullback by finality of the materialization, and the right
  square is a pullback because it is a pushout along a mono.  From the
  pasting lemma (pullback version) we can conclude that the composed
  square is a pullback as well.
	
  \noindent  
  \begin{minipage}{0.47\textwidth}
    \begin{equation}
      \xymatrix{
        &  \ar[dl]_{\phi}  L \ar@{ >->}[d]_{\eta_{\phi}} &
        L \ar@{ >->}[d]|{\bx{p}} \ar@{ >->}[l] & I \ar@{ >->}[d] 
        \ar@{ >->}[l]_{\phi_L} \\
        A &  \ar[l]_{\psi} {\mat{\phi}} & X \ar@/^1.2pc/[ll]^q
        \ar@{->}[l]_{g} 
        \itul{\rm\footnotesize\color{gray}(PB)}
        & C \ar@{ >->}[l]
        \itul{\rm\footnotesize\color{gray}\hspace*{0.1cm}(PB)}
      }
      \label{eq:pb-l-al-c}
    \end{equation} 
  \end{minipage}
  \quad
  \begin{minipage}{0.47\textwidth}
    \begin{equation}
      \xymatrix{
        & & I \ar@/_1.1pc/@{>->}[dll]_{\phi_L} \ar@{ >->}[ddd] \ar@{>->}[dl] \\ 
        L \ar@{>->}[d]_{\eta_{\phi}} & I \ar@{ >->}[d]
        \ar@{ >->}[l]_{\phi_L} \itul{\rm\footnotesize\color{gray}\hspace*{1cm}=} & \\
        {\mat{\phi}} & F \ar@{ >->}[l] \itul{\rm\footnotesize\color{gray}(FPBC)} & \\
        & & C \ar@{.>}[ul]^{\qquad \gamma} \ar@/^1.1pc/[ull]
      }
      \label{eq:fpbc-property}
    \end{equation} 
  \end{minipage}
	
  Combining the outer pullback of diagram (\ref{eq:pb-l-al-c}) with
  the final pullback complement of diagram (\ref{eq:fpbc-al-i}) we get
  diagram (\ref{eq:fpbc-property}). By Def.~\ref{def:fpbc} there
  exists a unique arrow $\gamma$ such that the diagram commutes (especially
  the lower triangle and the square to the right).
	
  By composing the arrows $\gamma \colon C \to F$ from diagram
  (\ref{eq:fpbc-property}) and
  $\beta \colon F \rat \rmat{\phi}{\phi_L}$ from diagram
  (\ref{eq:po-l-fpbc}) we get the arrow
  $c = \beta\circ\gamma \colon C \to \rmat{\phi}{\phi_L}$ shown in the
  commuting diagram (\ref{eq:uni-po-g}) where the right square is a
  pushout. The universal property of pushouts gives us a unique
  mediating arrow $\delta: X \to \rmat{\phi}{\phi_L}$.  To show that
  $\delta$ defines an arrow in $\matcat{\phi}$ from
  $L \oset{p}{\rat} X \oset{q}{\to} A$ to the rewritable
  materialization
  $L\oset{n_L}{\rat} \rmat{\phi}{\phi_L} \oset{\psi\circ
    \alpha}{\longrightarrow} A$ we need to prove that
  $q = \psi \circ \alpha \circ \delta$ (which is easily checked by
  diagram chasing) and that the left square is a pullback.
	
  \vspace*{-.2cm}
  \begin{minipage}{0.42\textwidth}
    \begin{equation}
      \xymatrix{
        L \ar@{>->}[d]_{n_L} & L \ar@{ >->}[d]|{\bx{p}} \ar@{ >->}[l]
        &
        I \ar@{ >->}[d] 
        \ar@{ >->}[l] \\
        {\rmat{\phi}{\phi_L}} & X \ar@{.>}[l]_(.4)\delta
        \itull{\rm\footnotesize\color{gray}\hspace*{.1cm}(?)} & C \ar@{ >->}[l]
        \ar@/^1.4pc/@{->}[ll]|{\bx{c}} 
        \itul{\rm\footnotesize\color{gray}\hspace*{0.1cm}(PO)}
      }
      \label{eq:uni-po-g}
    \end{equation} 
  \end{minipage}
  \qquad
  \begin{minipage}{0.47\textwidth}
    \begin{equation}
      \xymatrix{
        L \ar@{ >->}[d] & L \ar@{ >->}[d]|{\bx{n_L}} \ar@{ >->}[l] & 
        L \ar@{ >->}[d]|{\bx{p}}
        \ar@{ >->}[l] \\
        {\mat{\phi}}& {\rmat{\phi}{\phi_L}} \ar@{>->}[l]_{\alpha} \itul{\rm\footnotesize\color{gray}(PB) \hspace*{0.3cm}}
        & X \ar@{->}[l]_(.4)\delta \ar@/^1.4pc/@{->}[ll]|{\bx{g}}
        \itull{\rm\footnotesize\color{gray}\hspace*{0.1cm}(PB)}
      }
      \label{eq:a-abstract-g}
    \end{equation} 
  \end{minipage}
	
  In order to show that the square marked (?) is a pullback we
  consider diagram (\ref{eq:a-abstract-g}). The left square is a
  pullback as we have shown earlier, and the outer square is a
  pullback by Diagram~(\ref{eq:pb-l-al-c}).  From the pasting lemma
  (pullback version) we can conclude that the right square is a
  pullback. Also note that the diagram clearly commutes as the three
  arrows at the bottom are all unique.  \qed
\end{proof}

\vspace*{-.1cm}
\begin{proposition_for}{prop:language-rewriting-materialization}{\
    (Language of the rewritable materialization).}
  \languageRewritableMaterialization
\end{proposition_for}

\vspace*{-.6cm}
\begin{proof}
  We show that the two sets of arrows are included in one another:
  \begin{itemize}
  \item ($\supseteq$) Let $L\oset{m_L}{\rat} X$ such that there
    exists an arrow $\psi$ with $\phi = \psi\circ m_L$ and
    $X\oset{p,m_L}{\Longrightarrow}$. Then
    $L\oset{m_L}{\rat} X\oset{\psi}{\to} A$ is an object of
    the materialization category of rewritable objects (since the
    production can be applied, the pushout complement exists) and we
    obtain a unique arrow $X\to \rmat{\phi}{\phi_L}$ that creates a pullback
    $L, L, X, A$. Hence
    $m_L\in \mathcal{L}(L\oset{n_L}{\rat} \rmat{\phi}{\phi_L})$.
  \item ($\subseteq$) Now assume that
    $m_L\in \mathcal{L}(L\oset{n_L}{\rat} \rmat{\phi}{\phi_L})$. This
    implies the existence of an arrow $X\to \rmat{\phi}{\phi_L}$ such
    that the left square in Diagram~(\ref{eq:g-abstracted-by-a}) is a
    pullback. The arrow $\psi\colon X\to A$ is given by composing
    $X\to \rmat{\phi}{\phi_L} \oset{\alpha}{\rat} {\mat{\phi}} \to A$
    and by retracing the construction of $\rmat{\phi}{\phi_L}$ (see
    Prop.~\ref{prop:mat-rewritable}) it can be shown that
    $\phi = \psi\circ m_L$.
		
    Furthermore we constructed the outer square in
    Diagram~(\ref{eq:g-abstracted-by-a}) as a pushout, which is
    therefore also a pullback.
		
    \vspace*{-.85cm}
    \begin{minipage}{0.44\textwidth}
      \begin{equation}
      \scalebox{.95}{
        \xymatrix{
          L \ar@{>->}[d] & L \ar@{>->}^{m_L}[d] \ar@{>->}[l]
          & I \ar@{>->}[d] 
          \ar@{>->}^{\phi_L}[l] \\
          {\rmat{\phi}{\phi_L}} & X \ar@{->}[l] 
          \itull{\rm\footnotesize\color{gray}\hspace*{.1cm}(PB)}
          & F \ar@{>->}@/^1.1pc/[ll] 
        }
        \label{eq:g-abstracted-by-a}
      }
      \end{equation} 
    \end{minipage}
    \quad
    \begin{minipage}{0.47\textwidth}
      \begin{equation}
        \scalebox{.95}{
        \xymatrix{
          & & I \ar@/_1.1pc/@{>->}_{m_L \circ \phi_L}[dll]
          \ar@{>->}@/^1.1pc/[ddl] \ar@{>.>}[dl] \\ 
          X \ar@{->}[d] & C \ar@{->}[d] \ar@{>->}[l] \itul{\rm\footnotesize\color{gray}\hspace*{1cm}=} & \\
          {\rmat{\phi}{\phi_L}} & F \ar@{>->}[l] 
          \itull{\rm\footnotesize\color{gray}\hspace*{.1cm}(PB)} & 
          \itul{\rm\footnotesize\color{gray} = \hspace*{.2cm}} \\
        } }
        \label{eq:pb-g-f}
      \end{equation} 
    \end{minipage}
    \vspace*{.1cm}
    
    Now we take the pullback of $X \to \rmat{\phi}{\phi_L} \lat F$ and
    obtain the pullback object $C$ with the corresponding arrows (See
    Diagram~(\ref{eq:pb-g-f})). Since the outer square commutes, we
    get a unique arrow $I \rat C$ due to the property of
    pullbacks. Note that $I\rat C$ is a mono since $I\rat F$ is a
    mono. All we need to show is that $C$ is the pushout complement
    for our rewritable object $X$.
    
    \vspace*{.2cm}
    \noindent \begin{minipage}{0.62\textwidth} In order to show that
      it is a pushout we consider the diagram to the right. The bottom
      square is a Van Kampen square\footnotemark, furthermore the left
      square is trivially a pullback, the front square is a pullback
      according to Diagram~(\ref{eq:g-abstracted-by-a}) and the right
      square is a pullback by construction (see
      Diagram~(\ref{eq:pb-g-f})). Then it follows from classical
      pullback splitting that the back square is also a
      pullback. Finally it follows from the properties of adhesive
      categories that the top square is a pushout.
    \end{minipage}
    \ 
    \begin{minipage}{0.38\textwidth}
      \vspace*{-.3cm}
      \scalebox{.75}{
        \xymatrix{
          & I \ar@{>->}[rr] \ar@{>->}[ld] \ar@{>->}[dd] & & C 
          \ar@{>->}[ld] \ar[dd] \\
          L \ar@{>->}[rr] \ar@{>->}[dd] & & X \ar[dd] & \\
          & I \ar@{>->}[rr] \ar@{>->}[ld] & & F \ar@{>->}[ld] \\
          L \ar@{>->}[rr] & & {\rmat{\phi}{\phi_L}} & \\
        } 
      }
    \end{minipage}
    \footnotetext{Since every topos is
            adhesive, the Van Kampen square property holds. For more
            details see \cite{ls:adhesive-journal}.}
            
    \vspace{-.1cm}
    Therefore $X$ can be rewritten. The existence of the pushout
    complement is guaranteed using the described construction. This
    completes the proof. \qed
  \end{itemize}
\end{proof}

\begin{proposition_for}{prop:rewriting-materializations}{\
    (Rewriting abstract matches).}  \materializationRewriting
\end{proposition_for}

\vspace*{-.6cm}
\begin{proof}~
  \begin{itemize}
  \item ($\subseteq$) Assume that  $(L \oset{n_L}{\rat} \tilde{A}) 
  \oset{p}{\Rightarrow}
  (R\oset{n_R}{\rat} B)$ and
let $m_R\in \mathcal{L}(R\oset{n_R}{\rat} B)$ where
    $m_R\colon R\rat Y$.
 That is we have the diagram below, where the
    bottom squares are pushouts and the remaining squares are
    pullbacks (the squares in the back are actually pushouts as well).
    \vspace*{-.3cm}
    \[
      \scalebox{.95}{
      \xymatrix{
        & L \ar@{>->}[dd]^{\mathit{id}_L} & & I \ar@{>->}[ll]
        \ar@{>->}[rr] \ar@{>->}[dd]^{\mathit{id}_I} 
        & & R \ar@{>->}[dd]^{\mathit{id}_R} \ar@{>->}[ld]^{m_R} \\
        & & & & Y \ar[dd] & \\
        & L \ar@{>->}[ld]^{n_L} & & I \ar@{>->}[ll] \ar@{>->}[rr] \ar@{>->}[ld]
        & & R \ar@{>->}[ld]^{n_R} \\
        \tilde{A} & & C \ar@{>->}[ll] \ar@{>->}[rr] & & B
      } }
    \]
    Now take the pullback of $C\rat B$ and $Y\to B$, obtaining $Z$,
    which gives us $I\to Z$ as mediating arrow into the pullback
    object (see diagram below). In the right cube the right square is a
    pullback, the back square is trivially pullback and the front
    square is a pullback by construction. This means that the left
    square is also a pullback by pullback splitting. Due to the
    Van Kampen square property this implies that the top square is a
    pushout. Since all pushouts along monos are pullbacks in adhesive
    categories, the arrow $I\to Z$ must be a mono.
		
    Finally, take the pushout of $I\rat Z$ and $I\rat L$, resulting in
    $X$, which give us $X\to \tilde{A}$ as a mediating arrow.
    \vspace*{-.5cm}
    \[
     \scalebox{.95}{
      \xymatrix{
        & L \ar@{>->}[dd]^(.3){\mathit{id}_L} \ar@{>->}[ld]^{m_L}
        & & I \ar@{>->}[ll]
        \ar@{>->}[rr] \ar@{>->}[dd]^(.3){\mathit{id}_I} \ar@{>->}[ld]
        & & R \ar@{>->}[dd]^(.3){\mathit{id}_R} \ar@{>->}[ld]^{m_R} \\
        X \ar[dd] & & Z \ar@{>->}[ll]
        \ar@{>->}[rr] \ar[dd] & & Y \ar[dd] & \\
        & L \ar@{>->}[ld]^{n_L} & & I \ar@{>->}[ll] \ar@{>->}[rr] \ar@{>->}[ld]
        & & R \ar@{>->}[ld]^{n_R} \\
        \tilde{A} & & C \ar@{>->}[ll] \ar@{>->}[rr] & & B
      } }
    \]
    This illustrates that
    $(L\oset{m_L}{\rat} X)\oset{p}{\Rightarrow} (R\oset{m_R}{\rat}
    Y)$. Since in the left cube the back square is trivially a
    pullback and the right square is a pullback as well (see argument
    above), the front and left squares are pullbacks as well. This
    implies that
    $(L\oset{m_L}{\rat} X)\in \mathcal{L}(L\oset{n_L}{\rat}
    \tilde{A})$, as required.
		
  \item ($\supseteq$) Assume that
    $(L\oset{m_L}{\rat} X) \oset{p}{\Rightarrow}
    (R\oset{m_R}{\rat} Y)$ and that furthermore
    $(L\oset{m_L}{\rat} X)\in \mathcal{L}(L\oset{n_L}{\rat}
    \tilde{A})$. Together with the fact that
    $(L \oset{n_L}{\rat} \tilde{A}) \oset{p}{\Rightarrow}
    (R\oset{n_R}{\rat} B)$, this results in the  diagram
    below (without the dotted arrows), where the top and bottom
    squares of the cubes are all pushouts and the vertical squares are
    pullbacks.
    \vspace*{-.2cm}
    \[
      \scalebox{.95}{
      \xymatrix{
        & L \ar@{>->}[dd]^(.3){\mathit{id}_L} \ar@{>->}[ld]^{m_L}
        & & I \ar@{>->}[ll]
        \ar@{>->}[rr] \ar@{>->}[dd]^(.3){\mathit{id}_I} \ar@{>->}[ld]
        & & R \ar@{>->}[dd]^(.3){\mathit{id}_R} \ar@{>->}[ld]^{m_R} \\
        X \ar[dd] & & Z \ar@{>->}[ll]
        \ar@{>->}[rr] \ar@{.>}[dd] & & Y \ar@{.>}[dd] & \\
        & L \ar@{>->}[ld]^{n_L} & & I \ar@{>->}[ll] \ar@{>->}[rr] \ar@{>->}[ld]
        & & R \ar@{>->}[ld]^{n_R} \\
        \tilde{A} & & C \ar@{>->}[ll] \ar@{>->}[rr] & & B
      } }
    \]
    Due to the Van Kampen square property and the fact that pushout
    complements of mono arrows are unique, the object $Z$ can be
    constructed in two ways: either by taking the pullback of
    $X\to\tilde{A}$ and $C\rat\tilde{A}$ or by taking the pushout
    complement of $I\rat L$, $L\rat X$ as shown above. Hence there
    must be an arrow $Z\to C$ arising from the pullback and the front
    and right square of the left cube are pullbacks as
    well.

    Now the arrow $Y\to B$ is obtained as a mediating arrow into the
    pushout object and the front and right faces of the right cube are
    again pullbacks. This implies that
    $(R\oset{m_R}{\rat} Y)\in \mathcal{L}(R\oset{n_R}{\rat}
    B)$, as desired. \qed
  \end{itemize}
\end{proof}

\begin{corollary_for}{cor:rewriting-materializations}{\ (Co-match
    language of the rewritable materialization).}
  \corollaryCoMatch
\end{corollary_for}

\vspace*{-.6cm}
\begin{proof}
  Straightforward from
  Propositions~\ref{prop:language-rewriting-materialization}
  and~\ref{prop:rewriting-materializations}. \qed
\end{proof}

\vspace*{-.4cm}
\subsection{Annotated Objects}

\begin{lemma}
  \label{lem:global-annot-properties}
  The global annotation functor from Ex.~\ref{ex:global-annot}
  satisfies the homorphism property, the pushout property, the
  adjunction property, the Beck-Chevalley property and the isomorphism
  property.
\end{lemma}

\begin{proof}~
  
  \begin{description}
  \item[Homomorphism property:] Assume that $\phi\colon A\to B$ is an
    injective graph morphism.

    We first show that $\mathcal{B}^n_\phi$ preserves the unit, which
    is a map $a\colon V_A\cup E_A\to \mathcal{M}_n$ with $a(x) = 0$
    for all $x\in V_A\cup E_A$. For $y\in V_B\cup E_B$ we have
    $\mathcal{B}^n_\phi(a)(y) = \sum_{\phi(x)=y} a(x)$. Either $y$ has
    a unique preimage $x$ with $a(x) = 0$ and in this case the result
    is $0$. Or $y$ has no preimage, in which case we have the empty
    sum and the result is also $0$.

    Next, we show that $\mathcal{B}^n_\phi$ preservers the monoid
    operation: let $a_1,a_2\in V_A\cup E_A\to \mathcal{M}_n$. Then we
    have
    $\mathcal{B}^n_\phi(a_1+a_2)(y) = \sum_{\phi(x)=y}
    (a_1(x)+a_2(x))$. We distinguish two cases:
    \begin{itemize}
    \item Either $y$ has a unique preimage $x$ and then the result is
      \[ a_1(x)+a_2(x) = \sum_{\phi(x)=y} a_1(x) +
        \sum_{\phi(x)=y} a_2(x) = \mathcal{B}^n_\phi(a_1)(y) +
        \mathcal{B}^n_\phi(a_2)(y)  \]
    \item Or $y$ has no preimage under $\phi$ and we obtain
      \[ 0 = 0+0 = \sum_{\phi(x)=y} a_1(x) + \sum_{\phi(x)=y}
        a_2(x) = \mathcal{B}^n_\phi(a_1)(y) +
        \mathcal{B}^n_\phi(a_2)(y) \]
    \end{itemize}
    Preservation of subtraction can be shown analogously.

    Note that preservation of the monoid operation (but not
    preservation of subtraction) holds for any (also non-injective)
    graph morphism.
  \item[Adjunction property:] Assume that $\phi\colon A\to B$ is an
    injective graph morphism.
    \begin{itemize}
    \item We show that the right adjoint of
      $\mathcal{B}^n_\phi\colon \mathcal{B}^n(A) \to \mathcal{B}^n(B)$
      is
      $\mathit{red}_\phi \colon \mathcal{B}^n(B) \to \mathcal{B}^n(A)$
      where for $b\colon V_B\cup E_B\to \mathcal{M}_n$ we have
      $\mathit{red}_\phi(b)(x) = b(\phi(x))$ (for $x\in V_A\cup E_A$).
      Clearly, $\mathit{red}_\phi$ is monotone.

      Furthermore for $a\in \mathcal{B}^n(A)$ and $x\in V_A\cup E_A$
      we can show the following, using the fact that $\phi$ is
      injective:
      \[
        \mathit{red}_\phi(\mathcal{B}^n_\phi(a))(x) =
        \mathcal{B}^n_\phi(a)(\phi(x)) = \sum_{\phi(x')=\phi(x)} a(x') =
        a(x)
      \]
      Finally for $b\in \mathcal{B}^n(B)$ and $y\in V_B\cup E_B$ we
      have:
      \begin{eqnarray*}
        && \mathcal{B}^n_\phi(\mathit{red}_\phi(b))(y) =
        \sum_{\phi(x)=y}
        \mathit{red}_\phi(b)(x) = \sum_{\phi(x)=y} b(\phi(x)) \\
        & = & \left\{
            \begin{array}{ll}
              b(y) & \mbox{if $y\in\mathit{img}(\phi)$} \\
              0 & \mbox{otherwise}
            \end{array}
          \right\} \le b(y) 
      \end{eqnarray*}
    \item We have to show that $\mathit{red}_\phi$ is a monoid
      homomorphism that preserves subtraction.

      Let $b\colon V_B\cup E_B\to \mathcal{M}_n$ be the unit map that
      satisfies $b(y)=0$ for all $y\in V_B\cup E_B$. Then
      $\mathit{red}_\phi(b)(x) = b(\phi(x)) = 0$ for all
      $x\in V_A\cup E_A$, i.e., $\mathit{red}_\phi(b)$ is also the
      unit map.

      Furthermore for $b_1,b_2\colon V_B\cup E_B\to \mathcal{M}_n$ we
      have
      \begin{eqnarray*}
        && \mathit{red}_\phi(b_1+b_2)(x) = (b_1+b_2)(\phi(x)) =
        b_1(\phi(x)) + b_2(\phi(x)) \\
        & = & \mathit{red}_\phi(b_1)(x)+\mathit{red}_\phi(b_2)(x)
      \end{eqnarray*}
      Preservation of subtraction can be shown analogously.
    \item $\mathit{red}_\phi$ preserves standard annotations:
      $\mathit{red}_\phi(s_B)(x) = s_B(\phi(x)) = 1 = s_A(x)$.
    \end{itemize}
  \item[Pushout property:] Assume that we have a pushout as in
    Def.~\ref{def:prop-annotations} (pushout property) and let
    $d\in \mathcal{B}^n(D)$. We have to show that
    \[ d = \mathcal{B}^n_{\psi_1}(\mathit{red}_{\psi_1}(d)) +
      (\mathcal{B}^n_{\psi_2}(\mathit{red}_{\psi_2}(d)) -
      \mathcal{B}^n_{\eta}(\mathit{red}_\eta(d))) \]
    Let $y\in V_D\cup E_D$, then we obtain:
    \begin{eqnarray*}
      && \mathcal{B}^n_{\psi_1}(\mathit{red}_{\psi_1}(d))(y) +
      (\mathcal{B}^n_{\psi_2}(\mathit{red}_{\psi_2}(d))(y) -
      \mathcal{B}^n_{\eta}(\mathit{red}_\eta(d))(y)) \\
      & = & \sum_{\psi_1(x_1)=y} d(\psi_1(x_1)) +
      \big(\sum_{\psi_2(x_2)=y} d(\psi_2(x_2)) - \sum_{\eta(x_0)=y}
      d(\eta(x_0))\big) 
    \end{eqnarray*}
    We distinguish the following cases:
    \begin{itemize}
    \item $y$ has a (unique) preimage $x_1$ under $\psi_1$, but no
      preimage under $\psi_2$. This means that $y$ has no preimage
      under $\eta$ as well. In this case we obtain
      \[ \sum_{\psi_1(x_1)=y} d(\psi_1(x_1)) = d(y), \quad
        \sum_{\psi_2(x_2)=y} d(\psi_2(x_2)) = \sum_{\eta(x_0)=y}
        d(\eta(x_0)) = 0, \] from which the required equality follows.
    \item $y$ has a (unique) preimage $x_2$ under $\psi_2$, but no
      preimage under $\psi_1$. This case is analogous to the previous
      one.
    \item $y$ has a (unique) preimage $x_1$ under $\psi_1$ and a
      (unique) preimage $x_2$ under $\psi_2$. Hence it must also have
      a (unique) preimage $x_0$ under $\eta$ such that
      $\phi_1(x_0) = x_1$, $\phi_2(x_0) = x_2$. In this case we obtain
      \[ \sum_{\psi_1(x_1)=y} d(\psi_1(x_1)) = 
        \sum_{\psi_2(x_2)=y} d(\psi_2(x_2)) = \sum_{\eta(x_0)=y}
        d(\eta(x_0)) = d(y), \]
      yielding the result $d(y) + (d(y)-d(y)) = d(y)$.
    \end{itemize}
  \item[Beck-Chevalley property:] First, observe that since the square
    from Def.~\ref{def:prop-annotations} (Beck-Chevalley property) is
    a pullback, we can assume that the elements (vertices and edges)
    of $A$ are as follows:
    \[ V_A\cup E_A = \{(x_1,x_2)\mid x_1\in V_B\cup E_B, x_2\in
      V_C\cup E_C, \psi_1(x_1) = \psi_2(x_2) \} \]
    Now let $b\colon V_B\cup E_B\to \mathcal{M}_n$ and $x_2\in V_C\cup
    E_C$. Then we have:
    \begin{eqnarray*}
      && \mathcal{A}_{\phi_2}(\mathit{red}_{\phi_1}(b))(x_2) =
      \sum_{\phi_2((x'_1,x'_2))=x_2} b(\phi_1((x'_1,x'_2))) \\
      & = &
      \sum_{\psi_1(x'_1) = \psi_2(x_2)} b(\phi_1((x'_1,x_2))) =
      \sum_{\psi_1(x'_1)=\psi_2(x_2)} b(x_1) = 
      \mathcal{A}_{\psi_1}(b)(\psi_2(x_2)) \\
      & = &
      \mathit{red}_{\psi_2}(\mathcal{A}_{\psi_1}(b))(x_2)
    \end{eqnarray*}
  \item[Isomorphism property:] Assume that
    $\phi\colon X[s_X,s_X]\to Y[s_Y,s_Y]$ is a legal morphism. Then,
    since the standard annotation $s_Y$ is a lower and upper bound,
    every element $Y$ must have exactly one preimage in $X$ under
    $\phi$. This is equivalent to the fact that $\phi$ is an
    isomorphism. \qed
  \end{description}
\end{proof}

\begin{lemma}
  \label{lem:local-annot-properties}
  The local annotation functor from Ex.~\ref{ex:local-annot} satisfies
  the homorphism property and the pushout property for standard
  annotations.
\end{lemma}

\begin{proof}~
  
  \begin{description}
  \item[Homomorphism property:] Assume that $\phi\colon A\to B$ is an
    injective graph morphism.

    We first show that $\mathcal{S}^n_\phi$ preserves the unit, which
    is a map $a\colon V_A\to \mathcal{M}_n$ with $a(v) = 0$ for all
    $v\in V_A$. For $w\in V_B$ we have $\mathcal{S}^n_\phi(a)(w) =
    \bigvee_{\phi(v)=w} a(v)$. Either $w$ has a unique preimage $v$
    with $a(v)=0$ and in this case the result is $0$. Or $w$ has no
    preimage, in which case we have the empty supremum and the result
    is also $0$.

    Next, we show that $\mathcal{S}^n_\phi$ preservers the monoid
    operation: let $a_1,a_2\in V_A\to \mathcal{M}_n$. Then we have
    $\mathcal{S}^n_\phi(a_1+a_2)(w) = \bigvee_{\phi(v)=w}
    (a_1(v)+a_2(v))$. We distinguish two cases:
    \begin{itemize}
    \item Either $w$ has a unique preimage $v$ and then the result is
      \[ a_1(v)+a_2(v) = \bigvee_{\phi(v)=w} a_1(v) +
        \bigvee_{\phi(v)=w} a_2(v) = \mathcal{S}^n_\phi(a_1)(w) +
        \mathcal{S}^n_\phi(a_2)(w)  \]
    \item Or $w$ has no preimage under $\phi$ and we obtain
      \[ 0 = 0+0 = \bigvee_{\phi(v)=w} a_1(v) + \bigvee_{\phi(v)=w}
        a_2(v) = \mathcal{S}^n_\phi(a_1)(w) +
        \mathcal{S}^n_\phi(a_2)(w) \]
    \end{itemize}
    Preservation of subtraction can be shown analogously.
  \item[Pushout property for standard annotations:] In the following
    we will use $\mathit{out}\colon V\to \mathcal{M}_n$ as a function
    that assigns to a vertex $v\in V$ its out-degree, respectively
    $*$ if the out-degree is larger than $n$.

    Assume that we have a pushout as in
    Def.~\ref{def:prop-annotations} (pushout property). We have to
    show that
    \vspace*{-.1cm}
    \[ s_D = \mathcal{S}^n_{\psi_1}(s_B) +
      (\mathcal{S}^n_{\psi_2}(s_C) - \mathcal{S}^n_{\eta}(s_A)) \]
    Now let $w\in V_D$ and we distinguish the following cases:
    \begin{itemize}
    \item $w$ has a (unique) preimage under $\psi_1$, but no preimage
      under $\psi_2$. This means that $w$ has no preimage under $\eta$
      as well. In this case $\mathit{out}(w) = \mathit{out}(v)$ and we
      have:
      \vspace*{-.1cm}
      \[ s_D(w) = \mathit{out}(w) = \mathit{out}(v) = s_B(v) =
        \bigvee_{\psi_1(v) = w} s_B(v) = \mathcal{S}^n_\phi(s_B)(w)
      \]
      In addition $\mathcal{S}^n_{\psi_2}(s_C)(w) = 0$ and
      $\mathcal{S}^n_\eta(s_A)(w) = 0$ and this completes this case.
    \item $w$ has a (unique) preimage under $\psi_2$, but no preimage
      under $\psi_1$. This case is analogous to the previous case.
    \item $w$ has a (unique) preimage $v_1$ under $\psi_1$ and a
      (unique) preimage $v_2$ under $\psi_2$. Hence it must also have a
      (unique) preimage $v_0$ under $\eta$ such that $\phi_1(v_0) =
      v_1$, $\phi_2(v_0) = v_2$.

      Due to the properties of a pushout we have
      $\mathit{out}(w) = \mathit{out}(v_1) +
      (\mathit{out}(v_2)-\mathit{out}(v_0))$. (Note that due to the
      placement of the brackets, the left-hand side equals $*$ if and
      only if the right-hand side equals $*$.)

      Hence we obtain:
      \vspace*{-.2cm}
      \begin{eqnarray*}
        && s_D(w) = \mathit{out}(w) = \mathit{out}(v_1) +
        (\mathit{out}(v_2)-\mathit{out}(v_0)) \\
        & = & s_B(v_1) + (s_C(v_2) - s_A(v_0)) \\
        & = & \bigvee_{\psi_1(v)=w}
        s_B(v) +
        \big(\bigvee_{\psi_2(v)=w} s_C(v) - \bigvee_{\eta(v)=w} s_A(v)\big) \\
        & = & \mathcal{S}^n_{\psi_1}(s_B)(w) + (
        \mathcal{S}^n_{\psi_2}(s_C)(w) - \mathcal{S}^n_{\eta}(s_A)(w))
      \end{eqnarray*} 
    \end{itemize}
  \end{description}
  \qed
\end{proof}

\vspace*{-.4cm}
\begin{lemma}
  \label{lem:path-annot-properties}
  The path annotation functor from Ex.~\ref{ex:path-annot} satisfies
  the homorphism property and the pushout property for standard
  annotations.
\end{lemma}

\begin{proof}~
  
  \begin{description}
  \item[Homomorphism property:] Assume that $\phi\colon A\to B$ is an
    injective graph morphism.

    First observe that $\mathcal{T}_\phi(\emptyset) = \emptyset$.

    Now let $P_0,P_1\in \mathcal{T}(A)$, we have to show that
    $\mathcal{T}_\phi(P_0+P_1) = \mathcal{T}_\phi(P_0) +
    \mathcal{T}_\phi(P_1)$.
    \begin{description}
    \item[($\subseteq$)] Let $(w_0,w_n)\in \mathcal{T}_\phi(P_0+P_1)$
      where $w_0,w_n\in V_B$. Then $w_0,w_n$ have (unique) preimages
      $v_0,v_n\in V_A$ with $\phi(v_0) = w_0$, $\phi(v_n) = w_n$ and
      $(w_0,w_n)\in (P_0+P_1)$. Hence, by definition, there exist
      vertices $v_1,\dots,v_{n-1}\in V_A$ such that
      $(v_i,v_{i+1})\in P_{j_i}$, $j_i\in\{0,1\}$, $j_{i+1} = 1-j_i$,
      $i\in\{0,\dots,n-1\}$. This implies that
      $(\phi(v_i),\phi(v_{i+1})) = (w_i,w_{i+1})\in
      \mathcal{T}_\phi(P_{j_i})$. And hence, by definition of the
      monoid operation $+$ we have $(w_0,w_n)\in (\mathcal{T}_\phi(P_0) +
      \mathcal{T}_\phi(P_1))$.
    \item[($\supseteq$)] Let
      $(w_0,w_n)\in (\mathcal{T}_\phi(P_0) +
      \mathcal{T}_\phi(P_1))$. This implies that there exist
      $w_1,\dots,w_{n-1}\in V_B$ such that $(w_i,w_{i+1})\in
      \mathcal{T}_\phi(P_{j_i})$, $j_i\in\{0,1\}$, $j_{i+1}=1-j_i$,
      $i\in \{0,\dots,n-1\}$.

      Hence there are preimages
      $v^{j_0}_0,v^{j_0}_1,v^{j_1}_1,\dots,v^{j_{n-1}}_{n-1},v^{j_{n-1}}_n\in
      V_A$ of the $w_i$. In particular $\phi(v^j_i) = w_i$ and
      $(v^{j_i}_i,v^{j_i}_{i+1}) \in P_{j_i}$. Since
      $\phi(v^{j_i}_i) = w_i = \phi(v^{j_{i+1}}_i)$ and $\phi$ is
        injective, we can infer $v^{j_i}_i = v^{j_{i+1}}_i$. This means
        that $(v_0,v_n)\in (P_0+P_1)$ by definition of the monoid
        operation $+$. Finally, this implies that $(w_0,w_n) =
        (\phi(v_0),\phi(v_n)) \in \mathcal{T}_\phi(P_0+P_1)$.
    \end{description}
    
    Furthermore $\mathcal{T}_\phi$ trivially preserves subtraction:
    $\mathcal{T}_\phi(P_0-P_1) = \mathcal{T}_\phi(P_0) =
    \mathcal{T}_\phi(P_0) - \mathcal{T}_\phi(P_1)$.
  \item[Pushout property for standard annotations:] Consider the
    pushout of injective graph morphisms depicted below where
    $\eta = \psi_0\circ\phi_0 = \psi_1\circ\phi_1$:
    \[
      \xymatrix{ A \ar@{>->}[r]^{\phi_1} \ar@{>.>}[dr]_{\eta}
        \ar@{>->}[d]_{\phi_0} & B_1
        \ar@{>->}[d]^{\psi_1} \\
        B_0 \ar@{>->}[r]_{\psi_0} & D }
    \]
    We have to show that
    \[ s_D = \mathcal{T}_{\psi_1}(s_{B_0}) +
      (\mathcal{T}_{\psi_2}(s_{B_1}) - \mathcal{T}_\eta(s_A)) =
      \mathcal{T}_{\psi_1}(s_{B_0}) + \mathcal{T}_{\psi_2}(s_{B_1}) \]
    \begin{description}
    \item[($\subseteq$)] Let $(v_0,v_n)\in s_D$. This means that there
      exists a path in graph $D$, consisting of edges
      $e_0,\dots,e_{n-1}\in E_D$, from $v_0$ to $v_n$. In particular
      $s(e_i) = v_i$, $t(e_i) = v_{i+1}$.

      Since $D$ is a pushout, each edge has a preimage in $B_0$ or in
      $B_1$ (or in both). Hence we can group consecutive edges
      according to the origin of their preimages and we can (possibly
      non-uniquely) choose indices $i_0=0,\dots,i_k=n+1$ such that
      $e_{i_\ell},\dots,e_{i_{\ell+1}-1}$ have preimages in
      $B_{j_\ell}$ where $\ell\in\{0,\dots,k-1\}$, $j_\ell\in\{0,1\}$
      and $j_{\ell+1} = 1-j_\ell$.

      Now assume that the preimages of the $e_i$ are
      $f_0,\dots,f_{n-1}\in E_{B_0}\cup E_{B_1}$ where
      $\psi_0(f_i) = e_i$ and $\psi_1(f_i) = e_i$ whenever $\psi_0$
      respectively $\psi_1$ are defined on $f_i$.

      Since $\psi_0,\psi_1$ are injective, the edges
      $f_{i_\ell},\dots,f_{i_{\ell+1}-1}$ form a path in $B_{j_\ell}$,
      hence $(s(f_{i_\ell}),t(f_{i_{\ell+1}-1})) \in
      s_{B_{j_\ell}}$. This implies that
      \begin{eqnarray*}
        && (v_{i_\ell},v_{i_{\ell+1}}) =
        (s(e_{i_\ell}),t(e_{i_{\ell+1}-1})) =
        (s(\psi_{j_\ell}(f_{i_\ell})),t(\psi_{j_\ell}(f_{i_{\ell+1}-1}))) \\
        & = & (\psi_{j_\ell}(s(f_{i_\ell})),\psi_{j_\ell}(t(f_{i_{\ell+1}-1})))
        \in
        \mathcal{T}_{\psi_{j_\ell}}(s_{B_{j_\ell}})
      \end{eqnarray*}
      Hence, by the definition of the monoid operation $+$ we can
      infer that
      $(v_0,v_n)\in \mathcal{T}_{\psi_1}(s_{B_0}) +
      \mathcal{T}_{\psi_2}(s_{B_1})$.
    \item[($\supseteq$)] Let
      $(v_0,v_n)\in \mathcal{T}_{\psi_1}(s_{B_0}) +
      \mathcal{T}_{\psi_2}(s_{B_1})$. Hence there are vertices
      $v_1,\dots,v_{n-1}\in V_D$ such that
      $(v_i,v_{i+1})\in \mathcal{T}(s_{B_{j_i}})$, $j_i\in \{0,1\}$,
      $j_{i+1}=1-j_i$, $i\in \{0,\dots,n-1\}$.

      This means that there are preimages
      $w^{j_0}_0,w^{j_0}_1,w^{j_1}_1,\dots,w^{j_{n-1}}_{n-1},w^{j_{n-1}}_n$
      of the $v_i$. In particular $w^j_i\in V_{B_j}$ and
      $\psi_{j}(w^j_i) = v_i$. Furthermore there exists a path from
      $w^{j_i}_i$ to $w^{j_i}_{i+1}$ in $B_{j_i}$. Hence there must
      also be a path from $v_i = \psi^{j_i}(w^{j_i}_i)$ to
      $v_{i+1} = \psi^{j_i}(w^{j_i}_{i+1})$ in $D$. This in turn
      implies that there is a path from $v_1$ to $v_n$ in $D$ and
      hence $(v_1,v_n)\in D$.  \qed
    \end{description}
  \end{description}
 \end{proof}

\begin{lemma}~
  \label{lem:prop-red}
  \begin{enumerate}
  \item \label{it:prop-red-0} The pushout property for standard
    annotations implies that for every mono $\phi\colon A\rat B$ we
    have $\mathcal{A}_\phi(s_A) \le s_B$.
  \item \label{it:prop-red-1} The adjunction property and the
    Beck-Chevalley property imply that
    $\mathit{red}_\phi(\mathcal{A}_\phi(a)) = a$ for
  %%  $\mathcal{A}_\phi(\mathit{red}_\phi(a)) = a$ for
    $\phi\colon A\rat B$, $a\in \mathcal{A}(A)$.
  \item \label{it:prop-red-2} The pushout property and the adjunction
    property imply the pushout property for standard annotations.
  \item \label{it:prop-red-3} The adjunction property implies
    $\mathit{red}_{\phi\circ \psi} = \mathit{red}_\psi\circ
    \mathit{red}_\phi$ for
    $A\oset{\psi}{\rat} B\oset{\phi}{\rat} C$.
  \end{enumerate}
  \qed
\end{lemma}

\begin{proof}~
  \begin{enumerate}
  \item Consider the pushout below.
    \[
      \xymatrix{ A \ar@{>->}[r]^{\phi} \ar@{>.>}[dr]_{\phi}
        \ar@{>->}[d]_{\mathit{id}_A} & B
        \ar@{>->}[d]^{\mathit{id}_B} \\
        A \ar@{>->}[r]_{\phi} & B }
    \]
    According to the pushout property for standard annotations we have
    \[ s_B = \mathcal{A}_\phi(s_A) + (\mathcal{A}_{\mathit{id}_B}(s_B)
      - \mathcal{A}_\phi(s_A)) \ge \mathcal{A}_\phi(s_A), \] since
    $\mathcal{A}_{\mathit{id}_B}(s_B) - \mathcal{A}_\phi(s_A)\ge 0$
    ($0$ is the bottom element).
  \item First, consider the identity $\mathit{id}_A\colon A\rat A$:
    for $a\in \mathcal{A}(A)$ we have
    $a\le \mathit{red}_{\mathit{id}_A}(\mathcal{A}
    _{\mathit{id}_A}(a)) = \mathit{red}_{\mathit{id}_A}(a)$ and
    similarly
    $\mathit{red}_{\mathit{id}_A}(a) =
    \mathcal{A}_{\mathit{id}_A}(\mathit{red} _{\mathit{id}_A}(a)) \le
    a$. Hence $\mathit{red} _{\mathit{id}_A}(a) = a$.
		
    Since $\phi \colon A \to B$ is a mono, the following diagram is a
    pullback.
    \[
      \xymatrix{ A \ar@{>->}[r]^{\mathit{id}_A}
        \ar@{>->}[d]_{\mathit{id}_A} & A
        \ar@{>->}[d]^{\phi} \\
        A \ar@{>->}[r]_{\phi} & B \itul{\rm\footnotesize\color{gray} 
        (PB)} }
    \]
    From the Beck-Chevalley property it follows that
    \[ \mathit{red}_\phi(\mathcal{A}_\phi(a)) =
      \mathit{red}_{\mathit{id}_A}(\mathcal{A}_{\mathit{id}_A}(a))~=~a. \]
  \item Consider a pushout of $A,B,C,D$ as in the pushout property for
    standard annotations with
    $\eta = \psi_1\circ\phi_1 = \psi_2\circ\phi_2$. Due to the pushout
    property and the adjunction property we have
    \begin{eqnarray*}
      s_D & = & \mathcal{A}_{\psi_1}(\mathit{red}_{\psi_1}(s_D)) +
      (\mathcal{A}_{\psi_2}(\mathit{red}_{\psi_2}(s_D)) -
      \mathcal{A}_{\eta}(\mathit{red}_{\eta}(s_D))) \\
      & = & \mathcal{A}_{\psi_1}(s_B) + (\mathcal{A}_{\psi_2}(s_C) -
      \mathcal{A}_{\eta}(s_A))
    \end{eqnarray*}
  \item We have to show that $\mathit{red}_{\phi\circ \psi}$,
    $\mathit{red}_\psi\circ \mathit{red}_\phi$ are both left adjoints
    of $\mathcal{A}_{\phi\circ \psi}$, then the result follows from
    the fact that adjoints are unique. This is obvious for
    $\mathit{red}_{\phi\circ \psi}$ 
    and in the other case we obtain
 for $c\in \mathcal{A}(C)$:
    \begin{eqnarray*}
      \mathcal{A}_{\phi\circ
        \psi}(\mathit{red}_\psi(\mathit{red}_\phi(c))) & = &
      \mathcal{A}_{\phi}(\mathcal{A}_\psi(\mathit{red}_\psi(
      \mathit{red}_\phi(c)))) \\
      & \le & \mathcal{A}_{\phi}(\mathit{red}_\phi(c)) \\
      & \le & c
    \end{eqnarray*}
    and similarly for the other inequality.  \qed
  \end{enumerate}
 \end{proof}

\subsection{Abstract Rewriting of Annotated Objects}

\begin{proposition_for}{prop:morphism-with-mult-into-mat}{\ (Annotated
    rewritable materialization is terminal).}
  \annotMatMorphism
\end{proposition_for}

\begin{proof}
  The existence of the underlying arrow ${\zeta_A}$ follows from the
  fact that $L\rat \rmat{\phi}{\phi_L} \to A$ is the rewritable
  materialization (see
  Def.~\ref{def:materialization-category-rewritable}). This makes the
  left-hand square a pullback. We show that there exists a pair
  $(a'_1,a'_2)\in M$ (for $M$ as in
  Def.~\ref{def:annotated-rewritable-materialization}) for which
  $a'_1\le \mathcal{A}_{\zeta_A}(s_X) \le a'_2$.
	
  It holds that
  $\mathcal{A}_{\psi'}(\mathcal{A}_{\zeta_A}(s_X)) =
  \mathcal{A}_{\psi}(s_X) \ge a_1$ and
  $\mathcal{A}_{\psi'}(\mathcal{A}_{\zeta_A}(s_X)) \le a_2$.
  Furthermore
    $\mathcal{A}_{n_L}(s_L) =
    \mathcal{A}_{\zeta_A}(\mathcal{A}_{m_L}(s_L)) \le
    \mathcal{A}_{\zeta_A}(s_X)$ (using functoriality,
    Lem.~\ref{lem:prop-red}\ref{it:prop-red-0} and monotonicity).
  Then either
  $(\mathcal{A}_{\zeta_A}(s_X),\mathcal{A}_{\zeta_A}(s_X))\in M$ or it
  is subsumed by another, maximal, pair $(a'_1,a'_2)\in M$. In both
  cases this is the desired pair of annotations.\qed
\end{proof}

\begin{proposition_for}{prop:soundness}{\ (Soundness for $\leadsto$).}
  \soundnessProp
\end{proposition_for}

\begin{proof}
  Since $X\oset{p,m_L}{\Longrightarrow} Y$ we have that
  $(L\oset{m_L}{\rat} X) \oset{p}{\Rightarrow} (R\oset{m_R}{\rat}
  Y)$ for some co-match $m_R$. We set $\phi = \psi\circ m_L$ and
  Corollary~\ref{cor:rewriting-materializations} implies that
  $(R\oset{m_R}{\rat} Y) \in \mathcal{L}(R\oset{n_R}{\rat} B)$ where
  $(L \oset{n_L}{\rat} \rmat{\phi}{\phi_L}) \oset{p}{\Rightarrow}
  (R\oset{n_R}{\rat} B)$ and ${\rmat{\phi}{\phi_L}}$ is the rewritable
  materialization with
  $L\oset{n_L}{\rat} \rmat{\phi}{\phi_L} \oset{\psi'}{\to} A$ (such that
  $\psi'\circ\zeta_A = \psi$). This situation can be summarized in the
  diagram from the proof of
  Prop.~\ref{prop:rewriting-materializations} which is depicted below
  in a simplified form, but with added annotations.
  \[
    \xymatrix{
      L[s_L,s_L] \ar@{>->}[d]^{m_L} \ar@{>->}@/_2pc/[dd]_{n_L}
      & I[s_I,s_I] \ar@{>->}[l]_{\phi_L}
      \ar@{>->}[r]^{\phi_R} \ar@{>->}[d]^{m_I}  & 
      R[s_R,s_R] 
      \ar@{>->}[d]_{m_R} \ar@{>->}@/^2pc/[dd]^{n_R} \\
      X[s_X,s_X] \ar[d]^{\zeta_A} & Z[s_Z,s_Z] \ar@{>->}[l]_{\phi_X} 
      \ar@{>->}[r]^{\phi_Y} \ar[d]^{\zeta_C} & Y[s_{Y},s_{Y}] 
      \ar[d]_{\zeta_B} \\
      {\rmat{\phi}{\phi_L}}[a'_1,a'_2] & C[c_1,c_2] \ar@{>->}[l]_(.4){\phi_A}
      \ar@{>->}[r]^{\phi_B} & B[b_1,b_2]
    }  
  \]
  Due to Prop.~\ref{prop:morphism-with-mult-into-mat} there exists a
  pair of annotations $(a'_1,a'_2)\in M$  
  and a legal arrow $\zeta_A\colon X[s_X,s_X]\to 
  \rmat{\phi}{\phi_L}[a'_1,a'_2]$.  Furthermore
  we assume $c_1,c_2,b_1,b_2$ as in Def.~\ref{def:abstract-rewriting}.
	
  It is left to show that $\zeta_C$ and in particular $\zeta_B$ are
  legal morphisms.
	
  First, in order to show that $\zeta_C$ is legal, we observe that,
  due to functoriality, the homomorphism property and the pushout
  property for standard annotations, we have:
  \begin{eqnarray*}
    && \mathcal{A}_{\phi_A}(\mathcal{A}_{\zeta_C}(s_Z)) +
    (\mathcal{A}_{n_L}(s_L) - \mathcal{A}_{n_L\circ \phi_L}(s_I)) \\
    & = & \mathcal{A}_{\zeta_A}(\mathcal{A}_{\phi_X}(s_Z)) +
    (\mathcal{A}_{\zeta_A}(\mathcal{A}_{m_L}(s_L)) -
    \mathcal{A}_{\zeta_A}(\mathcal{A}_{m_L\circ \phi_L}(s_I))) \\
    & = & \mathcal{A}_{\zeta_A}(\mathcal{A}_{\phi_X}(s_Z) +
    (\mathcal{A}_{m_L}(s_L) - \mathcal{A}_{m_L\circ \phi_L}(s_I))) \\
    & = & \mathcal{A}_{\zeta_A}(s_X) 
  \end{eqnarray*}
  Since $a'_1\le \mathcal{A}_{\zeta_A}(s_X) \le a'_2$ we know from
  Def.~\ref{def:abstract-rewriting} that there is a (maximal)
  annotation $(c_1,c_2)$ satisfying the respective inequalities such
  that $c_1\le \mathcal{A}_{\zeta_C}(s_Z) \le c_2$, which implies that
  $\zeta_C$ is legal.
  
  Second, to show that $\zeta_B$ is legal, we observe that due to the
  pushout property for standard annotations, the homomorphism property
  and functoriality:
  \begin{eqnarray*}
    \mathcal{A}_{\zeta_B}(s_Y) & = &
    \mathcal{A}_{\zeta_B}(\mathcal{A}_{\phi_Y}(s_Z) +
    (\mathcal{A}_{m_R}(s_R) - \mathcal{A}_{m_R\circ \phi_R}(s_I))) \\
    & = &
    \mathcal{A}_{\zeta_B}(\mathcal{A}_{\phi_Y}(s_Z)) +
    (\mathcal{A}_{\zeta_B}(\mathcal{A}_{m_R}(s_R)) -
    \mathcal{A}_{\zeta_B}(\mathcal{A}_{m_R\circ \phi_R}(s_I))) \\
    & = &
     \mathcal{A}_{\phi_B}(\mathcal{A}_{\zeta_C}(s_Z)) +
     (\mathcal{A}_{n_R}(s_R) -
    \mathcal{A}_{n_R \circ \phi_R}(s_I)) \\
%    (\mathcal{A}_{\zeta_B}(\mathcal{A}_{m_R}(s_R)) -
%    \mathcal{A}_{\zeta_B}(\mathcal{A}_{n_R}(s_I))) \\
 \end{eqnarray*}
  Since $\zeta_C$ is legal and we have
  $c_1\le \mathcal{A}_{\zeta_C}(s_Z) \le c_2$, we obtain from the
  definition of $b_1,b_2$ and monotonicity that
  $b_1\le \mathcal{A}_{\zeta_B}(s_Y)\le b_2$.  \qed
\end{proof}

\begin{proposition_for}{prop:soundness-variant}{\ (Soundness for 
$\hookrightarrow$)).}
  \soundnessPropVariant
\end{proposition_for}

\begin{proof}
  We modify the proof of Prop.~\ref{prop:morphism-with-mult-into-mat},
  on which Prop.~\ref{prop:soundness} relies.  We have to show that
  there always exists a pair of annotations $(a'_1,a'_2)\in M$ for
  which we have a legal arrow
  $\zeta_A\colon X[s_X,s_X]\to \rmat{\phi}{\phi_L}[a'_1,a'_2]$. (The rest of
  the proof of Prop.~\ref{prop:soundness} proceeds as before.)
	
  As in Prop.~\ref{prop:morphism-with-mult-into-mat} we show that
  $(\mathcal{A}_{\zeta_A}(s_X),\mathcal{A}_{\zeta_A}(s_X))$ is an
  annotation $(a'_1,a'_2)$ which satisfies
  $a_1\le \mathcal{A}_\psi(a'_1)$ and $\mathcal{A}_\psi(a'_2)\le
  a_2$. Since the square consisting of $\mathit{id}_L,m_L,\zeta_A,n_L$
  is a pushout, we can use the Beck-Chevally property and the
  adjunction property to prove that
  $\mathit{red}_{n_L}(\mathcal{A}_{\zeta_A}(s_X)) =
  \mathcal{A}_{\mathit{id}_L}(\mathit{red}_{m_L}(s_X)) =
  \mathit{red}_{m_L}(s_X) = s_L$. Hence either
  $(\mathcal{A}_{\zeta_A}(s_X),\mathcal{A}_{\zeta_A}(s_X))$ or an
  annotation subsuming it is contained in the set $M$ of
  Def.~\ref{def:abstract-rewriting-variant}.\qed
\end{proof}

\begin{proposition_for}{prop:completeness}{\ (Completeness for 
$\hookrightarrow$).}
  \completenessProp
\end{proposition_for}

\begin{proof}
  Since there is a rewriting step from $A[a_1,a_2]$ to $B[b_1,b_2]$ we
  obtain $\rmat{\phi}{\phi_L}$ as the materialization (with
  $L\oset{n_L}{\rat} \rmat{\phi}{\phi_L} \oset{\psi'}{\to} A$ where
  $\phi = \psi'\circ n_L$) and the following two pushouts below.
  \[
    \xymatrix{ L[s_L,s_L] \ar@{>->}[d]_{n_L} & I[s_I,s_I]
      \ar@{>->}[l]_{\phi_L} \ar@{>->}[r]^{\phi_R} \ar@{>->}[d]^{n_I} &
      R[s_R,s_R] \ar@{>->}[d]^{n_R} & Y[s_Y,s_Y]
      \ar@{>->}[dl]^{\zeta_B}  \\
      \rmat{\phi}{\phi_L}[a'_1,a'_2] & C[c_1,c_2] \ar@{>->}[l]_(.4){\phi_A}
      \ar@{>->}[r]^{\phi_B} & B[b_1,b_2] }
  \]
  Furthermore $(a'_1,a'_2)\in M$ and
  \[ a'_1 \le \mathcal{A}_{\phi_A}(c_1) + (\mathcal{A}_{n_L}(s_L) -
    \mathcal{A}_{n_L\circ \phi_L}(s_I)) \quad
    \mathcal{A}_{\phi_A}(c_2) + (\mathcal{A}_{n_L}(s_L) -
    \mathcal{A}_{n_L\circ \phi_L}(s_I)) \le a'_2 \]
  \[ b_i = \mathcal{A}_{\phi_B}(c_i) + (\mathcal{A}_{n_R}(s_R) -
    \mathcal{A}_{n_R\circ \phi_R}(s_I)) \qquad \mbox{for
      $i\in\{1,2\}$} \] In addition $\zeta_B$ is a legal arrow that
  witnesses $Y\in \mathcal{L}(B[b_1,b_2])$, in particular
  $b_1\le \mathcal{A}_{\zeta_B}(s_{Y})\le b_2$.
	
  \begin{itemize}
  \item We first observe that there is a unique maximal pair
    $(c_1,c_2)$ satisfying the above inequalities, in particular
    $c_i = \mathit{red}_{\phi_A}(a'_i)$. We have
    \begin{align*}
      & a'_i \\
      & = \qquad \mbox{[PO property]} \\
      & \mathcal{A}_{\phi_A}(\mathit{red}_{\phi_A}(a'_i)) +
      (\mathcal{A}_{n_L}(\mathit{red}_{n_L}(a'_i)) -
      \mathcal{A}_{n_L\circ \phi_L}(\mathit{red}_{n_L\circ
        \phi_L}(a'_i)) \\ 
      & = \qquad \mbox{[$\mathit{red}_{n_L}(a'_i) = s_L$, Def. of $M$
        (from Def.~\ref{def:abstract-rewriting-variant})]} \\
      & \mathcal{A}_{\phi_A}(\mathit{red}_{\phi_A}(a'_i)) +
      (\mathcal{A}_{n_L}(s_L) - \mathcal{A}_{n_L\circ \phi_L}(s_I)) 
    \end{align*}
    Furthermore let $c_1$ be an annotation satisfying the above
    inequality. Then we obtain:
    \begin{align*}
      & \mathit{red}_{\phi_A}(a'_1) \\
      & \le \qquad \mbox{[Mon.]} \\
      & \mathit{red}_{\phi_A}(\mathcal{A}_{\phi_A}(c_1) +
      (\mathcal{A}_{n_L}(s_L) -
      \mathcal{A}_{n_L\circ \phi_L}(s_I))) \\
      & = \qquad \mbox{[Adj. prop.]} \\
      & \mathit{red}_{\phi_A}(\mathcal{A}_{\phi_A}(c_1)) +
      (\mathit{red}_{\phi_A}(\mathcal{A}_{n_L}(s_L)) -
      \mathit{red}_{\phi_A}(\mathcal{A}_{n_L\circ \phi_L}(s_I)))
      \\
      & = \qquad \mbox{[Lem.~\ref{lem:prop-red}\ref{it:prop-red-1}]} \\
      & c_1 + (\mathit{red}_{\phi_A}(\mathcal{A}_{n_L}(s_L)) -
      \mathit{red}_{\phi_A}(\mathcal{A}_{n_L\circ \phi_L}(s_I)))
      \\
      & = \qquad \mbox{[Funct.]} \\
      & c_1 + (\mathit{red}_{\phi_A}(\mathcal{A}_{n_L}(s_L)) -
      \mathit{red}_{\phi_A}(\mathcal{A}_{\phi_A\circ n_I}(s_I))) 
      \\
      & = \qquad \mbox{[Lem.~\ref{lem:prop-red}\ref{it:prop-red-1}]} \\
      & c_1 + (\mathit{red}_{\phi_A}(\mathcal{A}_{n_L}(s_L)) -
      \mathcal{A}_{n_I}(s_I)) \\
      & = \qquad \mbox{[Beck-Chevalley]} \\
      & c_1 + (\mathcal{A}_{n_I}(\mathit{red}_{\phi_L}(s_L)) -
      \mathcal{A}_{n_I}(s_I)) \\
      & = \qquad \mbox{[Adj. prop.]} \\
      & c_1 + (\mathcal{A}_{n_I}(s_I) - \mathcal{A}_{n_I}(s_I)) \\
      & = \qquad \mbox{[Subtr. well-behaved]} \\
      & c_1
    \end{align*}
    And similarly $\mathit{red}_{\phi_A}(a'_2) \ge c_2$ for an
    annotation $c_2$ satisfying the above equality.
  \item We will next show that there exists a mono $m_R\colon R\rat Y$
    such that
    $(R\oset{m_R}{\rat} Y)\in \mathcal{L}(R\oset{n_R}{\rat}
    B)$.  We do this by taking the pullback of the arrows
    $n_R,\zeta_B$, obtaining the following diagram.
    \[
      \xymatrix{ R'[s_R',s_R'] \ar@{>->}[d]_{m_R} \ar[r]^\iota &
        R[s_R,s_R]
        \ar@{>->}[d]^{n_R} \\
        Y[s_Y,s_Y] \ar[r]^{\zeta_B} & B[b_1,b_2] }
    \]
    According to the Beck-Chevalley property we have
    $$\mathcal{A}_\iota(s_{R'}) = \mathcal{A}_\iota(\mathit{red}_{m_R}(s_{Y}))
    = \mathit{red}_{n_R}(\mathcal{A}_{\zeta_B}(s_{Y})).$$ We know that
    $b_1\le \mathcal{A}_{\zeta_B}(s_{Y})\le b_2$ since $\zeta_B$ is
    legal and it follows with monotonicity of $\mathit{red}_{n_R}$
    that
    $$\mathit{red}_{n_R}(b_1) \le \mathcal{A}_\iota(s_{R'}) \le
    \mathit{red}_{n_R}(b_2).$$ Next, we show that
    $\mathit{red}_{n_R}(b_1) = \mathit{red}_{n_R}(b_2) = s_R$:
    \begin{align*}
      & \mathit{red}_{n_R}(b_i) \\
      & = \qquad \mbox{[Def.]} \\
      & \mathit{red}_{n_R}(\mathcal{A}_{\phi_B}(c_i) +
      (\mathcal{A}_{n_R}(s_R) - \mathcal{A}_{n_R\circ
        \phi_R}(s_I))) \\
      & = \qquad \mbox{[Adj. prop.]} \\
      & \mathit{red}_{n_R}(\mathcal{A}_{\phi_B}(c_i)) +
      (\mathit{red}_{n_R}(\mathcal{A}_{n_R}(s_R)) -
      \mathit{red}_{n_R}(\mathcal{A}_{n_R\circ\phi_R}(s_I)))
      \\
      & = \qquad \mbox{[Lem.~\ref{lem:prop-red}\ref{it:prop-red-1}]} \\
      & \mathit{red}_{n_R}(\mathcal{A}_{\phi_B}(c_i)) +
      (s_R - \mathcal{A}_{\phi_R}(s_I)) \\
      & = \qquad \mbox{[Beck-Chevalley]} \\
      & \mathcal{A}_{\phi_R}(\mathit{red}_{n_I}(c_i)) + (s_R -
      \mathcal{A}_{\phi_R}(s_I)) \\
      & = \qquad \mbox{[Adj. prop.]} \\
      &
      \mathcal{A}_{\phi_R}(\mathit{red}_{n_I}(\mathit{red}_{\phi_A}(a'_i)))
      + (s_R -
      \mathcal{A}_{\phi_R}(s_I)) \\
      & = \qquad \mbox{[Lem.~\ref{lem:prop-red}\ref{it:prop-red-3}]} \\
      &
      \mathcal{A}_{\phi_R}(\mathit{red}_{\phi_L}(\mathit{red}_{n_L}(a'_i)))
      + (s_R - \mathcal{A}_{\phi_R}(s_I)) \\
      & = \qquad \mbox{[$\mathit{red}_{n_L}(a'_i) = s_L$, Def. of
        $M$]} \\
      & \mathcal{A}_{\phi_R}(\mathit{red}_{\phi_L}(s_L)) + (s_R -
      \mathcal{A}_{\phi_R}(s_I)) \\
      & = \qquad \mbox{[Adj. prop.]} \\
      & \mathcal{A}_{\phi_R}(s_I) + (s_R -
      \mathcal{A}_{\phi_R}(s_I)) &  \\
      & = \qquad \mbox{[Subtr. well-behaved]} \\
      & s_R
    \end{align*}
    The last equality holds since $\mathit{red}_{\phi_R}(s_R) = s_I$ and
    hence
    $\mathcal{A}_{\phi_R}(s_I) =
    \mathcal{A}_{\phi_R}(\mathit{red}_{\phi_R}(s_R)) \le s_R$ (due to
    the adjunction property).
    
    This means that $\iota$ is a legal arrow and we can
    infer from the isomorphism property that it is an iso,
    without loss of generality we can assume that it is the
    identity.
    
    Hence
    $(R\oset{m_R}{\rat} Y)\in \mathcal{L}(R\oset{n_R}{\rat}
    B)$.
  \item Since
    $(L\oset{n_L}{\rat}\rmat{\phi}{\phi_L})\oset{p}{\Rightarrow}
    (R\oset{n_R}{\rat} B)$ we can infer from
    Corollary~\ref{cor:rewriting-materializations} that there exists a
    match $m_L\colon L\rat X$ where
    $(L\oset{m_L}{\rat} X)\in
    \mathcal{L}(L\oset{n_L}{\rat}\rmat{\phi}{\phi_L})$ and
    $(L\oset{m_L}{\rat} X)\oset{p}{\Rightarrow} (R\oset{m_R}{\rat}
    Y)$. This situation can be summarized in the diagram from the
    proof of Prop.~\ref{prop:rewriting-materializations} which is
    depicted below with added annotations.
    \[
      \xymatrix{ L[s_L,s_L] \ar@{>->}[d]^{m_L}
        \ar@{>->}@/_2pc/[dd]_{n_L} & I[s_I,s_I]
        \ar@{>->}[l]_{\phi_L} \ar@{>->}[r]^{\phi_R}
        \ar@{>->}[d]^{m_I} & R[s_R,s_R]
        \ar@{>->}[d]_{m_R} \ar@{>->}@/^2pc/[dd]^{n_R} \\
        X[s_X,s_X] \ar[d]^{\zeta_A} & Z[s_Z,s_Z]
        \ar@{>->}[l]_{\phi_X} \ar@{>->}[r]^{\phi_Y}
        \ar[d]^{\zeta_C} & Y[s_{Y},s_{Y}]
        \ar[d]_{\zeta_B} \\
        {\rmat{\phi}{\phi_L}}[a'_1,a'_2] & C[c_1,c_2]
        \ar@{>->}[l]_(.4){\phi_A} \ar@{>->}[r]^{\phi_B} &
        B[b_1,b_2] }
    \]
    It is left to show that $\zeta_C$ and in particular
    $\zeta_A$ are legal.
  \item For $\zeta_C$ we show that, due to the adjunction property,
    the Beck-Chevally property and monotonicity:
    \[ \mathcal{A}_{\zeta_C}(s_C) =
      \mathcal{A}_\zeta(\mathit{red}_{\phi_Y}(s_{Y})) =
      \mathit{red}_{\phi_B}(\mathcal{A}_{\zeta_B}(s_{Y}))
      \ge \mathit{red}_{\phi_B}(b_1)\] and similarly
    $\mathcal{A}_\zeta(s_C) =
    \mathit{red}_{\phi_B}(\mathcal{A}_{\zeta_A}(s_{Y}))
    \le \mathit{red}_{\phi_B}(b_2)$.
    
    Therefore,
    $\mathit{red}_{\phi_B}(b_1) \le \mathcal{A}_\zeta(s_C) \le
    \mathit{red}_{\phi_B}(b_2)$ holds and it is only left to show that
    $\mathit{red}_{\phi_B}(b_i) = c_i$ for $i \in \{1,2\}$. In
    particular, we have to show that
    $\mathit{red}_{\phi_B}(\mathcal{A}_{\phi_B}(c_i) +
    (\mathcal{A}_{n_R}(s_R) - \mathcal{A}_{n_L\circ \phi_R}(s_I)) =
    c_i$ and this is analogous to the proof concerning the left-hand
    square above.
  \item Now, we show that $\zeta_A$ is legal:
    \begin{align*}
      & \mathcal{A}_{\zeta_A}(s_X) \\
      & = \qquad \mbox{[PO prop. for std. ann.]} \\
      & \mathcal{A}_{\zeta_A}(\mathcal{A}_{\phi_X}(s_Z) +
      (\mathcal{A}_{m_L}(s_L) -
      \mathcal{A}_{m_L\circ\phi_L}(s_I))) \\
      & = \qquad \mbox{[Homom. prop.]} \\
      & \mathcal{A}_{\zeta_A}(\mathcal{A}_{\phi_X}(s_Z))
      + (\mathcal{A}_{\zeta_A}(\mathcal{A}_{m_L}(s_L)) -
      \mathcal{A}_{\zeta_A}(\mathcal{A}_{m_L\circ\phi_L}(s_I)))
      \\
      & = \qquad \mbox{[Funct.]} \\
      & \mathcal{A}_{\phi_A}(\mathcal{A}_{\zeta_C}(s_Z))
      + (\mathcal{A}_{n_L}(s_L) -
      \mathcal{A}_{n_L\circ\phi_L}(s_I)) \\
      & \ge \qquad \mbox{[Mon.]} \\
      & \mathcal{A}_{\phi_A}(c_1) +
      (\mathcal{A}_{n_L}(s_L) -
      \mathcal{A}_{n_L\circ\phi_L}(s_I)) \\
      & \ge \qquad \mbox{[Def. of $c_1$]} \\
      & a'_1 
    \end{align*}
    Similarly $\mathcal{A}_{\zeta_A}(s_X) \le a'_2$.
  \end{itemize}
  Hence we have found $m_L\colon L\rat X$ such that
  $X\in \mathcal{L}(\rmat{\phi}{\phi_L}[a'_1,a'_2])$ (witnessed by $\zeta_A$)
  and $X\oset{p,m_L}{\Longrightarrow} Y$. Since, due to the
  materialization $\psi'\colon \rmat{\phi}{\phi_L}[a'_1,a'_2]\to A[a_1,a_2]$ is
  a legal arrow, we have that $X\in \mathcal{L}(A[a_1,a_2])$,
  witnessed by $\psi := \psi'\circ \zeta_A$ and it holds that
  $\psi\circ m_L = \psi'\circ \zeta_A\circ m_L = \psi'\circ n_L =
  \phi$.  \qed
\end{proof}

\begin{corollary_for}{cor:strongest-postcondition}{\ (Strongest
    post-condition).}
  \corollaryStrongestPost
\end{corollary_for}

\begin{proof}
  Straightforward from Propositions~\ref{prop:soundness-variant}
  and~\ref{prop:completeness}. \qed
\end{proof}

\end{document}

%% file: defs.tex
\usepackage{amssymb,latexsym}
\usepackage{xifthen}

\usepackage[arrow,matrix,frame,curve,cmtip]{xy}\CompileMatrices
\newcommand{\Sct}{Sec}
\newdir{ >}{*{}!/-.2em/\dir{>}}
\newcommand{\dotarrow}{% to be used in math mode 
   \mathrel{\ooalign{\hss\raise1ex\hbox{\scalebox{1.25}{\normalfont .}}%
   \kern0.35ex\hss\cr$\rightarrow$}}}
\newcommand{\bx}[1]{\phantom{\big(}#1{\phantom{\big)}}}

\newcommand{\A}{}
\newcommand{\matcat}[2][]{%
\ensuremath{\mathbf{Mat}_{#2}^{\ifthenelse{\isempty{#1}}{}{{#1}}%
}}}
\newcommand{\mat}[1]{\ensuremath{\langle #1 \rangle}}
\newcommand{\rmat}[2]{\ensuremath{\llangle #1,#2 \rrangle}}
   
\usepackage{graphicx,epsfig,color}

\usepackage{amsmath}
\usepackage{wrapfig}
\usepackage{etoolbox}
\usepackage{xcolor}

\usepackage{tikz}
\usetikzlibrary{shapes.geometric}
\usetikzlibrary{calc,arrows,fit,backgrounds,patterns}
\usetikzlibrary{arrows.meta}
\usetikzlibrary{positioning}
\usetikzlibrary{decorations.pathreplacing}

\newcommand{\itul}[1]{{\save[]+<-0.7cm,0.65cm>*\txt{#1}\restore}}
\newcommand{\itull}[1]{{\save[]+<-0.9cm,0.65cm>*\txt{#1}\restore}}

\definecolor{dgreen}{rgb}{0,0.6,0}
\newcommand{\dgreen}[1]{{\color{dgreen}#1}}
\newcommand{\blue}[1]{{\color{blue}#1}}
\newcommand{\red}[1]{{\color{red}#1}}

\makeatletter
\newcommand\@optsub[2]{
  \ifstrempty{#2}{%
    #1%
  }{%
    #1_{#2}%
  }%
}
\newcommand\@optsup[2]{
  {#1}%
  \ifstrempty{#2}{}{^{#2}}%
}
\newcommand\@optapp[2]{
  {#1}%
  \ifstrempty{#2}{}{(#2)}%
}

\newcommand{\onemedunit}{2em}

\tikzset{
  mono/.style={>->},
  gnode/.style={circle,fill=black,inner sep=0mm,minimum 
  size=1.75mm,font=\scriptsize,text=white},
  gedge/.style={->,>=latex},
  arlab/.style={inner sep=1pt,font=\scriptsize},
  glab/.style={inner sep=1pt,font=\scriptsize},
  hyperedge/.style={shape=rectangle,draw,inner sep=0,
    minimum width=1cm,minimum height=.4cm},
  point/.style={shape=circle,inner sep=0,fill=black,
    minimum height=1pt,minimum width=1pt},
  gnoder/.style={circle,fill=red,inner sep=0mm,minimum 
  size=2mm,font=\scriptsize,text=white},
  gnodeb/.style={circle,draw,fill=white,inner sep=0mm,minimum 
  size=1.8mm,font=\scriptsize,text=white},
  gnodeblue/.style={circle,fill=blue,inner sep=0mm,minimum 
  size=1.8mm,font=\scriptsize,text=white},
  gnodew/.style={circle,fill=white,inner sep=0mm,minimum 
  size=2mm,font=\scriptsize,text=white},
}

\newcommand\graphbox[2][grbox]{
  \begin{pgfonlayer}{background}
    \node[fit=#2] (#1) {} ;
    \fill[black!20,rounded corners=2mm,postaction={draw,black}] 
          (#1.north west) -- (#1.north east) --
          (#1.south east) -- (#1.south west) -- cycle ;
  \end{pgfonlayer}
}

\newcommand\graphboxwhite[2][grbox]{
  \begin{pgfonlayer}{background}
    \node[fit=#2] (#1) {} ;
    \fill[white!20,rounded corners=2mm,postaction={draw,black}] 
          (#1.north west) -- (#1.north east) --
          (#1.south east) -- (#1.south west) -- cycle ;
  \end{pgfonlayer}
}

\newcommand\ghostgraphbox[2][grbox]{
  \begin{pgfonlayer}{background}
    \node[fit=#2] (#1) {} ;
    \fill[white!20,rounded corners=2mm,postaction={draw,white}] 
          (#1.north west) -- (#1.north east) --
          (#1.south east) -- (#1.south west) -- cycle ;
  \end{pgfonlayer}
}

\newcommand\interfacebox[2][ifbox]{
  \begin{pgfonlayer}{background}
    \node[fit=#2] (#1) {} ;
    \draw[dashed,rounded corners=2mm] 
          (#1.north west) -- (#1.north east) --
          (#1.south east) -- (#1.south west) -- cycle ;
  \end{pgfonlayer}
}

\newcommand\substy[2]{
  ($ (0,0)!#1!(1,0) + (0,#2) $)
}

\newcommand{\boundellipse}[3]% center, xdim, ydim
{(#1) ellipse (#2 and #3)
}

\renewcommand{\phi}{\varphi}

\newcommand{\lat}{\leftarrowtail}
\newcommand{\rat}{\rightarrowtail}

\makeatletter
\providecommand*{\cupdot}{%
  \mathbin{%
    \mathpalette\@cupdot{}%
  }%
}
\newcommand*{\@cupdot}[2]{%
  \ooalign{%
    $\m@th#1\cup$\cr
    \hidewidth$\m@th#1\cdot$\hidewidth
  }%
}

\newcommand{\GR}[1][]{{\ensuremath{\mathbf{Graph}}}}

\newcommand{\C}[1][]{{\ensuremath{\mathbf{C}}}}
\newcommand{\CMono}[1][]{{\ensuremath{\mathbf{CMono}}}}
\newcommand{\D}[1][]{{\ensuremath{\mathbf{D}}}}

\newcommand{\Mon}{{\ensuremath{\mathbf{Mon}}}}

\newcommand\sLab[1][]{\@optsub{\mathit{\ell}}{#1}}
\newcommand\sSrc[1][]{\@optsub{\mathit{src}}{#1}}
\newcommand\sTgt[1][]{\@optsub{\mathit{tgt}}{#1}}

\newcommand\fLab[2][]{\sLab[#1](#2)}
\newcommand\fSrc[2][]{\sSrc[#1](#2)}
\newcommand\fTgt[2][]{\sTgt[#1](#2)}

\newcommand\sFlower[1][]{\@optsub{\mbox{\ding{82}}}{#1}}
\newcommand\sFlowerM[1][]{\@optsub{\mathit{fl}}{#1}}

\newcommand{\mytilde}{{\raise.17ex\hbox{$\scriptstyle\mathtt{\sim}$}}\xspace}

\newcommand{\oset}[2]{
  \mathrel{\stackrel{#1}{\mbox{\raisebox{-0.9pt}{\ensuremath{#2}}}}}}

\newcommand{\bkchange}[1]{\red{#1}}
\renewcommand{\bkchange}[1]{#1}
\newcommand{\dnchange}[1]{\dgreen{#1}}
\renewcommand{\dnchange}[1]{#1}

%% file: theorem-eng.tex
\newenvironment{corollary_for}[2]{\noindent{\bf Corollary~\ref{#1}#2}\it}{}
\newenvironment{proposition_for}[2]{\noindent{\bf Proposition~\ref{#1}#2}\it}{}

%% file: ex-edge.tex
\begin{picture}(0,0)%
\includegraphics{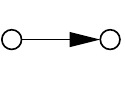}%
\end{picture}%
\setlength{\unitlength}{4144sp}%
\begingroup\makeatletter\ifx\SetFigFont\undefined%
\gdef\SetFigFont#1#2#3#4#5{%
  \reset@font\fontsize{#1}{#2pt}%
  \fontfamily{#3}\fontseries{#4}\fontshape{#5}%
  \selectfont}%
\fi\endgroup%
\begin{picture}(556,399)(3818,-1732)
\end{picture}%